\documentclass[prd,aps,twocolumn,a4paper,floatfix]{revtex4-1}
\usepackage{hyperref}
\usepackage{color}

\newcommand{\hateq}{\,\,\hat{=}\,\,}
\newcommand{\re}{\mbox{Re}}
\newcommand{\im}{\mbox{Im}}
\newcommand{\Lie}{{\cal L}}

\newcommand\p{\partial}
\newcommand\talpha{\tilde{\alpha}}
\newcommand\tbeta{\tilde{\beta}}
\newcommand\tgamma{\tilde{\gamma}}
\newcommand\tT{\tilde{\Theta}}
\newcommand\tZ{\tilde{Z}}
\newcommand\tnu{\hat{\nu}}
\newcommand\tomega{\hat{\omega}}
\newcommand\mbeta{\mathring\beta}

\newcommand\tg{\tilde{g}}

\def\half{\frac{1}{2}}

\newcommand{\sDel}{\Delta\mkern-12mu /}

\usepackage{nicefrac}
\usepackage{mathrsfs}
\usepackage{graphicx,psfrag}
\usepackage{amsmath,amsfonts,amssymb,amsthm}
\newtheorem*{Thm}{Theorem}
\newtheorem*{Def}{Definition}

\newcommand{\plot}[1]{     
  \begin{minipage}[t]{0.49\textwidth}
    \includegraphics[scale=0.32]{#1}
    \medskip
  \end{minipage}
}

\begin{document}

\title{The initial boundary value problem for free-evolution 
formulations of General Relativity}
\author{David Hilditch$^1$}
\author{Milton Ruiz$^{2,3,4}$}
\affiliation{$^1$
Theoretical Physics Institute, University of Jena, 07743 Jena, Germany\\
$^{2}$Department of Physics, University of Illinois at Urbana-Champaign, Urbana, IL 61801\\
$^{3}$Escuela de F{\'\i}ısica, Universidad Industrial de Santander, Ciudad Universitaria, Bucaramanga 680002, Colombia\\
$^{4}$Departament de F{\'\i}sica, Universitat de les Illes Balears, Palma de Mallorca, E-07122, Spain}

\begin{abstract}
We consider the initial boundary value problem for free-evolution  
formulations of general relativity coupled to a parametrized 
family of coordinate conditions that includes both the moving 
puncture and harmonic gauges. We concentrate primarily on 
boundaries that are geometrically determined by the outermost 
normal observer to spacelike slices of the foliation. We present 
high-order-derivative boundary conditions for the gauge, constraint 
violating and gravitational wave degrees of freedom of the formulation. 
Second order derivative boundary conditions are presented in terms of 
the conformal variables used in numerical relativity simulations. 
Using Kreiss-Agranovich-M\'etivier theory  we demonstrate, in the 
frozen coefficient approximation, that with sufficiently high order 
derivative boundary conditions the initial boundary value problem can be rendered 
boundary stable. The precise number of derivatives required depends on the gauge. 
For a choice of the gauge condition that renders the system strongly hyperbolic 
of constant multiplicity, well-posedness of the initial boundary value problem 
follows in this approximation. Taking into account the theory of pseudo-differential 
operators, it is expected that the nonlinear problem is also well-posed locally in time.
\end{abstract}


\maketitle

\tableofcontents

\section{Introduction}             
\label{section:Introduction}        

For standard applications in numerical relativity we are 
forced to consider the mathematical properties of the initial 
boundary value problem (IBVP) for general relativity. An 
essential property of the IBVP is that it should be well-posed. 
The requirement of well-posedness is three-fold. We require 
that a solution exists, is unique, and depends continuously
on given initial and boundary data~\cite{Kre78,KreLor89}.

There are further complications. Formulations 
of general relativity (GR) typically have constraints 
which must be satisfied in order to recover a full solution 
of the Einstein equations. If the boundary conditions (BCs) 
are not constraint preserving then, even if the IBVP is 
well-posed, as illustrated for example in~\cite{MilGreSue03a,BucSar06,RuiHilBer10}, 
constraint violations will enter through the boundary and render 
the solution of the partial differential equation (PDE) system 
unphysical. Furthermore, since we are often interested in solutions 
that are asymptotically flat, we would like the BCs to be as 
transparent as possible to outgoing radiation, be it physical or 
gauge, in the sense that these conditions do not introduce large 
spurious reflections from the boundary. Such reflections would 
either be unphysical, or simply produce undesirable gauge dynamics. 
A general discussion of non-reflecting BCs of the wave problem in 
applied mathematics and engineering can be found in~\cite{Giv91}.
Two formulations of GR are currently known to admit a well-posed 
IBVP with constraint preserving boundary conditions
(CPBCs)~\cite{FriNag99,KreWin06,Rin06a,KreReuSar07,RuiRinSar07,KreReuSar08}.
They are the generalized harmonic gauge (GHG)~\cite{Fri85,Fri86,Gar01} 
and Friedrich-Nagy formulations~\cite{FriNag99}. Of these, GHG has 
been used widely in numerical relativity 
simulations~\cite{Pre04,Pre05,LinSchKid05,BoyLinPfe06,PfeBroKid07}.
Boundary conditions employed in GHG numerical simulations are 
described, for instance, in ~\cite{Rin06a,SeiSziPol08,HilWeyBru15}. 
On the other hand, many numerical relativity groups use formulations 
involving a conformal decomposition of the field equations, such as the 
Baumgarte-Shapiro-Shibata-Nakamura-Oohara-Kojima (BSSNOK) 
formulation~\cite{BauSha98,ShiNak95,NakOohKoj87} or a conformal 
decomposition of the Z4 formulation~\cite{BonLedPal03,BonLedPal03a} 
as developed 
in~\cite{BerHil09,RuiHilBer10,WeyBerHil11,AliBonBon11,CaoHil11,AliKasRez13}.
These formulations are normally used in combination with the moving 
puncture gauge 
condition~\cite{BonMasSei94a,Alc02,BakCenCho05,CamLouMar05,MetBakKop06,GunGar06}. 

The IBVP for these `conformal' formulations is less well understood.
The key difficulty, as we shall see, is the complicated structure 
of the principal part of the equations with the moving puncture gauge. 
Thus most codes use so-called radiative boundary conditions on every evolved 
field~\cite{Alc08}, which overdetermine the IBVP and therefore are expected 
to render it ill-posed. These conditions do not preserve 
the constraints. Well-posedness of the IBVP of BSSNOK has been studied in 
a number of places. For instance, in~\cite{BeySar04} the dynamical BSSNOK 
system is recast as a first order symmetric hyperbolic system and the 
corresponding IBVP shown to be well-posed through a standard energy 
method. However, the boundary conditions presented in~\cite{BeySar04} do 
not preserve the constraints, and  the analysis of the IBVP does not 
include the moving puncture gauge condition. In~\cite{NunSar09} 
constraint preserving boundary conditions for the BSSNOK formulation 
were shown to give a well-posed IBVP when the system is linearized 
around flat-space. These conditions have not yet been tested in numerical 
relativity simulations. A numerical implementation of CPBCs in spherical
symmetry for the above system were presented in Appendix B of~\cite{RuDeAlNuSa12},
and extensively tested in~\cite{AlToe14}. The key point of this implementation
is to numerically construct the outgoing and incoming modes, and to express the 
latter in terms of the constraints where possible. BCs are then set to enforce 
that the incoming modes do not introduce spurious reflections. For a detailed 
discussion of the IBVP in GR, see the review~\cite{SarTig12}. For the Z4 formulation 
CPBCs are straightforward, since the constraint subsystem consists entirely of wave 
equations, whereas the BSSNOK constraint subsystem contains a characteristic variable 
with vanishing speed. Using this fact, CPBCs were implemented, in explicit 
spherical symmetry, and shown very effective at absorbing constraint 
violations~\cite{RuiHilBer10}. Moreover BCs compatible with the constraints 
for a symmetric hyperbolic first order reduction of Z4 were specified and 
studied in numerical applications in~\cite{BonLedLuq04a,BonBon10}. The 
conditions are of the maximally dissipative type and so well-posedness of 
the resulting IBVP could be shown with a standard energy estimation, although 
harmonic slicing and normal, or vanishing shift, coordinates were employed, and 
it is not clear how generally the results can be extended to other gauge choices. 
Full 3D numerical relativity simulations using Z4c and radiation controlling, 
CPBCs were presented~\cite{HilBerThi12}. But no attempt was made to analyze 
well-posedness of the IBVP.

In this work, we therefore attempt to complete the theoretical story, 
in the sense that we prove well-posedness of the IBVP, in the frozen 
coefficient approximation, of particular formulations of GR coupled to
a parametrized family of gauge conditions including both the harmonic and 
moving puncture gauges. Our discussion will focus primarily on the formulation 
of~\cite{HilRic13}. From the PDEs point of view this is the preferred choice of
formulation because it decouples the gauge and constraint violating degrees 
of freedom to the greatest degree possible for the live gauges under consideration. 
This formulation has not yet been used in numerical relativity but is expected to 
have all of the advantages of Z4 over BSSNOK, most notably propagating constraints,  
whilst simultaneously avoiding possible breakdown of hyperbolicity associated with 
the clash of gauge and constraint violating characteristic speeds. The Mathematica 
notebooks that accompany the paper can be modified to treat the Z4 and BSSNOK 
formulations. By the theory of pseudo-differential operators, our calculations are 
expected to extend locally in time to the original nonlinear 
equations~\cite{Es73,KreLor89}. 

We begin in section~\ref{section:HR} with a summary of the formulation, a 
geometric formulation of the problem and the identification of the BCs taken 
in the subsequent analysis. Our geometric formulation fixes the outer boundary
to be that timelike surface generated by the outermost observers in the initial 
data as they are Lie-dragged up the foliation by the timelike normal vector. This 
results in an outer boundary that may drift in local coordinates. The numerical 
relativist interested in implementing a basic approximation to our conditions 
need only concern themselves with sections~\ref{section:Conformal_decomposition}
and~\ref{section:Conformal_BCs}. Section~\ref{section:WP} contains our well-posedness 
results with high order BCs, and discussion of the difficulties that arise if we 
try to fix the coordinate position of the outer boundary, plus gauge conditions in 
which this is straightforward, and in which the fewer derivatives are required 
to achieve boundary stability. We conclude in section~\ref{section:Conclusion}. 

\section{Formulation of the IBVP}  
\label{section:HR}            

In this section, we summarize the geometrical setup of the 
IBVP, present the formulation of~\cite{HilRic13} in the ADM and 
conformal variables and discuss the high-order BCs analyzed in 
section~\ref{section:WP}. Finally, we display the second order 
special case of the BCs in terms of the conformal variables that 
are used in standard numerical applications. Here `order' 
refers to the highest derivative of either the metric, lapse 
or shift components appearing in the boundary condition.

\subsection{Analytical Setup}  
\label{section:Analytic}            


\begin{figure}[t!]
\plot{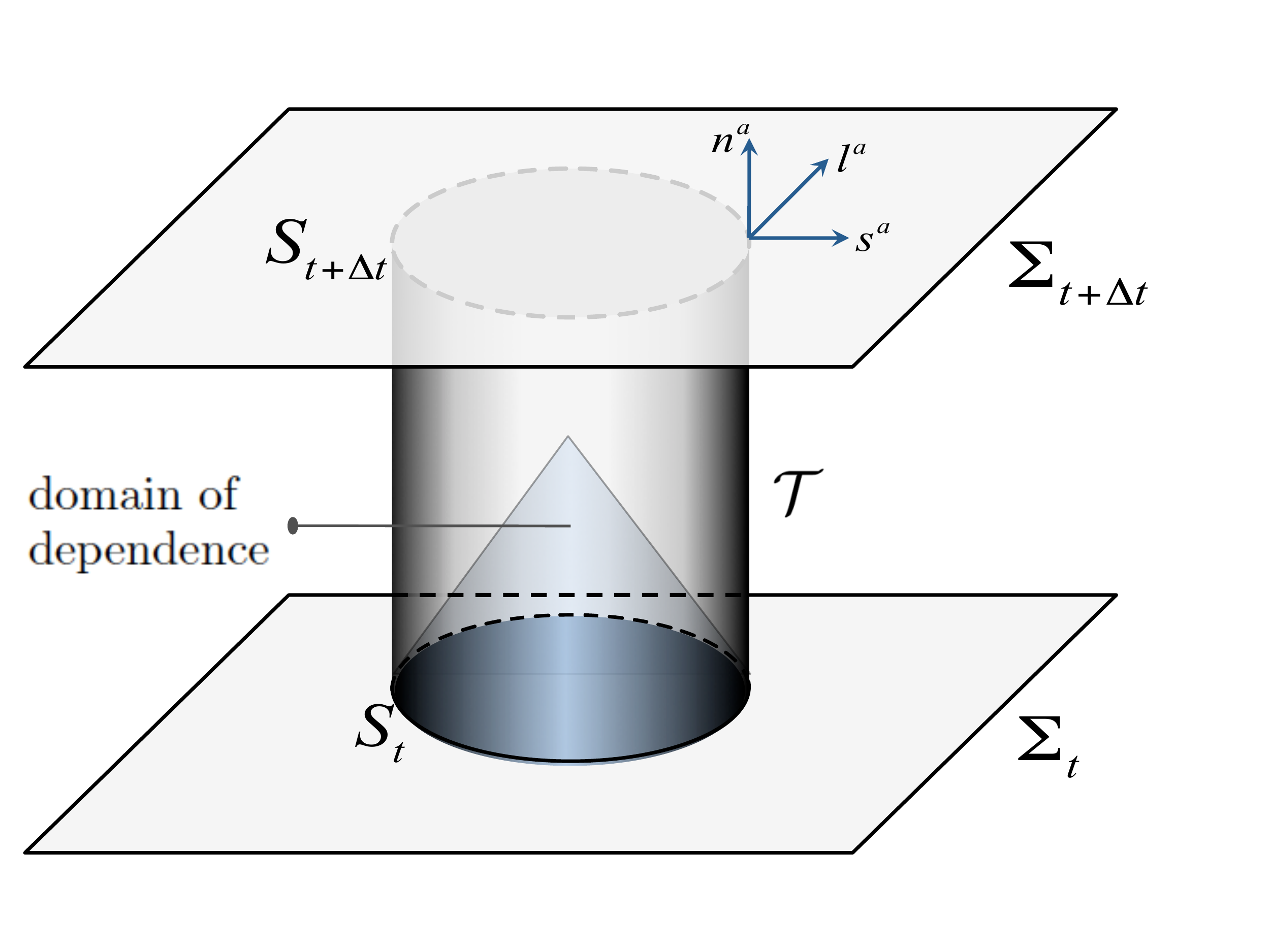}
\caption{\label{fig:world_tube}
Manifold setup for the IBVP. The manifold is foliated by three 
dimensional surfaces~$\Sigma_t$. We impose a timelike boundary 
condition~$S_t$ in a compact region of each surface $\Sigma_t$, which 
restricts the domain of dependence of the initial data to the 
inner conical region of the timelike tube~$\mathcal{T}$.}
\end{figure}

\paragraph*{Manifold structure and geometry of the boundary:} We 
investigate the evolution equations on a manifold~$M=[0,T]\times\Sigma$. 
The three dimensional compact manifold~$\Sigma$ has smooth 
boundary~$\p\Sigma$. We assume that the gravitational field is weak 
near the boundary so that the boundary of the full 
manifold~$\mathcal{T}=[0,T]\times\p\Sigma$ is timelike and the 
three dimensional slices~$\Sigma_t=\{t\}\times\Sigma$ are spacelike 
as shown in Figure~\ref{fig:world_tube}. The boundary of a spatial 
slice is denoted~$S_t=\{t\}\times\partial\Sigma$. We define~${n}^a$, 
the future pointing unit normal to the slices~$\Sigma_t$, and similarly 
employ the standard notation for the induced metric~$\gamma_{ab}$ and 
extrinsic curvature~$K_{ab}$ of the foliation.  The spatial covariant 
derivative is denoted~$D$. Initial data will be specified on some 
constant~$t$ slice, and boundary conditions, yet to be determined, 
on~$\mathcal{T}$. The outer boundary~$\mathcal{T}$ can be characterized 
as the level set of a scalar field~$r=r_B$, defined at least in a neighborhood 
of~$\mathcal{T}$. We may then perform a~$2+1$ split relative to the 
unit spatial vector, 
\begin{align}
s^a&=LD^ar\,,
\end{align}
where we define the length scalar~$L^{-2}=\gamma^{ab}D_arD_br$,
to study the geometry of the boundary. We will however only introduce the 
quantities to be employed in the boundary conditions. The vector~$s^a$ 
is thus the unit normal to the two-surface~$\{t\}\times\p\Sigma$ as 
embedded in~$\Sigma_t$. The standard approach in numerical relativity is
to take~$r$ to be a radial-type coordinate built in the normal way from the
asymptotically Cartesian coordinates defining the tensor basis used to
represent the evolved variables. In this case we have~$\p_tr=0$ and so the
coordinate position of the outer boundary is fixed in time. Perhaps a more
geometrically natural condition is to insist that the future pointing
normal~$n^a$ to slices of the foliation point directly up the boundary. This
can be achieved by requiring instead~$~\Lie_nr=0$, which must be solved at least
in a neighborhood of the outer boundary. One may then think of~$r$ as a natural
radial coordinate of normal observers to the slice. When working under this
assumption we say that we work ``under the boundary orthogonality condition''. 
Notice that this leads to a hyperbolic equation of motion,
\begin{align}
\p_t(\p_ir)&=\beta^j\p_j(\p_ir)+(\p_jr)\p_i\beta^j\,,
\end{align}
for the appropriate components of the Jacobian mapping between the two coordinate
systems, since the second term is non-principal, as it may be replaced by a
first-order reduction variable in any such reduction. The numerical implementation
of this idea is left to future work, but we note that the approach fits naturally
within the dual foliation formalism~\cite{Hil15}. A consequence of insisting on working
with the boundary orthogonality condition is that the 
outer boundary will drift in local coordinates.
Geometrically this condition is the same as that for the longitudinal component of
the shift in~\cite{NunSar09} for BSSNOK. But now~$r$ is not one of our coordinates,
and nor is the associated vector~$(\tfrac{\p}{\p r})^a$ necessarily a member of the
tensor basis in which we work for the~$3+1$ evolution. The motivation for choosing this
orthonormality in the BSSNOK case was that in this way the number of incoming characteristic
fields at the outer boundary can be fixed, removing the need to treat various special 
cases. With the present formulation that motivation is absent because there 
are no shift-speed characteristic variables. This imposes a major difference 
in our analysis as compared to the standard boundary treatment in numerical 
relativity, where the outer boundary remains at fixed coordinates. We expect that
this complication can be sidestepped by working with the dual-foliation formalism,
but this will be investigated elsewhere. The problems that arise in the PDEs
analysis if we do not work with the boundary orthogonality condition are discussed in
section~\ref{section:Why_orthogonal?}.

\paragraph*{Newman-Penrose null tetrad:} 
The previous vector fields allow us to introduce, for later convenience, 
the following Newman-Penrose null vectors,
\begin{align}
l^a&= \frac{1}{\sqrt{2}}\,\left(n^a+s^a\right)\,,\quad
k^a= \frac{1}{\sqrt{2}}\,\left(n^a-s^a\right)\,,\nonumber\\
m^a&= \frac{1}{\sqrt{2}}\,\left(\iota^a+i\,\upsilon^a\right)\,,\quad
\bar{m}^a= \frac{1}{\sqrt{2}}\,\left(\iota^a-i\,\upsilon^a\right)\,,\label{eqn:dyadnull}
\end{align}
where~$\iota^a$ and~$\upsilon^a$ are spatial unit vectors mutually 
orthogonal to both~$n^a$, $s^a$ and each other. 

\paragraph*{Equations of motion:} 
Following~\cite{HilRic13}, in which the formulation was first 
presented, we replace the Einstein equations with the expanded set 
of equations,
\begin{align}
\p_t\gamma_{ij}&=-2\alpha K_{ij}+\Lie_\beta\gamma_{ij}\,,\nonumber\\
\p_tK_{ij}&=-D_iD_j\alpha+\alpha[R_{ij}-2K_i{}^kK_{kj}+KK_{ij}
\nonumber\\
&+2\hat{D}_{(i}Z_{j)}-\kappa_1(1+\kappa_2)\gamma_{ij}\Theta]\nonumber\\
\label{eqn:K_dot}
&+4\pi\alpha[\gamma_{ij}(S-\rho)-2 S_{ij}]
+\Lie_\beta K_{ij}\,, 
\end{align}
where~$\Theta$ and~$Z_i$ are a set of four variables defining 
an expanded phase space in which our PDEs analysis is performed, and we
must have~$\Theta=Z_i=0$ to recover solutions of GR. The equations of
motion for these variables are given momentarily. We write,
\begin{align}
\hat{D}_{i}Z_{j}\equiv
\gamma^{-\frac{1}{3}}\gamma_{kj}\p_{i}\tilde{Z}^k\,,
&\qquad\qquad 
\tilde{Z}^i=\gamma^{\frac{1}{3}}Z^i\,.
\end{align}
The free parameters~$\kappa_1$ and~$\kappa_2$ serve to parametrize 
the strength of constraint damping in the evolution 
equations~\cite{GunGarCal05}. These terms were not included in the 
discussion of~\cite{HilRic13} and, as non-principal terms will play
no fundamental role in the discussion of boundary stability, but are 
expected to effectively damp away constraint violation in numerical 
applications. Here we also modify the constraint addition as compared 
with~\cite{HilRic13} so that the equations of motion look as natural 
as possible when written in terms of the conformal variables. The 
dynamical ADM equations are of course recovered when the 
constraints~$\Theta$ and~$Z_i$ vanish. 

\paragraph*{Constraints:} The set of constraints~$\Theta,Z_i$ 
are completed by the Hamiltonian and momentum constraints, 
\begin{align}
H &\equiv R-K_{ij}K^{ij}+K^2-16\pi\rho=0\,,\nonumber\\
M_i&\equiv D^j\left(K_{ij}-\gamma_{ij}K\right)-8\pi S_i=0\,.
\end{align}
Their equations of motion are,
\begin{align}
\p_t\Theta&=\alpha\,\left[\frac{1}{2} H + \hat{D}^iZ_i
-\kappa_1(2+\kappa_2)\Theta\right]+\Lie_\beta\Theta\,,\nonumber\\
\p_tZ_i&=\alpha\,\left[M_i+\frac{1}{3}\,\big(4-\eta_\chi\big)\,
D_i\Theta-\kappa_1Z_i\right]
\nonumber\\
&\quad
+\gamma^{\frac{1}{3}}Z^j
\p_t\left[\gamma^{-\frac{1}{3}}\gamma_{ij}\right]+\beta^j\hat{D}_jZ_i\,,
\label{eqn:Theta_Z_dot}
\end{align}
where the scalar~$\eta_\chi$ is determined by the gauge choice
as discussed below. The time dependence of the constraints can be 
computed from~\eqref{eqn:K_dot}, and is found to be,
\begin{align}
\label{eqn:Ham_dot}
\p_tH=&-2\alpha D^iM_i-4\,M_iD^i\alpha+2\,\alpha\,K\,H\nonumber\\
&+ 2\,\alpha\,\left(2\,K\,\gamma^{ij}-K^{ij}\right)
\hat{D}_{(i}Z_{j)}\nonumber\\
&-4\,\kappa_1\,(1+\kappa_2)\,\alpha\,K\,\Theta+\Lie_\beta H\,,
\end{align}
for the Hamiltonian constraint and
\begin{align}
\label{eqn:Mom_dot}
\p_tM_i=&-\half\alpha D_i H + \alpha\,K\,M_i 
-(D_i\alpha)\,H \nonumber\\
&+D^j\left(2\,\alpha\,\hat{D}_{(i}Z_{j)}\right)
-D_i\left(2\,\alpha\,\gamma^{kl}\,\hat{D}_{(k}Z_{l)}\right)\nonumber\\
&+2\kappa_1\,(1+\kappa_2)\,D_i(\alpha\,\Theta)+\Lie_\beta M_i\,,  
\end{align}
for the momentum constraint. It is clear that this formulation 
is a mild modification of the Z4c system, the only difference in 
the principal part occurring in~\eqref{eqn:Theta_Z_dot}.

\paragraph*{Gauge conditions:} We close the evolution
system with a parametrized gauge condition, consisting of 
the Bona-Mass\'o lapse condition~\cite{BonMas94} and the 
shift condition,
\begin{align}
\p_t\alpha&=-\alpha^2\,\mu_L\,\hat{K}+\beta^i\,\p_i\alpha\,,\nonumber\\
\p_t\beta^i&=\alpha^2\,\mu_S\,\chi\,\Big[\tilde{\Gamma}^i+\tfrac{1}{2}\,\eta_\chi
\tilde{\gamma}^{ij}\,\p_j\ln\chi\Big]-\alpha\,\eta_L\,\chi\,\tilde{\gamma}^{ij}
\p_j\alpha\nonumber\\
&\quad-\eta\,\beta^i+\beta^j\,\p_j\beta^i\,,
\label{eqn:gauge}
\end{align}
where~$\hat{K}=K-2\,\Theta$, the contracted conformal 
Christoffel is a shorthand for,
\begin{align}
\label{eqn:G_shorthand}
\tilde{\Gamma}^i&= \gamma^{\frac{1}{3}}\,\gamma^{ij}
\Big[2\,Z_j +\gamma^{kl}(\p_k\gamma_{lj}
-\frac{1}{3}\,\p_j\gamma_{kl})\Big]\,,
\end{align}
and the conformal metric is defined by~$\tilde{\gamma}_{ij}=\chi\,\gamma_{ij}$,
with~$\chi=\gamma^{-1/3}$. The harmonic gauge is recovered with the 
choice~$\mu_L=\eta_\chi=1$, $\mu_S=\eta_L=1$, and~$\eta=0$.
The standard moving puncture gauge choice is the ``1+log'' variant
of the Bona-Mass\'o condition,~$\mu_L=2/\alpha$, combined with the 
Gamma-driver shift~\cite{AlcBruDie02}, with~$\eta_\chi=\eta_L=0$, 
and various choices for~$\mu_S$. The effect of the gauge damping 
term~$\eta$ on numerical simulations with the Gamma-driver shift 
has been studied in~\cite{Sch10,MueBru09,AliRezHin10}.

\paragraph*{Projection operators:} We define the projection 
operators into directions tangential to the boundary $S_t$, 
and onto the ``physical'' degrees of freedom by,
\begin{align}
q^i{}_j&=\delta^i{}_j-s^is_j\,,
\quad q^{(P)}{}^{ij}{}_{kl}=q^i{}_{(k}q^j{}_{l)}
-\frac{1}{2}q_{kl}q^{ij}\,,
\end{align}
respectively. We use the notation,
\begin{align}
D_sD_s\alpha\equiv s^i\,s^j\,D_iD_j\alpha\,,
\end{align}
for longitudinal derivatives; we do not commute the spatial 
normal vector with {\it any} derivative operator. Likewise, we 
never commute the projection operator with {\it any} derivative 
operator, so for example,
\begin{align}
D_AD_B\alpha\equiv q^i{}_A\,q^j{}_B\,D_iD_j\alpha\,,
\end{align}
where we use upper case Latin letters to denote indices that 
have been projected into the directions tangential to~$S_t$.

\subsection{Boundary conditions}   
\label{sec:BCs}

We want to impose BCs on the formulation. 
Following~\cite{Rin06,Rin06a}, these conditions should satisfy 
the following conditions:
\begin{description}

\item[Well-posedness] The IBVP must be well-posed. Without 
this requirement, existence of a solution, even locally in 
time, is not guaranteed. Without continuous dependence on 
given data at the continuum level, no numerical method can 
converge to the continuum solution. Furthermore, in principle
without continuous dependence the PDE formulation of the 
physical problem has no predictive power.

\item[Constraint preservation] The conditions should be 
constraint preserving. Otherwise the physical solution will be 
compromised as soon as it is reached by the constraint violations 
propagating from the outer boundary into the domain.

\item[Radiation control] The BCs should minimize spurious  
reflections and allow us to control the incoming gravitational radiation. 
Without this property, the solution can not necessarily be viewed 
as an isolated body unperturbed by incoming waves. Note that this
characterization relies on the assumption that the gravitational 
field near the boundary is weak. 
\end{description}

With these considerations in mind, we propose the following set
of BCs:

\paragraph*{Gauge boundary conditions:} Following~\cite{RuiHilBer10}, 
for the lapse we choose the boundary condition,
\begin{align}
\boxed{\Big(r^2\,i^a_{\mu_L}\p_a\Big)^{L+1}\alpha\,\,\hat{=}\,\,
(r^2\,\Lie_n)^{L+1}h_L\,,}\label{eqn:bc_alpha}
\end{align}
where $i^a_{\mu_L}$ the vector pointing along the outgoing characteristic 
surfaces of the Bona-Mass\'o lapse condition, defined according to, 
\begin{align}
i^a_{\mu}=\frac{1}{\sqrt{2}}\,\big(n^a+\sqrt{\mu}\,s^a\big)\,,
\end{align}
a shorthand valid for arbitrary~$\mu>0$, and $\Lie_n$ the derivative 
along the $n^a$ direction. Here, and in what follows,~$\hateq$ 
denotes an equality which holds only in the boundary~$S_t$. We take~$L$
to be a natural number, and~$h_L$ an arbitrary smooth scalar function 
in the boundary which can be interpreted as the given boundary data. 

Next, in order to specify BCs on the components~$\beta^i$, define the 
shorthands~$\mu_{S_L}=(4-\eta_\chi)\,\mu_S/3$ and,
\begin{align}
B^s&=i_{\mu_{S_L}}^a\p_a(\p_i\beta^i)
-\left(\tfrac{\eta_L\mu_L-\mu_{S_L}}{\mu_L-\mu_{S_L}}\right)
\,\alpha\,i_{\mu_{S_L}}^a\p_a\hat{K}\,.
\end{align}
We emphasize that this variable has nothing to do with the standard  
reduction variable~``$B^i$'' used sometimes with the moving-puncture gauge.  
The reason for choosing this particular  combination will become clear during 
the following analysis. We choose the BC,
\begin{gather}
\boxed{
\begin{split}
r^4\,\Big(r^2\,i_{\mu_{S_L}}^a\p_a\Big)^{L-1}B^s\,\hat{=}\,
(r^2\Lie_n)^{L+1}h_{{S_L}}\,,\label{eqn:bc_betas}
\end{split}}
\end{gather}
for the longitudinal component of the shift. The given data here is
the scalar~$h_{\beta^s}$. Next we define the shorthand,
\begin{align}
B^A&=\gamma^{ik}s_{[k}\,q_{j]}{}^A\p_i\beta^j\,.
\end{align}
For the transverse components of the shift we choose,
\begin{align}
\boxed{\Big(r^2\,\,i_{\mu_{S}}^a\p_a\Big)^{L}
B^A\,\hat{=}\,(\Lie_n)^{L-1}\big(\,\Lie^2_n-\mu_S\sDel\,\big)h_{\mu_S}^A\,.}
\label{eqn:bc_betaA}
\end{align}
The given data~$h_{\mu_S}^A$ are to be treated as two smooth 
scalar functions in the boundary. The operator~$\sDel$ is the two
dimensional Laplacian associated with the induced metric~$q_{AB}$.
The inclusion of this made in order to cancel bad terms in the following
Laplace-Fourier analysis. Note that from the point of view of absorption of
outgoing gauge waves this condition is not optimal, but since 
we are also concerned with minimizing the number of derivatives in 
the conditions, we accept this potential shortcoming. We will see
in the following analysis that the complicated characteristic 
structure of the gauge conditions forces us to take high order
BCs~($L=4$) so that we can obtain boundary stability in the 
analysis. The key point is to choose given data containing
particular combinations of derivatives. To obtain boundary stability in the
rest of the formulation we need only take~$L=1$. We can adjust the gauge so
that there too, only~$L=1$ is required. For details see~\cite{Hil17} and the
Mathematica notebooks that accompany the paper.

\paragraph*{Constraint preserving boundary conditions:} 
In~\cite{RuiHilBer10}, we studied high order BCs for the 
constraints~$\Theta$ and~$Z_i$ for the Z4c formulation. Here we 
are forced to modify those conditions because the characteristic
structure of the constraint subsystem for the present formulation 
is slightly more complicated than that of Z4c. First for the 
scalar constraint~$\Theta$ we choose,
\begin{align}
\label{eqn:CPBCs_Theta}
\boxed{\,
r^2\,\Big(r^2\,\,i_{\mu_C}^a\p_a\Big)^L\Theta\,\hat{=}\,
\Big(r^2\,\Lie_n\Big)^{L+1}h_\Theta\,,}
\end{align}
where we have defined~$\mu_C=\mu_{S_L}/\mu_S=(4-\eta_\chi)/3$ and 
choose given data~$h_\Theta$ which will be taken to vanish
in applications. For the lowest derivative order~$L=1$ boundary we 
choose,
\begin{align}
l^a\p_a\tilde{Z}^i\,\hat{=}\,\Lie^2_n\tilde{h}_{Z}^i\,,
\label{eqn:CPBCs_Z_Low}
\end{align}
where we write~$\tilde{Z}^{i}=\tilde{\gamma}^{ij}Z_j$ and
$\tilde{h}_{Z}^i=\tilde{\gamma}^{ij}h_{Z_j}$. This choice is made so that 
the boundary conditions become more convenient when written in terms of 
the conformal variables used in numerical applications (see 
Sec.~\ref{section:Conformal_decomposition}). For higher order 
conditions, however, it turns out to be more natural to make 
some adjustment. We use the shorthands,
\begin{align}
\tilde{X}^i=\Lie_n\tilde{Z}^i-\mu_C\,\tilde{\gamma}^{ij}D_j\Theta\,.
\end{align}
The remaining constraint conditions are then,
\begin{align}
\label{eqn:CPBCs_Z}
\boxed{
\,\Big(r^2\,l^a\p_a\Big)^{L-1}\tilde{X}^i\,\hat{=}\,
(r^2\,\Lie_n)^{L-1}\big(\,\Lie^2_n-\sDel\,\big)\tilde{h}_{Z}^i\,.}
\end{align}
Again the given data~$\tilde{h}_{Z}^i$ will typically be taken to 
vanish in applications, but we have to include it 
to show estimates in the free-evolution approach. 

\paragraph*{Radiation controlling boundary conditions:} A standard 
BC for the GHG formulation that controls the incoming gravitation 
radiation is the $\Psi_0$-{\it freezing condition}
\cite{FriNag99,SarTig04,LinSchKid05,Rin06,RinLinSch07,RuiRinSar07,KreReuSar08,KreWin06,Rin06a}
which serves as a good first approximation to an absorbing 
condition~\cite{BucSar06,BucSar07}. In particular, 
freezing~$\Psi_0$ to its initial value allows the absorption of 
outgoing gravitational waves by minimizing spurious reflections. It 
has been shown analytically~\cite{BucSar06} that the spurious 
reflections from the freezing-$\Psi_0$ condition decay as fast 
as~$(k\,R)^{-4}$, for monochromatic radiation with wavenumber~$k$ 
and for an outer boundary with areal radius~$R$. This condition 
has also been considered with the BSSNOK formulation~\cite{NunSar09}. 

To impose~$\Psi_0$-freezing conditions, we take the electric and 
magnetic parts of the Weyl tensor~\cite{Alc08},
\begin{align}
E_{ij}&=\left[R_{ij}+K\,K_{ij}-K^l{}_iK_{il}
+2\,\hat{D}_{(i}Z_{j)}-4\,\pi\,S_{ij}\right]^{\textrm{TF}},
\nonumber\\
B_{ij}&=\epsilon_{(i|}{}^{kl}D_k K_{l|j)}\,.
\label{eqn:electric_magnetic}
\end{align}
The Weyl scalar~$\Psi_0$ is given by, 
\begin{align}
\Psi_0&=\left(E_{mm}-i\,B_{mm}\right)\,,
\label{eqn:psi0}
\end{align}
where the index~$m$ refers to contraction with the null vector $m^a$.
To motivate our choice of given data recall that, for 
linear plane gravitational waves propagating on flat space, we 
have~\cite{Alc08}
\begin{align}
\Psi_0&=-\frac{1}{4}\big(\p^2_th^{+}+2\p_t\p_rh^{+}+\p^2_rh^{+}\big)
\nonumber\\
&\quad-\frac{i}{4}\big(\p^2_th^{\times}+2\p_t\p_rh^{\times}
+\p^2_rh^{\times}\big)\,,
\end{align}
with~$h^{+}$ and~$h^{\times}$ the independent components of the 
transverse-traceless part of the metric perturbation. Assuming that we 
have an incoming gravitational wave, then~$h^+\sim h^{\times}\sim h(t+r)$
and then,
\begin{align}
\Psi_0&=-\p^2_th^{+}-i\,\p^2_th^{\times}\,.
\end{align}
Thus, for the lowest order boundary we choose,
\begin{align}
\Psi_0\,\hat{=}\,(r^2\Lie_n)^{2}h_{\Psi_0}\,,
\end{align}
where~$h_{\Psi_0}$ is smooth given data at the boundary. For higher order BCs,
One naively could hit the left-hand side of the above condition by a Sommerfeld boundary
operator as many times as is desired. However since~$\Psi_0$, depending on the particular 
gauge, satisfies in the principal part a wave equation only up to a coupling with~$\Theta$,
the necessary analysis for arbitrary values of~$L$ becomes messy. To avoid this we choose,
\begin{align}
\boxed{
\,r^4\,\Big(r^2\,l^a\p_a\Big)^{L-2}\hat\Psi_0
\hat{=}\,(r^2\Lie_n)^{L+1}h_{\Psi_0}\,,
\label{eqn:bc_psi0}}
\end{align}
for~$L\geq 2$, where the shorthand~$\hat\Psi_0$ is given by
\begin{align}
\hat\Psi_0 =\Lie_n\Psi_0-2\,{\mu_C}\,D_m\,D_m\Theta\,.
\nonumber  
\end{align}

\subsection{Conformal decomposition}
\label{section:Conformal_decomposition}

For numerical integration favorable PDE properties, such as 
well-posedness, may not be enough to guarantee robust evolution. 
It is therefore common to work with conformally decomposed 
variables. We define the variables~\cite{BerHil09},
\begin{align}
&\tilde{\gamma}_{ij} = \gamma^{-\frac{1}{3}}\,\gamma_{ij}\,,\quad
\chi = \gamma^{-\frac{1}{3}}, \nonumber\\
&\hat{K} = \gamma^{ij}\,K_{ij} - 2\,\Theta\,,\quad
\tilde{A}_{ij}=\gamma^{-\frac{1}{3}}\,(K_{ij}-\frac{1}{3}\,
\gamma_{ij}\,K)\,,\nonumber\\
&\tilde{\Gamma}^{i} = 2\,\tilde{\gamma}^{ij}\,Z_j + \tilde{\gamma}^{ij}
\,\tilde{\gamma}^{kl}\,\p_l\tilde{\gamma}_{jk}\,,\quad
(\tilde{\Gamma}_{\textrm{d}})^i
=\tilde{\gamma}^{jk}\,\tilde{\Gamma}^{i}{}_{jk}\,,
\end{align}
the idea of which is to make as many variables as possible 
non-singular, so that for example puncture black holes can 
be treated numerically. Variations on this decomposition have 
been studied in the literature~\cite{WitHilSpe10,PolReiSch11}, but here we will be satisfied 
with the vanilla form. Note that the definition of~$\tilde{\Gamma}^i$ is 
compatible with the shorthand given in~\eqref{eqn:G_shorthand}. Under 
this change of variables the equations of motion become,
\begin{align}
\p_t \chi &= \frac{2}{3}\,\chi\,
\left[\alpha\,(\hat{K}+2\Theta) - D_i\beta^i\right]\,,\nonumber\\
\p_t \tilde{\gamma}_{ij} &= -2\,\alpha\,\tilde{A}_{ij}+\beta^k\p_k
\tilde{\gamma}_{ij}+2\,\tilde{\gamma}_{k(i}\p_{j)}\beta^k\nonumber\\
&\quad-\frac{2}{3}\,\tilde{\gamma}_{ij}\p_k\beta^k\,,
\end{align}
for the metric and, 
\begin{align}
\p_t\hat{K}&= -D^iD_i\alpha + \alpha\,\left[\tilde{A}_{ij}\tilde{A}^{ij}
+\frac{1}{3}(\hat{K}+2\Theta)^2\right]\nonumber\\
&+4\,\pi\,\alpha\,\left[S+\rho\,\right]
+\alpha\,\kappa_1\,(1-\kappa_2)\,\Theta+\beta^i\p_i\hat{K}\,,
\nonumber\\
\p_t \tilde{A}_{ij} &= \chi\,\big[-D_iD_j\alpha
+\alpha\,(R_{ij}-8\,\pi\,S_{ij})\big]^{\textrm{tf}}\nonumber\\
& +\alpha\,\left[(\hat{K}+2\,\Theta)\tilde{A}_{ij}-
2\,\tilde{A}^k{}_i\tilde{A}_{kj}\right]\nonumber\\
& +\beta^k\,\p_k\tilde{A}_{ij}+2\,\tilde{A}_{k(i}\,\p_{j)}\beta^k
-\frac{2}{3}\,\tilde{A}_{ij}\,\p_k\beta^{k}\,, 
\end{align}
for the extrinsic curvature. For the contracted conformal Christoffels 
we have,
\begin{align}
\p_t\tilde{\Gamma}^{i} &= -2\,\tilde{A}^{ij}\,\p_j\alpha+2\,\alpha
\left[\tilde{\Gamma}^i{}_{jk}\,\tilde{A}^{jk}
-\frac{3}{2}\,\tilde{A}^{ij}\,\p_j\ln(\chi)\right.
\nonumber\\
&\quad\left.-\frac{2}{3}\,\tilde{\gamma}^{ij}\,\p_j\,\hat{K}
-8\,\pi\,\tilde{\gamma}^{ij}\,S_j\right]
+\tilde{\gamma}^{jk}\,\p_j\p_k\beta^i\nonumber\\&\quad
+\frac{1}{3}\,\tilde{\gamma}
^{ij}\p_j\p_k\beta^k+\beta^j\,\p_j\tilde{\Gamma}^i
-(\tilde{\Gamma}_{\textrm{d}})^j\,\p_j\beta^i\nonumber\\
&\quad+\frac{2}{3}\,(\tilde{\Gamma}_{\textrm{d}})^i\,\p_j\beta^j
-2\,\alpha\,\kappa_1\,\big[\tilde{\Gamma}^i
-(\tilde{\Gamma}_{\textrm{d}})^i\big]\,.\label{eqn:Gtilde_dot}
\end{align}
The difference between Z4c and the present formulation, displayed 
in~\eqref{eqn:Theta_Z_dot}, propagates through the change of variables
resulting in the disappearance of the~$\Theta$ constraint from 
this equation. Finally we have,
\begin{align}
\p_t\Theta &=\frac{1}{2}\,\alpha\,\big[R 
- \tilde{A}_{ij}\,\tilde{A}^{ij}
+\frac{2}{3}\,(\hat{K}+2\,\Theta)^2\big]\nonumber\\
&\quad-\alpha\,\big[8\,\pi\,\rho+\kappa_1\,(2+\kappa_2)\,
\Theta\big]+\beta^i\p_i\Theta\,.
\end{align}
This system can be trivially implemented in a moving puncture code as 
a modification of either the Z4c or BSSNOK formulations. Within this 
decomposition the intrinsic curvature is written as,
\begin{align}
R_{ij} &= R^{\chi}{}_{ij} + \tilde{R}_{ij}\,,\nonumber\\
\tilde{R}^{\chi}{}_{ij}&=
\frac{1}{2\chi}\tilde{D}_i\tilde{D}_j\chi+\frac{1}{2\chi}
\tilde{\gamma}_{ij}\,\tilde{D}^l\tilde{D}_l\chi\nonumber\\
&-\frac{1}{4\chi^2}\,\tilde{D}_i\chi\tilde{D}_j\chi-\frac{3}{4\,\chi^2}
\tilde{\gamma}_{ij}\,\tilde{D}^l\chi\tilde{D}_l\chi\,,\nonumber\\
\tilde{R}_{ij} &=
 - \frac{1}{2}\,\tilde{\gamma}^{lm}\,\p_l\p_m\tilde{\gamma}_{ij} 
+\tilde{\gamma}_{k(i}\,\p_{j)}\tilde{\Gamma}^k+(\tilde{\Gamma}_{\textrm{d}})^k
\tilde{\Gamma}_{(ij)k}\nonumber\\
&+\tilde{\gamma}^{lm}\,\left(2\tilde{\Gamma}^k{}_
{l(i}\,\tilde{\Gamma}_{j)km}+\tilde{\Gamma}^k{}_{im}\,
\tilde{\Gamma}_{klj}\right)\,.\label{eqn:Conf_D_Ric}
\end{align}
The equations above are constrained by two algebraic expressions,
$\ln(\det\tilde{\gamma})=0$ and~$\tilde{\gamma}^{ij}\tilde{A}_{ij}=0$,
which we stress {\it must} be explicitly imposed in numerical
applications if the analysis contained in this work is to be valid.

\subsection{Second order boundary conditions on the conformal 
variables}\label{section:Conformal_BCs}

Suitably constructed high order BCs, namely those in which~$L$ is
taken to be a large number, are expected to more efficiently absorb
outgoing gauge, constraint violating, and gravitational
waves~\cite{BayTur80,BucSar06,BucSar07,RuiRinSar07}.  Unfortunately,
their implementation requires the definition of auxiliary fields
confined to the boundary~$S_t$, which is an involved technical
exercise. The improved absorption properties of high order conditions
has been demonstrated in an implementation for a first order reduction
of the GHG formulation~\cite{RinBucSch08}. For the GHG system the task
is made more straightforward by the simple characteristic structure of
the formulation. As a compromise we start by considering the simple
case~$L=1$, the highest order BCs that do not require the definition
of auxiliary variables for implementation. These conditions have the
advantage that they can be easily implemented in a code, but the
serious disadvantage that we can not show estimates for the initial
boundary value problem.  They are however constraint preserving, and
in some approximation do minimize spurious reflections of
gravitational waves from the outer boundary.  We will see in the
analysis that the failure to obtain estimates with low order
derivative boundary conditions is caused primarily by the complicated
characteristic structure of the gauge conditions.  The boundary
orthogonality condition adds another unwanted complication to the
implementation of the boundary conditions. We are thus interested here
in giving a prescription to implement an approximation to our {\it
true} conditions easily in a standard numerical relativity code,
which we hope can serve as a holdover giving improved behavior until
the boundary orthogonality condition can be properly managed and our
higher order conditions can be employed. Therefore we also modify the
conditions by lower order terms, and adjust the given data so as to
drop the boundary orthogonality condition.

\paragraph*{Gauge boundary conditions:} We assume in this section 
that~$\eta_\chi=\eta_L=0$. We start with the lapse 
condition~\eqref{eqn:bc_alpha} with~$L=1$, which becomes, 
\begin{align}
\p_t\hat{K}\,&\hat{=}-\alpha\,\sqrt{\mu_L}\,\p_s\hat{K}
-\tfrac{1}{2}\p^A\p_A\alpha+\alpha\,\p_t^2h_\alpha+\beta^i\p_i\hat{K}\,,
\end{align}
for the extrinsic curvature. Note that in this equation we have adjusted the 
expressions by non-principal terms, and redefined the given data. 
Altering these terms does not affect well-posedness of the IBVP. We 
have chosen this type of condition because it minimizes the number of 
derivatives required to show boundary stability. Numerically, however, 
these conditions have been found to cause a drift of the lapse. 
Therefore, in practice, it may be more useful to use similar high-order 
conditions, but with the~$i_{\mu_L}^a\p_a$ operator applied to~$\hat{K}$. 

Next is the boundary condition for the longitudinal component of the shift.
Using the equations of motion~\eqref{eqn:gauge} 
and~\eqref{eqn:Gtilde_dot} we arrive at,
\begin{align}
\label{eqn:BC_first}
\p_t\tilde{\Gamma}^s&\hat{=}-\alpha\,\sqrt{\mu_{S_L}}\,\p_i\tilde{\Gamma}^i
+\chi^{-1}\p^A(\p_A\beta^s-\p_s\beta_A)\nonumber\\
&\quad-\tfrac{4\alpha}{3\chi(\mu_L-\mu_{S_L})}\left(
\sqrt{\mu_{S_L}}\Lie_n\hat{K}+\mu_L\Lie_s\hat{K}
\right)\nonumber\\
&\quad+\alpha\,\p_t^2h_{S_L}+\beta^i\p_i\tilde{\Gamma}^s\,.
\end{align}
The~$\Lie_n\hat{K}$ term can be substituted from the lapse boundary condition.
Here we have dropped several non-linear terms, but also terms involving the 
gamma-driver damping term~$\eta$. For applications one will have to experiment
with including this term to be sure that the longitudinal part of the shift does
not grow in an uncontrolled way.

The remaining two BCs for the gauge conditions are,
\begin{align}
\p_t\tilde{\Gamma}^A\,&\hat{=}-\alpha\,\sqrt{\mu_S}\,
\left[\p_s\tilde{\Gamma}^A-\p^A\tilde{\Gamma}^s\right]
-\frac{4\,\alpha}{3\,\chi}\,\p^A\hat{K}
+\frac{1}{\chi}\,\p^B\p_B\beta^A\nonumber\\
&\quad+\frac{4}{3\,\chi}\,\p^A\p_s\beta^s+\frac{1}{3\,\chi}\,
\p^A\p_B\beta^B
+\beta^i\p_i\tilde{\Gamma}^A\,,
\end{align}
in the vector sector. Here we have dropped non-principal terms and
set the given data to vanish. 

\paragraph*{Constraint preserving boundary conditions:} In 
terms of the conformal variables, the constraint preserving 
conditions for~$\Theta$ with~$L=1$ can be written,
\begin{align}
\p_t\Theta&\,\hat{=}\,-\alpha\,\sqrt{\mu_C}
\,\big(\p_s\Theta+\tfrac{1}{r}\Theta\big)
+\beta^i\p_i\Theta\,.
\end{align}
The longitudinal part of the~$Z_i$ boundary 
condition~\eqref{eqn:CPBCs_Z_Low} is given by,
\begin{align}
\p_t\tilde{A}_{ss}&\,\hat{=}\,
-\alpha\,\chi\,\left\{2\,\tilde{D}^i\tilde{A}_{is}
-\frac{4}{3}\,\tilde{D}_s\hat{K}
-\frac{2}{3}\,R_{ss}\right.\nonumber\\
&+\frac{2}{3}\chi\,\p_s\left[\tilde{\Gamma}^s
-(\tilde{\Gamma}_{\textrm{d}})^s\right]
-\frac{1}{3}\,\chi\,\p_A\left[\tilde{\Gamma}^A
-(\tilde{\Gamma}_{\textrm{d}})^A\right]
\nonumber\\
&\left.+\frac{1}{3}\,R_{qq}-3\,\tilde{D}^i(\ln\chi)\tilde{A}_{is}
-\kappa_1\,\left[\tilde{\Gamma}_s-(\tilde{\Gamma}_{\textrm{d}})_s
\right]\right\}
\nonumber\\
&+\alpha\,\left[\tilde{A}_{ss}\,(\hat{K}+2\,\Theta)
-2\,\tilde{A}^i{}_{s}\,\tilde{A}_{is}\right]
-\frac{2}{3}\,\chi\,D_sD_s\alpha\nonumber\\
&+\frac{1}{3}\,\chi\,D^AD_A\alpha+\Lie_\beta\tilde{A}_{ss}\,,
\end{align}
in the scalar sector. In the vector sector, the low order 
conditions~\eqref{eqn:CPBCs_Z_Low} become,
\begin{align}
\p_t\tilde{A}_{sA}&\,\hat{=}\,-\alpha\,\chi
\left\{\tilde{D}^i\tilde{A}_{iA}
-\frac{2}{3}\,\tilde{D}_A\hat{K}-R_{sA}\right.\nonumber\\
&-\frac{3}{2}\,\tilde{D}^i(\ln\chi)\,\tilde{A}_{iA} 
-\frac{1}{2}\,\kappa_1\,\left[\tilde{\Gamma}_A
-(\tilde{\Gamma}_{\textrm{d}})_A\right]
\nonumber\\
&\left.+\frac{1}{2}\chi\,q_{Ai}\,
\p_s\left[\tilde{\Gamma}^i
-(\tilde{\Gamma}_{\textrm{d}})^i\right]\right\}-\chi\,
D_AD_s\alpha\nonumber\\
&+\alpha\,\left[\tilde{A}_{sA}\,(\hat{K}+2\,\Theta)
-2\,\tilde{A}^i{}_A\tilde{A}_{is}\right]
+\Lie_\beta\tilde{A}_{sA}\,.
\end{align}
In the conformal decomposition of these BCs, it is important 
to keep all of the non-principal terms. Otherwise, the BCs will 
not be truly constraint preserving. Note that we are assuming 
compact support, away from the boundary of matter fields. 

\paragraph*{Radiation controlling boundary conditions:} After 
the conformal decomposition, lengthy calculations reveal that 
the~$L=1$ radiation controlling condition is 
\begin{align}
\p_t\tilde{A}^{\textrm{TF}}_{AB}&\,\hat{=}\,-\alpha\Big[
\tilde{D}_s\tilde{A}_{AB}-\tilde{D}_{(A}\tilde{A}_{B)s}
+\frac{1}{2}\tilde{A}_{s(A}\tilde{D}_{B)}(\ln\chi)
\nonumber\\
&-\frac{1}{2}\,\tilde{A}_{AB}\tilde{D}_s(\ln\chi)
+\tilde{A}^i{}_A\,\tilde{A}_{iB}
-\frac{2}{3}\,\tilde{A}_{AB}\,(\hat{K}+2\Theta)\Big]^{\textrm{TF}}
\nonumber\\
&+\alpha\,\chi\,\Big[(\iota_A\,\iota_B-\upsilon_A\,\upsilon_B)
\,\re(\p_t^2h_{\Psi_0})
\nonumber\\
&+2\,\iota_{(A}\,\upsilon_{B)}\,\im(\p_t^2h_{\Psi_0})\Big]
-\chi\,D_AD_B^{\textrm{TF}}\alpha
+\Lie_\beta\tilde{A}_{AB}^{\textrm{TF}}\,,\label{eqn:BC_last}
\end{align}
where~$\re(h_{\Psi_0})$ and~$\im(h_{\Psi_0})$ denote the 
real and imaginary parts of the boundary data~$h_{\Psi_0}$, 
respectively. Similarly to the constraint preserving conditions, 
for true control of the Weyl scalar~$\Psi_0$, all of the 
non-principal terms are required in these conditions. Note that 
in this subsection the spatial Ricci tensor as given in~\eqref{eqn:Conf_D_Ric}
should be evaluated {\it without} using the evolved contracted 
conformal Christoffels~$\tilde{\Gamma}^i$, but 
rather with~$(\tilde{\Gamma}_{\textrm{d}})^i$. This happens because we use
the boundary conditions to manipulate the equations of motion. 

\paragraph*{Implementation:} Remarkably, these expressions for 
the BCs suggest a natural generalization to  three-dimensions 
of the approach used for implementation inside
a numerical relativity code in spherical 
symmetry~\cite{RuiHilBer10}. Given a smooth boundary, the recipe 
is to populate as many ghostzones as required to compute 
finite differences and artificial dissipation at the boundary 
as in the bulk of the computational domain. Then, the standard 
evolution equations are used to update the metric components 
at the boundary, whilst the remaining variables are updated 
with~(\ref{eqn:BC_first}-\ref{eqn:BC_last}). This recipe has 
been used successfully in the evolution of blackhole and 
neutron star spacetimes~\cite{RuiHilBer10} in spherical 
symmetry. Similar conditions were also used in full 3D numerical 
relativity simulations of compact binary objects with the Z4c 
formulation, so there is reason to be optimistic that the 
recipe will work, although naturally a proof of numerical stability 
is desirable, at least for the linearized problem.

\section{Well-posedness analysis}
\label{section:WP}            

To prove that the resulting IBVP with the proposed BCs, namely 
Eqs.\eqref{eqn:bc_alpha}, \eqref{eqn:bc_betas}, \eqref{eqn:CPBCs_Theta},
\eqref{eqn:CPBCs_Z} and~\eqref{eqn:bc_psi0}, 
is well-posed, we work in the frozen coefficient approximation, 
where one considers small amplitude, high-frequency perturbations 
of a smooth background solution~\cite{KreLor89,GusKreOli95}. As 
pointed out before, this is the regime important for continuous 
dependence of the solution on the given data. It is expected that 
if the resulting problem is well-posed in this approximation the 
original nonlinear system will also be locally well-posed~\cite{Kre78,KreLor89}.

\subsection{Basic strategy}
\label{sec:Strategy}

Since there are a number of different ingredients in the 
analysis, we begin by summarizing our basic strategy. There 
are six key points. First we make a gauge choice that renders 
the PDE system strongly hyperbolic of constant multiplicity, 
which guarantees applicability of the Kreiss-Agranovich-M\'etivier 
theory. Second, to apply the theory we work in the linear 
high-frequency frozen coefficient approximation. Third, we perform 
the Laplace-Fourier transform, and make a pseudo-differential reduction 
to first order, resulting in a first order ODE system. Fourth, to 
represent the general solution of the system in a convenient form 
we choose dependent variables in which the equations of motion 
have a particular structure. This choice enables us to compute 
the solution easily in computer algebra (see Mathematica 
notebooks~\cite{Hilwebsite}). With the solution in hand 
we transform back to the original variables. Fifth, we express the 
high order boundary conditions in an algebraic form. Finally we 
substitute the general solution into the boundary conditions and 
solve in order to show boundary stability. 

\subsection{Strong hyperbolicity and multiplicity 
of speeds}\label{sec:Strong}

To apply the theory outlined in the following subsection we need 
conditions under  which the system is strongly hyperbolic
of constant multiplicity. Choosing an arbitrary unit spatial 
vector~$s^i$, not to be confused with the outward pointing 
normal used elsewhere in the paper, the principal symbol of the 
system coupled to the puncture gauge can be trivially read 
off from the principal part of the equations of motion under 
a~$2+1$ decomposition against~$s^i$ and discarding transverse 
derivatives. For convenience in this section we denote,
\begin{align}
\hat{\Gamma}^i=\chi\,\tilde{\Gamma}^i
+\tfrac{1}{2}\,\eta_\chi\,\tilde{\gamma}^{ij}\p_j\chi\,.
\end{align}
In the scalar sector we have,
\begin{align}
\p_t\alpha&\simeq-\alpha^2\,\mu_L\,\hat{K}+\beta^s\,\p_s\alpha\,,
\nonumber\\
\p_t\hat{K}&\simeq-\p_s\p_s\alpha+\beta^s\,\p_s\hat{K}\,,\nonumber\\
\p_t\beta^s&\simeq\alpha^2\,\mu_S\,\hat{\Gamma}^s
-\alpha\,\eta_L\,\p_s\alpha+\beta^s\,\p_s\beta^s\,,
\nonumber\\
\p_t\hat{\Gamma}^s&\simeq\mu_C\,\p_s\p_s\beta^s
-\alpha\,\mu_C\,\p_s\,\hat{K}+\beta^s\,\p_s\hat{\Gamma}^s\,,\nonumber\\
\p_t\gamma_{qq}&\simeq -2\,\alpha\,K_{qq} + \beta^s\,\p_s\gamma_{qq}\,,
\nonumber\\
\p_tK_{qq}&\simeq -\tfrac{1}{2}\,\alpha\,\p_s\p_s\gamma_{qq}
+\beta^s\,\p_sK_{qq}\,,\nonumber\\
\p_t\Theta&\simeq-\tfrac{1}{2}\,\alpha\,\p_s\p_s\gamma_{qq}+\alpha\,\p_sZ_s
+\beta^s\,\p_s\Theta,\nonumber\\
\p_tZ_s&\simeq-\alpha\,\p_sK_{qq}+\alpha\,\mu_C\,\p_s\Theta+\beta^s\,\p_sZ_s\,.
\end{align}
where~$\simeq$ denotes equality up to transverse derivatives and non-principal 
terms. In the vector sector, 
\begin{align}
\p_t\beta^A&\simeq\alpha^2\,\mu_S\,\hat{\Gamma}^A
+\beta^s\,\p_s\beta^A\,,\nonumber\\
\p_t\hat{\Gamma}^A&\simeq\p_s\p_s\beta^A
+\beta^s\,\p_s\tilde{\Gamma}^A\,,\nonumber\\
\p_tK_{sA}&\simeq\alpha\,\p_sZ_A+\beta^s\,\p_sK_{sA}\,,\nonumber\\
\p_tZ_A&\simeq\alpha\,\p_sK_{sA}+\beta^s\,\p_sZ_A\,.
\end{align}
Finally, in the tensor sector
\begin{align}
\p_t\gamma_{AB}^{\textrm{TF}}&\simeq-2\,\alpha\,K_{AB}^{\textrm{TF}}
+\beta^s\,\p_s\gamma_{AB}^{\textrm{TF}}\,,\nonumber\\
\p_tK_{AB}^{\textrm{TF}}&\simeq-\tfrac{1}{2}\,\alpha\,
\p_s\p_s\gamma_{AB}^{\textrm{TF}}
+\beta^s\,\p_sK_{AB}^{\textrm{TF}}.
\end{align}
Strong hyperbolicity, that is the existence of a pseudo-differential reduction to
first order possessing a principal symbol with a complete set of eigenvectors and 
imaginary eigenvalues~\cite{NagOrtReu04}, is equivalent to the existence of a 
complete set of characteristic variables~\cite{GunGar05} subject to a suitable 
uniformity condition. Except in special cases discussed below the equations of motion 
are strongly hyperbolic. The characteristic variables of the scalar sector are,
\begin{align}
u_{\pm\mu_L}&=\hat{K}\pm\tfrac{1}{\mu_L}\,\p_s\ln\alpha\,,\nonumber\\
u_{\pm\mu_{S_L}}&=\hat{\Gamma}^s\pm\tfrac{\sqrt{\mu_{S_L}}}
{\alpha\,\mu_S}\p_s\beta^s\nonumber\\
&-\tfrac{\sqrt{\mu_{S_L}}}{\mu_{S}(\mu_L-\mu_{S_L})}
\big[\sqrt{\mu_{S_L}}(1-{\eta}_L)\,\p_s\ln\alpha\nonumber\\
&\mp(\mu_{S_L}-{\eta}_L\,\mu_L)\hat{K}\big]\nonumber\\
u_{\pm1\,H,M}&= K_{qq}\pm\tfrac{1}{2}\,\p_s\gamma_{qq}\,,\nonumber\\
u_{\pm1\,\Theta,Z}&= -\tfrac{1}{2}\,\p_s\gamma_{qq}\pm\Theta+Z_s\,,
\end{align}
with speeds seen by the normal observer in the 
foliation~$\mp\sqrt{\mu_L}\,,\pm\sqrt{\mu_{S_L}}\,,\mp1$ and~$\pm1$. 
These variables are degenerate when~$\mu_{S_L}=\mu_L$, unless the harmonic gauge 
is chosen. The characteristic variables in the vector sector are, 
\begin{align}
u_{\pm\mu_S}^A&=\hat{\Gamma}^A
\pm\tfrac{1}{\alpha\sqrt{\mu_{S}}}\p_s\beta^A\,,\nonumber\\
u_{A\pm1\,Z,M}&= Z_A\pm K_{sA}\,,
\end{align}
with speeds~$\pm\sqrt{\mu_{S}}$ and~$\pm1$. In the 
tensor sector we have characteristic variables
\begin{align}
u_{\pm1\,AB}^{\textrm{TF}}&=\p_s\gamma_{AB}^{\textrm{TF}}
\pm \tfrac{1}{2} K_{AB}^{\textrm{TF}}\,,
\end{align}
with speeds~$\pm1$.

In typical evolutions of asymptotically flat data we have that 
$0\le\alpha\lesssim3/2$ and~$\gamma\geq 1$. Therefore, by 
choosing~$\mu_S$ sufficiently large we may expect to avoid
the degenerate special case mentioned above, and clashing
speeds so that for example either~$\mu_L<\mu_{S}=\mu_{S_L}$
or~$\mu_L<\mu_S<\mu_{S_L}$.

\subsection{Kreiss-Agranovich-M\'etivier Theory}
\label{sec:KA-theory}

In order to prove that the resulting IBVP of the system is 
well-posed, we use a theory developed by Kreiss~\cite{Kre70} 
which gives us necessary  and sufficient conditions for the 
well-posedness of the IBVP for {\it strictly hyperbolic systems}. 
Agranovich has extended this theory to the case in which the 
system is {\it strongly hyperbolic and the eigenvalues have 
constant multiplicity}~\cite{Agr72}. A more recent, and more 
digestible, demonstration of the theory can be found 
in~\cite{Met00}, although there the terminology differs slightly 
from ours. Here we briefly review this theory. 

\paragraph*{Basic system:} Consider a hyperbolic first order 
system 
\begin{align}
\partial_t {{u}}&=A^i\,\p_i{u}+{F}\nonumber\\&
= {A}^x\,\partial_{x}\, {u}
+\sum_{A=2}^d {A}^A\,\partial_A {u}+{F}\,,
\label{eqn:def-system}
\end{align}
with variable coefficients on the half-space~$t\ge 0$, 
$x\ge 0$ and~$-\infty < x^A<\infty$, where the index~$A\in [2,\cdots,d]$, 
where ${u}$ is an~$d$-dimensional vector, $ {A^x}$ and  
$ {A}^A$ are~$d\times d$ matrices and~${F}$ 
is a source term. We assume that~(\ref{eqn:def-system}) is strongly 
hyperbolic with constant multiplicity. This means that the principal 
symbol~$P=A^i\,s_i$, where~$s_i$ is an arbitrary spatial 
vector at any point in space, has a complete set of eigenvectors, 
which depend smoothly on~$s_i$, such that the number of coincident 
eigenvalues is constant over~$s_i$ and in space. With this assumption 
we furthermore restrict our attention to an arbitrary point on the 
boundary and work in the frozen coefficient approximation, so from 
here we assume that~$A^i$ is constant. 

\paragraph*{Boundary conditions:} Assuming that~$ {A^x}$ is 
non-singular, it can be rewritten in the form,
\begin{equation}
 {A^x}=\left(
\begin{array}{cc}
- {\Lambda}^I&0\\
0& {\Lambda}^{II}\\
\end{array}
\right)\,,
\label{eqn:formA}
\end{equation}
with $ {\Lambda}^I$ and $ {\Lambda}^{II}$ real and
positive definite diagonal matrices of order~$m$ and~$d-m$, 
respectively. We impose~$m$ BCs at~$x=0$ in the form
\begin{equation}
\left. {L}^I\, {u}^I(t,x)\right|_{x=0}\hateq 
\left. {L}^{II}\, {u}^{II}(t,x)\right|_{x=0}+
{g}(t,x^A)\,,
\label{eqn:BC}
\end{equation}
where~$ {L}^I$ and~$L^{II}$ are~$d\times m$ and~$d\times (d-m)$ constant 
matrices, respectively,  and~$ {g}= {g}(t,x^A)$ is given boundary 
data vector. Finally, we consider trivial  initial 
data~$ {u}(0,x,x^A)=0$.

\paragraph*{Laplace-Fourier transform:} In the following, we 
solve the above IBVP by performing a Laplace-Fourier (LF) 
transformation with respect to the directions~$t$ and~$x^A$ 
tangential to the boundary~$x=0$. Let~$\tilde{ {u}}
=\tilde{{u}}(s,x,\omega^A)$ denote the LF transformation 
of~$ {u}(t,x)$. Then, $\tilde{ {u}}$ satisfies the ordinary 
differential system
\begin{eqnarray}
&&\hspace{-1cm}
\partial_x\tilde{ {u}}= {M}(s,\omega)\,\tilde{{u}}+
\tilde{{F}}\,,
\qquad\textrm{on}\,x\in(0,\infty)\,,
\nonumber\\
&&\hspace{-1cm}
{L}^I\tilde{ {u}}^I
\hateq{L}^{II}\,\tilde{u}^{II}+
\tilde{g}\,,
\qquad\textrm{at}\,x\hateq0\,,
\label{eqn:BC-LF}
\end{eqnarray}
where~$\tilde{{g}}$ and~$\tilde{ {F}}$ denote the LF  
transformation of~$ {g}$ and~$ {F}$, respectively. 
In applications boundary conditions typically contain derivatives, 
but after LF transform we see that such conditions can nevertheless 
be written in this form, although we need then to take care of the 
norms in which estimates can be obtained. The matrix~${M}$ is given 
by,
\begin{equation}
 {M}(s,\omega)=  ({A^x})^{-1}\,(s\, {\mathbb{I}}_{d\times d}
+i\,\omega_A\,  {A}^A)\,,
\end{equation}
and~$\mathbb{I}_{m \times m}$ is the identity matrix. 

\paragraph*{General solution and theorems:} If~$\tau_i$ 
and~$e_i(s,\omega)$ are the corresponding eigenvalues, with 
negative real part, and eigenvectors of~${M}$ respectively then, 
assuming that~$\tilde{{F}}$ vanishes, the~$L_2$ solution of the 
above ODE system is given by,
\begin{equation}
\tilde{{u}}=\sum_{i=1}^m\sigma_i\, {e}_i(s,\omega)\,
\,\textrm{exp}(\tau_i\,x)\,,
\label{eqn:solution_LF}
\end{equation}
where~$\sigma_i$'s are complex integration constants which are
determined by the boundary conditions. In the case that~$M$ is
missing eigenvectors the general solution is modified in a standard way 
by a polynomial expression in~$x$ and using generalized eigenvectors. By 
substituting~(\ref{eqn:solution_LF}) into the 
expression~(\ref{eqn:BC-LF}) we obtain a system of~$m$ linear 
equations for the unknown~$\sigma_i$'s. 

\begin{Def}
The IBVP above system is called boundary stable 
if, for all~$\re(s)>0$ and~$\omega\in\mathbb{R}$, 
there is a positive constant~$C$ which does not depend on~$s$,~$\omega$ 
and~$\tilde{ {g}}$ such that
\begin{equation}
|\tilde{ {u}}(s,0,\omega)|\leq C\,|\tilde{ {g}}(s,\omega)|\,.
\label{eqn:bound-stab-cond}
\end{equation}
\end{Def}
It is straightforward to show that boundary stability is a necessary 
condition for well-posedness~\cite{GusKreOli95}. Agranovich showed that 
if the system is strongly hyperbolic with eigenvalues of constant multiplicity 
and boundary stable then there exists a smooth 
symmetrizer~$\hat{R}=\hat{R}(s,\omega)$ with the following properties~\cite{Agr72}:
\begin{itemize}
\item $\hat{R}$ is a Hermitian matrix,
\item there is a positive constant $C_1$ such that
\begin{equation*}
\hat{R}\,M_I+M^*_I\,\hat{R}\geq C_1\,\re(s)\,\mathbb{I}_{m\times m}\,,
\end{equation*}
\item for all $\tilde{u}$ which satisfy the boundary 
conditions~(\ref{eqn:BC-LF}), there are positive constants~$C_2$ 
and~$C_3$ such that
\begin{equation*}
\left<\hat{R}\,\tilde{u},\tilde{u}\right>+C_2\,|\tilde{g}|
\geq C_3\,|\tilde{u}|^2\,,\qquad\textrm{at}\, x=0\,,
\end{equation*}
\end{itemize}
where~$\left<\cdot,\cdot\right>$ and $|\cdot|$ denote the scalar product 
in $\mathbb{C}^d$ and the corresponding norm, respectively. Therefore, 
using this symmetrizer, the well-posedness of the above IBVP can be 
established via a standard energy estimation in the frequency domain.
By inverting the LF transformation, one can show 
that~\cite{Kre70,Agr72,KreWin06}
\begin{Thm}
If the above IBVP is boundary stable then it is  strongly well-posed in 
the generalized sense. The solution~$u=u(t,x^i)$ satisfies the estimation
\begin{eqnarray}
\int_0^t \| {u}(\cdot,\tau)\|^2_\Sigma\,d\tau + 
\int_0^t \| {u}(\cdot,\tau)\|^2_{\p\Sigma}\,d\tau\nonumber\\
\leq K_T\,\left\{ \int_0^t \| {F}(\cdot,\tau)\|^2_\Sigma d\tau +
\int_0^t \| {g}(\cdot,\tau)\|^2_{\p\Sigma}\,d\tau\right\}\,,
\label{eqn:kreiss-estim}
\end{eqnarray}
in the interval $0\leq{t}\leq{T}$ for a positive constant $K_T$ which does 
not depend on~$F$ and~$g$. Here $\|\cdot\|_\Sigma,\|\cdot\|_{\p\Sigma}$ 
denote the $L_2$ norm with respect to the half-space and the boundary 
surface, respectively.
\end{Thm}
As pointed out earlier (see for instance~\cite{KreLor89}), 
using pseudo-differential operators and the symmetrizer~$\hat{R}$, 
well-posedness can be established in the variable coefficient and
quasilinear case. 

\paragraph*{Second order systems:} The equations of motion are 
not a first order system of the form~\eqref{eqn:def-system}, but 
fortunately this issue can be side-stepped by 
following~\cite{KreWin06}. Since the theory summarized here is 
developed with pseudo-differential calculus, the results carry 
over to hyperbolic systems of higher order by working with an 
appropriate first order pseudo-differential reduction of the 
form~\eqref{eqn:BC-LF}, which is the strategy we adopt. 

\subsection{Laplace-Fourier transformed system}
\label{sec:LFofsEoM}

In the frozen coefficient approximation, only the principal part 
of the equations of motion is considered and the  coefficient 
appearing in front of any operator is frozen to its 
value at an arbitrary point~$p$. By performing a suitable coordinate 
transformation which leaves the  foliation
$\Sigma_t = \{ t \} \times \Sigma$ invariant, it is possible to bring 
the background metric into the form~\cite{RuiRinSar07}, 
\begin{equation}
ds^2(p)|_p = -dt^2 + (dx + \mbeta\, dt)^2 + dy^2 + dz^2\,,
\label{eqn:FrozenMetric}
\end{equation}
where~$\mbeta$ is a constant, which we will assume to be smaller 
than one in magnitude. This is a condition which holds near the
boundary since the boundary surface~$\mathcal{T}$ is, by assumption, 
time-like. If, as will typically be the case, we insist on imposing 
boundary conditions under the boundary orthogonality condition 
we have~$\mbeta=0$. We will, nevertheless, keep track of the background 
shift for as long as possible to help clarify the resulting difficulties.

The non-linear IBVP for the formulation is thus reduced to a
linear constant coefficient problem on the manifold
$\Omega=(0,\infty)\times\Sigma$, where
$\Sigma=\{(x,y,z)\,\in\,\mathbb{R}^3:x>0\}$ is the half-plane. 
Restricting our attention to the high-frequency frozen 
coefficient limit, and performing the LF transform, we define a 
triad from the vectors~$\hat{x}^i,\hat{\omega}^A,\hat{\nu}^A$, 
where~$\hat{x}^i=-s^i$ with~$s^i$ the unit normal to the 
boundary as before, $\omega^A$ is the wave vector from the 
Fourier transform, and~$\omega^A=\omega\,\hat{\omega}^A$ 
with~$\omega=\sqrt{\omega^A\,\omega_A}$. Note again that these 
quantities are now defined with respect to the background 
metric. We form a projection operator into the boundary from 
the two members of the basis, 
\begin{align}
q_{ij}&=\hat{\omega}_i\hat{\omega}_j+\hat{\nu}_i\hat{\nu}_j,
\end{align}
which is compatible with the projection operator used 
in the strong hyperbolicity analysis. For later convenience, 
we introduce the normalized quantities~$\omega'=\omega/\kappa$ 
and~$s'=s/\kappa$ with~$\kappa=\sqrt{|s|^2+\omega^2}$. We 
decompose the resulting ODE system against the triad as,
\begin{align}
\tgamma_{ij}&=\hat{x}_i\,\hat{x}_j\tgamma_{\hat{x}\hat{x}}
+\tfrac{1}{2}\,q_{ij}\,\tgamma_{qq}
+2\,\hat{x}_{(i}\,\hat{\omega}_{j)}\,\tgamma_{\hat{x}\hat{\omega}}\nonumber\\
&+2\,\hat{x}_{(i}\,\hat{\nu}_{j)}\,\tgamma_{\hat{x}\hat{\nu}}
+2\,\hat{\omega}_{(i}\,\hat{\nu}_{j)}\,\tgamma_{\hat{\omega}\hat{\nu}}
+\hat{\nu}_i\,\hat{\nu}_j\,\tgamma_{\hat{\nu}\hat{\nu}}\,,
\end{align}
where here and in what follows, lapse, shift and metric 
components marked with a tilde denote the corresponding Laplace, 
with respect to~$t$, and Fourier transformed, with respect 
to~$y$ and~$z$, quantity, and are not to be confused with 
the conformal metric used in numerical applications. For details 
on the LF approach please refer to e.g.~\cite{SarTig12}.
This decomposition results in the second order ODE system,
\begin{align}
\label{eqn:LF-first}
\kappa^2\Lie_0^2\,\talpha&=\mu_L\,(\p_x^2-\omega^2)\,\talpha\,,
\nonumber\\
\kappa^2\Lie_0^2\,\tbeta_{\hat{x}}&=
\mu_{S}\,(\mu_C\,\p_x^2-\omega^2)\,\tbeta_{\hat{x}}+\mu_S\,(\mu_C-1)
\,i\,\omega\,\p_x\tbeta_{\hat{\omega}}\nonumber\\
&+\Big(\frac{\mu_{S_L}}{\mu_L}-{\eta}_L\Big)\,\kappa\,\Lie_0\,\p_x\talpha\,,
\nonumber\\
\kappa^2\Lie_0^2\,\tbeta_{\hat{\omega}}&=\mu_S\,(\p_x^2-\mu_C\,\omega^2)\,
\tbeta_{\hat{\omega}}+\mu_S\,(\mu_C-1)\,i\,\omega\,\p_x\tbeta_{\hat{x}}\
\nonumber\\
&+\Big(\frac{\mu_{S_L}}{\mu_L}-\eta_L\Big)\,i\,\omega\,\kappa\,\Lie_0\talpha\,,\nonumber\\
\kappa^2\Lie_0^2\,\tbeta_{\hat{\nu}}&=\mu_S\,(\p_x^2-\omega^2)\,
\tbeta_{\hat{\nu}}\,,
\end{align}
for the gauge variables and 
\begin{align}
\kappa^2\Lie_0^2\,\tgamma_{\hat{x}\hat{x}}&=
\,(\p_x^2-\omega^2)\,\tgamma_{\hat{x}\hat{x}}
+\tfrac{1}{3}\,(1-\eta_\chi)\,\p_x^2\,(\tgamma_{\hat{x}\hat{x}}+\tgamma_{qq})\nonumber\\
&\hspace{-.75cm}+2\,\Big(1-\frac{{\eta}_L}{\mu_S}\Big)\,\p_x^2\talpha
+2\,\Big(1-\frac{1}{\mu_S}\Big)\,\kappa\,\Lie_0\,\p_x\tbeta_{\hat{x}}\,,
\nonumber\\
\kappa^2\Lie_0^2\,\tgamma_{qq}&=\,(\p_x^2-\omega^2)\,\tgamma_{qq}
-\tfrac{1}{3}(1-\eta_\chi)\,\omega^2\,(\tgamma_{\hat{x}\hat{x}}+\tgamma_{qq})\nonumber\\
&\hspace{-.75cm}-2\,\Big(1-\frac{{\eta}_L}{\mu_S}\Big)\,\omega^2\,\talpha
+2\,\Big(1-\frac{{\eta}_L}{\mu_S}\Big)
\,i\,\omega\,\kappa\,\Lie_0\,\tbeta_{\hat{\omega}}\,,\nonumber\\
\kappa^2\Lie_0^2\,\tgamma_{\hat{x}\hat{\omega}}&=
\,(\p_x^2-\omega^2)\,\tgamma_{\hat{x}\hat{\omega}}
+\tfrac{1}{3}(1-\eta_\chi)\,i\,\omega\,\p_x(\tgamma_{\hat{x}\hat{x}}+\tgamma_{qq})
\nonumber\\
&+2\,\Big(1-\frac{{\eta}_L}{\mu_S}\Big)\,i\,\omega\,\p_x\talpha
\nonumber\\
&+2\,\Big(1-\frac{1}{\mu_S}\Big)\,\kappa\,\Lie_0\,
(\p_x\tbeta_{\hat{\omega}}+i\,\omega\,\beta_{\hat{x}})
\,,\nonumber\\
\kappa^2\Lie_0^2\,\tgamma_{\hat{x}\hat{\nu}}&=
\,(\p_x^2-\omega^2)\,\tgamma_{\hat{x}\hat{\nu}}
+\Big(1-\frac{1}{\mu_S}\Big)\,\kappa\,\Lie_0\,
\p_x\tbeta_{\hat{\nu}}
\,,\nonumber\\
\kappa^2\Lie_0^2\,\tgamma_{\hat{\omega}\hat{\nu}}&=
\,(\p_x^2-\omega^2)\,\tgamma_{\hat{\omega}\hat{\nu}}
+\Big(1-\frac{1}{\mu_S}\Big)\,i\,\omega\,
\kappa\,\Lie_0\,\tbeta_{\hat{\nu}}\,,\nonumber\\
\kappa^2\Lie_0^2\,\tgamma_{\hat{\nu}\hat{\nu}}&=(\p_x^2-\omega^2)\,
\tgamma_{\hat{\nu}\hat{\nu}}\,,
\label{eqn:LF-last}
\end{align}
for the metric, where we use the shorthand~$\Lie_0=s'-\kappa^{-1}\,\mbeta\,\p_x$. 
To reduce the system to first order we use the normalized 
pseudo-differential reduction variables,
\begin{align}
d\talpha&=\kappa^{-1}\p_x\talpha\,,\qquad 
d\tbeta_i=\kappa^{-1}\p_x\tbeta_i\,,\nonumber\\
d\tgamma_{ij}&=\kappa^{-1}\p_x\tgamma_{ij}\,,
\label{eqn:Red-LF}
\end{align}
and decompose them as above. Substituting these definitions 
into~(\ref{eqn:LF-first}-\ref{eqn:LF-last}), we can solve for the
LF equations of motion for the new variables. The 
reduction is crucial for the application of the 
Kreiss-Agranovich-M\'etivier theory. We suppress the equations to 
avoid repetition, but they can be found in the Mathematica notebooks 
that accompany the paper. The symbol~$M(s,\omega)$ of the ODE system
resulting from the LF transform can be straightforwardly read 
off from the reduced equations.

\subsection{$L_2$ solution of the reduction}
\label{sec:L2Soln}

\paragraph*{Change of variables:} To construct the general~$L_2$ 
solution of the first order reduction, we begin by transforming 
to a convenient choice of variables, which we find greatly speeds 
up the calculations in computer algebra. We remove,
\begin{align}
\{\tbeta_{\hat{x}},\tgamma_{\hat{x}\hat{x}},\,\tgamma_{qq},
\,\tgamma_{\hat{x}\hat{\omega}},\,\tgamma_{\hat{x}\hat{\nu}},\,
\tgamma_{\hat{\nu}\hat{\nu}}\,\}\nonumber
\end{align}
and their corresponding first derivative reduction variable 
from the state vector and replace them with the variables,
\begin{align}
\label{eqn:change_first}
\tilde{\Lambda}&=\tgamma_{\hat{x}\hat{x}}+\tgamma_{qq}
+2\frac{1-{\eta}_L}{\mu_{S_L}-\mu_L}\talpha,\nonumber\\
\tT&= \frac{1}{2\,\mu_L}\Lie_0\talpha
-\frac{1}{4}\,\mathcal{L}_0(
\tgamma_{\hat{x}\hat{x}}+\tgamma_{qq})\nonumber\\
&+\frac{1}{2}(d\tbeta_{\hat{x}}+\,i\,\omega'\tbeta_{\hat{\omega}})\,,
\nonumber\\
\tZ_{\hat{x}}&=\frac{1}{2\,\mu_{S}}\,\mathcal{L}_0\tbeta_{\hat{x}}
+\frac{{\eta}_L}{2\,\mu_S}\,d\talpha
-\frac{1}{4}\,\mu_C\,d\tgamma_{\hat{x}\hat{x}}
\nonumber\\
&+\frac{1}{4}\,(2-\mu_C)\,d\tgamma_{qq}
-\frac{i}{2}\,\omega'\,\tgamma_{\hat{x}\hat{\omega}}\,,
\nonumber\\
\tilde{Z}_{\hat{\omega}}&=
\frac{1}{2\,\mu_{S}}\,\mathcal{L}_0\tbeta_{\hat{\omega}}
+\frac{{\eta}_L}{2\,\mu_{S}}\,i\,\omega'\,\talpha
\nonumber\\
&+\frac{1}{4}(2-\mu_C)
\,i\,\omega'\,\tgamma_{\hat{x}\hat{x}}
-\frac{1}{4}\,\mu_C\,i\,\omega'\,
\tgamma_{qq}\nonumber\\
&+\frac{i}{2}\,\omega'\,\tgamma_{\hat{\nu}\hat{\nu}}
-\frac{1}{2}d\tgamma_{\hat{x}\hat{\omega}}\,,
\nonumber\\
\tilde{Z}_{\hat{\nu}}&=\frac{1}{2\,\tilde{\mu}_{S}}\,
\mathcal{L}_0\tbeta_{\hat{\nu}}
-\frac{1}{2}\,i\,\omega'\tgamma_{\hat{\omega}\hat{\nu}}
-\frac{1}{2}\,d\gamma_{\hat{x}\hat{\nu}},\nonumber\\
\tgamma_{\hat{\omega}\hat{\omega}}&=\tgamma_{qq}-\tgamma_{\hat{\nu}\hat{\nu}}\,,
\end{align}
and also,
\begin{align}
D\tilde{\Lambda}&=\Lie^{\mu_{S_L}}_x\tilde{\Lambda}\,,\quad&
D\tT=\Lie_x^{\mu_{C}}\tT\,,\nonumber\\
D\tZ_{\hat{x}}&=\Lie_x\tZ_{\hat{x}}\,,\quad&
D\tilde{Z}_{\hat{\omega}}=\Lie_x\tZ_{\hat{\omega}}\,,\nonumber\\
D\tilde{Z}_{\hat{\nu}}&=\Lie_x\tZ_{\hat{\nu}}\,,\quad&
D\tgamma_{\hat{\omega}\hat{\omega}}=\Lie_xD\tgamma_{\hat{\omega}\hat{\omega}}\,.
\label{eqn:change_last}
\end{align}
Here we have defined,
\begin{align}
\Lie^\mu_x&=\kappa^{-1}\,\p_x+\gamma_\mu^2\,\mbeta\,s'\,,
\end{align}
and write~$\Lie^1_x=\Lie_x$. We furthermore introduce the 
shorthand~$\gamma_\mu^{-2}=\mu-\mbeta^2$. Note that~$\gamma$ 
in this section is not to be confused with the determinant of the spatial 
metric, which is fixed in the frozen coefficient approximation. We also use,
\begin{align}
\lambda_\mu&=\sqrt{s'^2+\gamma_\mu^{-2}\omega'^2}\,,\nonumber\\
\tau_{\mu\pm}&=-\kappa\,\gamma_\mu^2\,(s'\,\mbeta\,\mp\,\sqrt{\mu}\,\lambda_\mu)\,,
\end{align}
and write~$\tau_{\mu\pm}'=\tau_{\mu\pm}/\kappa$. In the definition 
of~$\lambda_\mu$ we take the square root to have positive real part. We 
likewise write~$\gamma_1=\gamma$, $\lambda_1=\lambda$ 
and~$\tau_{1\pm}=\tau_{\pm}$. To further simplify the form of the ODE system we 
replace~$d\talpha,\,d\tbeta_{\hat{\omega}},\,d\tbeta_{\hat{\nu}},\,
d\tgamma_{\hat{\omega}\hat{\nu}}$, with
\begin{align}
D\talpha&=\Lie^{\mu_L}_x\talpha\,,\quad&
D\tbeta_{\hat{\omega}}=\Lie^{\mu_{S}}_x\tbeta_{\hat{\omega}}\,,\nonumber\\
D\tbeta_{\hat{\nu}}&=\Lie^{\mu_{S}}_x\tbeta_{\hat{\nu}}\,,\quad&
D\tgamma_{\hat{\omega}\hat{\nu}}=\Lie_x\tgamma_{\hat{\omega}\hat{\nu}}\,.
\label{eq:reduction_vbeta}
\end{align}
The choice of variables here seems natural except that one would naively
prefer to use~$\tbeta_{\hat{x}}$ rather than~$\tilde{\Lambda}$ 
and~$\tgamma_{qq}-2\tgamma_{\hat{\nu}\hat{\nu}}$ rather 
than~$\tgamma_{\hat{\omega}\hat{\omega}}$. Indeed, when working
under the boundary orthogonality condition
this is possible, but if~$\mbeta\ne0$ the resulting transformation is not 
invertible for some~$s'$ with positive real part. Therefore, we make this 
minor compromise so that we can construct the general~$L_2$ solution 
easily in the more general case as well. The composite transformation has 
determinant,
\begin{align}
\frac{\gamma_{\mu_C}^2\,\tau_+'^3\,\tau_-'^3\,\tau'_{\mu_{S}+}\,
\tau'_{\mu_{S}-}}{256\,\gamma_{\mu_S}^2\,\mu_{S}},\nonumber
\end{align}
and since the real part of~$s'$ is greater than zero 
the transformation is always invertible. We do not require any
boundedness property on this transformation. We use it only to arrive at 
equations of motion with the convenient lower block diagonal form, which 
allows us to easily construct the general solution to the ODE system
in computer algebra. Once we have the various eigenvectors we immediately 
transform back to the original variables. Note that the constraint violating 
variables are the LF transform of the constraint violations 
normalized by a factor of~$\kappa$. 

\paragraph*{Reduced equations of motion:} In terms of 
these variables, the system splits into a number of decoupled
or closed subsystems, starting with the Laplace-Fourier 
transformed constraint subsystem,
\begin{align}
\Lie_x\tT&=D\tT\,,\quad\quad\Lie_xD\tT=\mu_C\,\gamma_{\mu_C}^4
\,\lambda_{\mu_C}^2\,\tT\,,
\nonumber\\
\Lie_x\tZ_{\hat{x}}&=D\tZ_{\hat{x}}\,,\quad\quad
\Lie_x\tZ_{\hat{\omega}}=D\tZ_{\hat{\omega}}\,, 
\nonumber\\
\Lie_xD\tZ_{\hat{x}}&=\lambda^2\,\tZ_{\hat{x}}
+\gamma^2\,(\mu_C-1)\,\kappa^{-1}\Lie_0\p_x\tT\,,\nonumber\\
\Lie_xD\tZ_{\hat{\omega}}&=\lambda^2\,\tZ_{\hat{\omega}}+\gamma^2\,(\mu_C-1)
\,i\,\omega'\Lie_0\tT
\,,\nonumber\\
\Lie_x\tZ_{\hat{\nu}}&=D\tZ_{\hat{\nu}}\,,\quad\quad
\Lie_xD\tZ_{\hat{\nu}}=-\lambda^2\,\tZ_{\hat{\nu}}\,.
\label{eqn:LF_constraints}
\end{align}
which is coupled to the equations for the gauge variables,
\begin{align}
\Lie^{\mu_L}_x\talpha&=D\talpha\,,\quad
\Lie^{\mu_L}_xD\talpha=\mu_L\,\gamma_{\mu_L}^4\,\lambda_{\mu_L}^2\,\talpha\,,\nonumber\\
\Lie^{\mu_{S_L}}_{x}\tilde{\Lambda}&=D\tilde{\Lambda}\,,
\nonumber\\
\Lie^{\mu_{S_L}}_{x}D\tilde{\Lambda}&=\mu_{S_L}
\gamma_{\mu_{S_L}}^4\,\lambda_{\mu_{S_L}}^2
\,\tilde{\Lambda}-4\,\gamma_{\mu_{S_L}}^2\,(\mu_S-1)\,\Lie_0\tT\,,\nonumber\\
\Lie^{\mu_{S}}_{x}\tbeta_{\hat{\omega}}&=D\tbeta_{\hat{\omega}}\,,
\nonumber\\
\Lie^{\mu_{S}}_{x}D\tbeta_{\hat{\omega}}&=
\mu_S\,\gamma_{\mu_S}^4\,\lambda_{\mu_{S}}^2\,\tbeta_{\hat{\omega}}
\nonumber\\
&\hspace{-.8cm}+\gamma_{\mu_{S}}^2\frac{(\mu_L-\mu_{S})
(\eta_L\mu_L-\mu_{S_L})}
{\mu_L(\mu_L-\mu_{S_L})}\,i\,\omega'\,\Lie_0\talpha\nonumber\\
&\hspace{-.8cm}+\frac{1}{2}\,\gamma_{\mu_{S}}^2
\,(\mu_{S}-\mu_{S_L})\,i\,\omega'\,\big(\Lie_0\tilde{\Lambda}+4\,\tT\big)\,,
\label{eqn:LF_gauge}
\end{align}
and the metric components,
\begin{align}
\Lie_x\tgamma_{\hat{\omega}\hat{\omega}}&=D\tgamma_{\hat{\omega}\hat{\omega}}\,,
\nonumber\\
\Lie_xD\tgamma_{\hat{\omega}\hat{\omega}}&=\gamma^4\lambda^2\,\tgamma_{\hat{\omega}\hat{\omega}}
+\gamma^2(\mu_C-1)\,\omega'^2\,\tilde{\Lambda}
\nonumber\\
&\hspace{-1.3cm}+\frac{2\,\gamma^2}{\mu_L-\mu_{S_L}}\Big[
\mu_L-\mu_{S_L}+\mu_C-1
+\frac{\eta_L}{\mu_S}(\mu_L-\mu_S)\Big]\,\omega'^2\,\talpha\nonumber\\
&\hspace{-1.3cm}
+2\,\gamma^2\frac{1-\mu_S}{\mu_S}\,i\,\omega'
\Lie_0\tbeta_{\hat{\omega}}\,.\label{eqn:LF_gomom}
\end{align}
The second subsystem is completely decoupled, and is formed from the 
remaining shift and metric components,
\begin{align}
\Lie^{\mu_S}_{x}\tbeta_{\hat{\nu}}&=D\tbeta_{\hat{\nu}}\,,\quad
\Lie^{\mu_S}_{x}D\tbeta_{\hat{\nu}}
=\mu_S\,\gamma_{\mu_S}^4\,\lambda_{\mu_S}^2\,\tbeta_{\hat{\nu}}\,,\nonumber\\
\Lie_x\tgamma_{\hat{\omega}\hat{\nu}}&=D\tgamma_{\hat{\omega}\hat{\nu}}\,,\nonumber\\
\Lie_xD\tgamma_{\hat{\omega}\hat{\nu}}&=\gamma^4\lambda^2\,\tgamma_{\hat{\omega}\hat{\nu}}
+\gamma^2\frac{1-\mu_S}{\mu_S}\,i\,\omega'
\Lie_0\tbeta_{\hat{\nu}}\,.\label{eqn:LF_second_sub}
\end{align}

\paragraph*{Properties of the symbol:} The two decoupled
subsystems~\eqref{eqn:LF_constraints}-\eqref{eqn:LF_gomom}
and~\eqref{eqn:LF_second_sub} can be written in the form,
\begin{align}
\p_x\tilde{u}&=\kappa\,M\tilde{u}\,.
\label{eq:EqoMMatrix}
\end{align}
Ordering the state vector according to 
equations~\eqref{eqn:LF_constraints}-\eqref{eqn:LF_gomom}
and~\eqref{eqn:LF_second_sub}, the symbol of these two subsystems 
has a lower block diagonal form, a familiar structure as identified 
in~\cite{HilRic13},
\begin{align}
\label{eqn:low_block}
M&=\left(\begin{array}{cc}
A & 0 \\
B & C
\end{array}\right).
\end{align}
In the first decoupled subsystem~\eqref{eqn:LF_constraints}-\eqref{eqn:LF_gomom}
there are in fact two natural places for such a partition, namely 
after~$D\tZ_{\hat{\nu}}$ and similarly after 
after~$D\tbeta_{\hat{\omega}}$ in the state vector. For the second decoupled 
subsystem~\eqref{eqn:LF_second_sub} the partition lies after~$D\tbeta_{\hat{\omega}}$. 
The upper left block of the first system, corresponding to the constraint
subsystem, has eigenvalues~$\tau'_{\mu_C\pm}$, and~$\tau'_{\pm}$ of multiplicity three, 
and a complete set of eigenvectors for every~$s'$ and~$\omega'$. The central block 
of~\eqref{eqn:LF_constraints}-\eqref{eqn:LF_gomom}, corresponding to part of the 
pure gauge subsystem, has eigenvalues~$\tau'_{\mu_L\pm},\tau'_{\mu_{S_L}\pm}\tau'_{\mu_S\pm}$,
each of multiplicity one and likewise a complete set of eigenvectors for every 
frequency. The lower right block has eigenvalues~$\tau'_{\pm}$ and a complete set of 
eigenvectors. The decoupled subsystem~\eqref{eqn:LF_second_sub} has 
eigenvalues~$\tau'_{\mu_C\pm},\tau'_{\pm}$ and again a complete set of eigenvectors 
at every frequency. The eigenvalues of the full principal symbol are simply the union 
of those of the various subsystems. For a generic gauge condition, the full principal 
symbol of the subsystem~\eqref{eqn:LF_constraints}-\eqref{eqn:LF_gomom}
is diagonalizable unless~$s'=\pm\mbeta\,\omega'$. Diagonalizability 
when~$s'=\pm\mbeta\,\omega'$ is restored by restricting the gauge choice to,
\begin{align}
\label{eqn:still_diag}
\mu_{S_L}=\mu_S={\eta}_L\,,
\end{align}
a special case that includes the harmonic gauge. 
Since the square root in~$\lambda_\mu$ has positive 
real part for~$\re(s')>0$, 
\begin{align}
\re(\lambda_\mu)\geq\re(s')\,,
\end{align}
and since~$\re(s')$ is a strictly positive parameter it follows 
that~$\re(\tau_{\mu-})<0< \re(\tau_{\mu+})$. So all of the eigenvalues 
with~``-'' have negative real part and have corresponding solutions which 
are~$L_2$. For~$s'\ne-\mbeta\,\omega'$ the 
eigenvalues~$\tau_{\mu_{L}-},\tau_{\mu_{S_L}-},
\tau_{\mu_{S-}},\tau_-$ are pairwise distinct, and the full 
principal symbol has a complete set of eigenvectors, thus the~$L_2$ 
solution of the IBVP is of the type~\eqref{eqn:solution_LF}.
When~$s'=-\mbeta\,\omega',$ all of the eigenvalues with negative 
real part clash, with value~$-\omega'$, and the full principal 
symbol is missing two eigenvectors, so a polynomial ansatz is 
needed for the associated eigensolutions.

\paragraph*{General Solution for~$s'\,\ne\,-\mbeta\,\omega'$:} The 
general~$L_2$ solution can be computed from the eigenvectors 
of~$M$. In practice to do this we work with the matrices 
described in the last section and then transform back to the
original variables. We now define the abbreviation,
\begin{align}
\chi_\mu&=(\mbeta\,\lambda_\mu+\sqrt{\mu}\,s')\,\gamma_\mu^2\,\sqrt{\mu}\,.
\end{align}
For~$s'\,\ne\,-\mbeta\,\omega'$, the solution 
at the boundary~$x=0$ is given by the remarkably simple 
expressions,
\begin{align}
\talpha&=\sigma_{\talpha}\,,\nonumber\\
\tbeta_{\hat{x}}&=\frac{\mu_{S_L}\,\tau_{\mu_{S_L-}}'}
{2\,\chi_{\mu_{S_L}}}\,\sigma_{\tilde{\Lambda}}
-\frac{(\mu_{S_L}-\eta_L\,\mu_L)\,\tau_{\mu_{L-}}'}
{(\mu_{S_L}-\mu_L)\,\chi_{\mu_L}}\,
\,\sigma_{\talpha}
-\frac{i\,\omega'}{\tau_{\mu_{S-}}'}
\,\sigma_{\tbeta_{\hat{\omega}}}\,,\nonumber\\
\tbeta_{\hat{\omega}}&=\sigma_{\tbeta_{\hat{\omega}}}
+\frac{\mu_{S_L}}{2\,\chi_{\mu_{S_L}}}
\,i\,\omega'\,\sigma_{\tilde{\Lambda}}-\frac{\mu_{S_L}-\eta_L\,\mu_L}
{(\mu_{S_L}-\mu_L)\,\chi_{\mu_L}}\,i\,\omega'\,\sigma_{\talpha}
\,,\nonumber\\
\tbeta_{\hat{\nu}}&=\sigma_{\tbeta_{\hat{\nu}}}\,,
\end{align}
for the gauge variables restricted to the boundary. For the metric 
we find,
\begin{align}
\tgamma_{\hat{x}\hat{x}}&=-\frac{\omega'^2}{\tau_-'^2}\,\sigma_{\tgamma_{\hat{\omega}\hat{\omega}}}
-2\,\frac{(1-\eta_L)\,\mu_L\,\tau_{\mu_{L-}}'^2}{(\mu_{S_L}-\mu_L)\,\chi_{\mu_L}^2}\,\sigma_{\talpha}
+\frac{\mu_{S_L}\,\tau_{\mu_{S_L-}}'^2}{\chi_{\mu_{S_L}}^2}\,\sigma_{\tilde{\Lambda}}
\nonumber\\
&\quad-\frac{2\,i\,\omega'}{\chi_{\mu_{S}}}\sigma_{\tbeta_{\hat{\omega}}}
-4\,\frac{\mu_C\,\tau_{\mu_{C-}}'^2}{\chi_{\mu_C}^3}\,\sigma_{\tT}
-\frac{2}{\tau_-'}\,\sigma_{\tZ_{\hat{x}}}
+\frac{2\,i\,\omega'}{\tau_-'^2}\,\sigma_{\tZ_{\hat{\omega}}}\nonumber\\
\tgamma_{qq}&=\frac{\omega'^2}{\tau_-'^2}\,\sigma_{\tgamma_{\hat{\omega}\hat{\omega}}}
+2\,\frac{(1-\eta_L)\,\mu_L\,\omega'^2}{(\mu_{S_L}-\mu_L)\,\chi_{\mu_L}^2}\,\sigma_{\talpha}
-\frac{\mu_{S_L}\,\omega'^2}{\chi_{\mu_{S_L}}^2}
\,\sigma_{\tilde{\Lambda}}\nonumber\\
&\quad+\frac{2\,i\,\omega'}{\chi_{\mu_{S}}}\,\sigma_{\tilde{\beta}_{\hat{\omega}}}
-4\frac{\mu_C\,\omega'^2}{\chi_{\mu_C}^3}\sigma_{\tT}
+\frac{2}{\tau_-'}\sigma_{\tZ_{\hat{x}}}-\frac{2\,i\,\omega'}{\tau_-'^2}\sigma_{\tZ_{\hat{\omega}}}\,,
\end{align}
for the components that would appear in the scalar sector of the principal symbol in the~$\hat{x}$
direction. Next we have,
\begin{align}
\tgamma_{\hat{x}\hat{\omega}}&=
-\frac{i\,\omega'}{\tau_-'}\,\sigma_{\tgamma_{\hat{\omega}\hat{\omega}}}
-2\,\frac{(1-\eta_L)\,\mu_L\,\tau_{\mu_{L-}}'}
{(\mu_{S_L}-\mu_L)\,\chi_{\mu_L}^2}\,i\,\omega'\,\sigma_{\talpha}
\nonumber\\
&\quad+\frac{\mu_{S_L}\,\tau_{\mu_{S_L-}}'}{\chi_{\mu_{S_L}}^2}
\,i\,\omega'\,\sigma_{\tilde{\Lambda}}
+\frac{\tau_{\mu_{S_L-}}'^2+\mu_{S}\omega'^2}
{\mu_{S}\,\tau_{\mu_{S_L-}}'\,\chi_{\mu_{S}}}\,\sigma_{\tilde{\beta}_{\hat{\omega}}}
\nonumber\\
&\quad+4\,\frac{\mu_C\,\tau_{\mu_{C-}}'}
{\chi_{\mu_C}^3}\,i\,\omega'\,\sigma_{\tT}
-\frac{2}{\tau_-'}\,\sigma_{\tZ_{\hat{\omega}}}\,,\nonumber\\
\tgamma_{\hat{\nu}\hat{\nu}}&=
-\frac{\chi^2}{\tau_-'^2}\,\sigma_{\tgamma_{\hat{\omega}\hat{\omega}}}
+\frac{2}{\tau_-'}\,\sigma_{\tZ_{\hat{x}}}
-\frac{2\,i\,\omega'}{\tau_-'^2}\,\sigma_{\tZ_{\hat{\omega}}}\,,
\end{align}
and finally,
\begin{align}
\tgamma_{\hat{x}\hat{\nu}}&=
-\frac{i\,\omega'}{\tau_-'}\,\sigma_{\tgamma_{\hat{\omega}\hat{\nu}}}
+\frac{\tau_{\mu_{S-}}'}{\chi_{\mu_{S}}}\,\sigma_{\tbeta_{\hat{\nu}}}
-\frac{2}{\tau_-'}\,\sigma_{\tZ_{\hat{\nu}}}\,,\nonumber\\
\tgamma_{\hat{\omega}\hat{\nu}}&=
\sigma_{\tgamma_{\hat{\omega}\hat{\nu}}}+\frac{i\,\omega'}{\chi_{\mu_{S}}}
\,\sigma_{\tbeta_{\hat{\nu}}}\,,
\end{align}
for the remaining components. Here the~$\sigma$'s are 
complex constants to be determined by substituting the general 
solution into the boundary conditions. The solution for the 
reduction variables such as the ones in~(\ref{eq:reduction_vbeta}) 
are given by taking the expression for the corresponding metric 
component and replacing,
\begin{align}
\sigma_{\talpha}&\to\tau_{\mu_{L-}}'\sigma_{\talpha}\,,
\quad&\sigma_{\tilde{\Lambda}}\to\tau_{\mu_{S_L-}}'
\sigma_{\tilde{\Lambda}}\,,\nonumber\\
\sigma_{\tilde{\beta}_{\hat{\omega}}}&\to\tau_{\tilde{\mu}_{S-}}'
\sigma_{\tilde{\beta}_{\hat{\omega}}}\,,\quad&
\sigma_{\tilde{\beta}_{\hat{\nu}}}\to\tau_{\mu_{S-}}'
\sigma_{\tilde{\beta}_{\hat{\nu}}}\,,\nonumber\\
\sigma_{\tT}&\to\tau_{\mu_{C-}}'
\sigma_{\tT}\,,\quad&
\end{align}
and~$\sigma\to\tau_-'\sigma$ for the remaining free parameters.
One can easily show that this functional form for the reduction 
variables follows for such a pseudo-differential reduction of  
a second order system.

\paragraph*{General Solution for the special 
case~$s'=-\mbeta\,\omega'$:} In the special 
case, the eigenvectors associated with the 
parameters~$\sigma_{\tZ_{\hat{x}}},\sigma_{\tZ_{\hat{\omega}}},
\sigma_{\tZ_{\hat{\nu}}},\sigma_{\tgamma_{\hat{\omega}\hat{\omega}}}$ 
and~$\sigma_{\tgamma_{\hat{\omega}\hat{\nu}}}$ are unaltered, and can be 
obtained just by taking the generic solution at the special 
frequency. On the other hand, at least for generic gauge choices, 
the eigenvectors associated with the 
parameters~$\sigma_{\tbeta_{\hat{\omega}}},\sigma_{\tbeta_{\hat{\nu}}}$ 
must be replaced by eigenvectors of a different form. All three of the 
vectors associated with~$\sigma_{\talpha},\sigma_{\tilde{\Lambda}}$ 
and~$\sigma_{\tT}$ are replaced by vectors of a different form; two 
are generalized eigenvectors, the other a true eigenvector. Since 
this part of the solution will not be used in what follows we do 
not give details.

\paragraph*{Solution with the restricted gauge~\eqref{eqn:still_diag} and~$s\ne-\mbeta\,\omega$:} 
Employing the restricted gauge~\eqref{eqn:still_diag}, the natural form of the solution for 
general frequencies is altered slightly because we can take linear combinations of the previous
eigenvectors which now have shared eigenvalues in order to simplify the expressions.
This amounts to a redefinition of the~$\sigma$ parameters. The components~$\talpha,\tbeta_{\hat{\nu}},
\tgamma_{\hat{x}\hat{\nu}}$ and~$\tgamma_{\hat{\omega}\hat{\nu}}$ are unaffected by the restriction, and can 
be evaluated just by taking the appropriate parameters in the earlier expressions. The 
remaining components are modified, and become
\begin{align}
\tbeta_{\hat{x}}&=\frac{\chi_{\mu_{S_L}}}
{2\,\tau_{\mu_{S-}}'}\,\sigma_{\tilde{\Lambda}}
+\frac{(\mu_L-1)\,\mu_{S}\,\tau_{\mu_{L-}}'}
{(\mu_{S}-\mu_L)\,\chi_{\mu_L}}\,
\,\sigma_{\talpha}
-\frac{i\,\omega'}{\tau_{\mu_{S-}}'}
\,\sigma_{\tbeta_{\hat{\omega}}}\,,\nonumber\\
\tbeta_{\hat{\omega}}&=\sigma_{\tbeta_{\hat{\omega}}}
+\frac{(\mu_L-1)\,\mu_{S}}
{(\mu_{S_L}-\mu_L)\,\chi_{\mu_L}}\,i\,\omega'\,\sigma_{\talpha}
\,,\nonumber\\
\tgamma_{\hat{x}\hat{x}}&=
\sigma_{\tbeta_{\hat{x}}}-\frac{2}{\tau_-'}\,\sigma_{\tZ_{\hat{x}}}
+2\frac{(\mu_{S}-1)\mu_L\tau_{\mu_L-}'^2}
{(\mu_{S}-\mu_L)\chi_{\mu_L}}\,\sigma_{\talpha}
+\frac{2}{\tau_-'^2}i\,\omega'\,\sigma_{\tZ_{\hat{\omega}}}\nonumber\\
&\quad-\frac{2\,i\,\omega'}{\chi_{\mu_{S}}}\,\sigma_{\tbeta_{\hat{\omega}}}
-\frac{\omega'^2}{\tau_-'^2}\,\sigma_{\tgamma_{\hat{\omega}\hat{\omega}}}
-2\,\frac{\tau_-'^2+\omega'^2}{\chi\,\tau_-'^2}\,\sigma_{\tT}\,,\nonumber\\
\tgamma_{qq}&=
\frac{2}{\tau_-'}\,\sigma_{\tZ_{\hat{x}}}-2\,\frac{\chi}{\tau_-'^2}\,\sigma_{\tT}
-\frac{2}{\tau_-'^2}\,i\,\omega'\,\sigma_{\tZ_{\hat{\omega}}}
+\frac{2\,i\,\omega'}{\chi_{\mu_{S}}}\,\sigma_{\tbeta_{\hat{\omega}}}\nonumber\\
&\quad+\frac{\omega'^2}{\tau_-'^2}\sigma_{\tgamma_{\hat{\omega}\hat{\omega}}}
+2\frac{\mu_{S}-1}{(\mu_{S}-\mu_L)\,\chi_{\mu_L}^2}
\,\omega'^2\,\mu_L\sigma_{\talpha}\,,\nonumber\\
\tgamma_{\hat{x}\hat{\omega}}&=
-\frac{2}{\tau_-'}\,\sigma_{\tZ_{\hat{\omega}}}
-\frac{i\,\omega'}{\tau_-'}\,
\sigma_{\tgamma_{\hat{\omega}\hat{\omega}}}
+\frac{i'\,\omega'}{2\,\tau_{\mu_{S-}}'}\sigma_{\tbeta_{\hat{x}}}
-\frac{2\,i\,\omega'}{\chi\,\tau_-'}\sigma_{\tT}\nonumber\\
&\!\!\!\!\!\!\!\!\!-\frac{2\,(\mu_{S}-1)\,\mu_L\,\tau_{\mu_L-}'}{(\mu_{S}-\mu_L)
\,\chi_{\mu_L}^2}i\,\omega'\,\sigma_{\talpha}
+\Big(\frac{2\,\tau_{\mu_{S-}}'}{\chi_{\mu_{S}}}
-\frac{\chi_{\mu_{S}}}{\mu_{S}\tau_{\mu_{S-}}'}\Big)
\sigma_{\tbeta_{\hat{\omega}}}\,,\nonumber\\
\tgamma_{\hat{\nu}\hat{\nu}}&=
\frac{2}{\tau_-'}\,\sigma_{\tZ_{\hat{x}}}
-\frac{2\,\chi}{\tau_-'^2}\,\sigma_{\tT}
-\frac{2\,i\,\omega'}{\tau_-'^2}\sigma_{\tZ_{\hat{\omega}}}
-\frac{\chi^2}{\tau_-'^2}\sigma_{\tgamma_{\hat{\omega}\hat{\omega}}}\,.
\end{align}
The~`$d$' reduction variables can be evaluated as before, again adjusting the parameters appropriately.
Note that with the restriction~\eqref{eqn:still_diag} the formulation is really the same as the Z4 system 
coupled to our particular condition on the lapse and shift.

\paragraph*{Solution with the restricted gauge~\eqref{eqn:still_diag} for 
the special case~$s=-\mbeta\,\omega$:} Using the restricted gauge the symbol~$M$ remains 
diagonalizable in the special case~$s=-\beta\,\omega$, but some of the eigenvectors do take a different 
form. The solutions for~$\talpha$ and~$\tbeta_{\hat{\nu}}$ are once again unaffected and can be obtained 
by evaluating the standard previous expressions at the particular frequency. The remaining components are 
modified. The interested reader is directed to the Mathematica notebooks that accompany the paper. To show 
boundary stability we must demonstrate both that the solution is well-behaved at generic 
frequencies and with this form at this special frequency.

\paragraph*{The harmonic gauge:} For the harmonic gauge a possible 
approach to the IBVP is instead to put Sommerfeld boundary conditions on the 
combinations, see for example equations (33-35) in~\cite{RuiRinSar07},
\begin{align}
&-\talpha+\tbeta_{\hat{x}}\,+\tfrac{1}{2}
\tgamma_{\hat{x}\hat{x}}\,,&\quad&\tbeta_{\hat{\omega}}
+\tgamma_{\hat{x}\hat{\omega}}\,,\nonumber\\
&-\talpha-\tfrac{1}{2}\tgamma_{\hat{x}\hat{x}}\,,&\quad&
\tbeta_{\hat{\nu}}+\tgamma_{\hat{x}\hat{\nu}}\,.\nonumber
\end{align}
These conditions seem a little unnatural from the point of view 
of the physicist, who may view the lapse and shift as encoding the 
coordinate choice and prefer to specify boundary conditions on them 
directly. Nevertheless, the issue does not pose any mathematical 
problem because in the harmonic gauge these combinations also 
satisfy wave-equations, and a cascade structure of boundary 
conditions~\cite{KreWin06} is obtained. It may be possible to extend 
this construction to a larger class of gauge conditions, but 
here we are primarily concerned with generic members of the 
family~\eqref{eqn:gauge}, 
and so will not attempt to do so. The price we will pay for 
treating generic gauges is that boundary stability can only 
be obtained by taking high order derivative conditions, where 
as with the cascade structure first derivatives suffice.

\paragraph*{$L_2$ solution for Laplace-Fourier transformed Z4:}
In the notebooks that accompany the paper~\cite{Hilwebsite} we
construct for completeness also the general~$L_2$-solution for the Z4
formulation in the approximation treated here. This should allow the
interested reader to investigate boundary stability for a variety of
different boundary conditions for that formulation.

\subsection{Laplace-Fourier transformed boundary conditions with the 
boundary orthogonality condition}\label{sec:LFBCs}

We perform a LF transformation of the high order BCs, 
Eqs.~(\ref{eqn:bc_alpha}), (\ref{eqn:bc_betas}-\ref{eqn:CPBCs_Theta}), 
\eqref{eqn:CPBCs_Z}, and~(\ref{eqn:bc_psi0}).  
Following~\cite{RuiRinSar07,RuiHilBer10},  we  rewrite  these conditions  
in a suitable algebraic form  which  allows one to write down the 
resulting IBVP for the system as in~(\ref{eqn:BC-LF}). Defining the 
linear operator
\begin{align}
\mathcal{L}_{\mu}&=\sqrt{\mu}\,s'-\frac{1}{\kappa}\,\mu\,\p_x\,,
\end{align}
it turns out that the high order BCs can be rewritten as follows:

\paragraph*{Lapse condition:}
The BC~(\ref{eqn:bc_alpha}) becomes 
\begin{align}
\mathcal{L}_{\mu_L}^{L+1}{\talpha}\hateq
s'^{L+1}\,\tilde{g}_L\,,
\label{eqn:LFbc_alpha}
\end{align}
with $\tilde{g}_L={{\mu_L}^{(L+1)/2}}\,\tilde{h}_L$
the LF transformation of the boundary data $g_\alpha$. 
Following~\cite{RuiRinSar07,RuiHilBer10}, it can be shown 
show that, using the equations of motion~(\ref{eqn:LF_gauge}),
the above condition with $L=0$ can be written as,
\begin{align}
\mathcal{L}_{\mu_L}\left(\begin{array}{c}
\talpha\\D\talpha
\end{array}
\right)=A\,
\left(\begin{array}{c}
\talpha\\D\talpha
\end{array}
\right)\,,
\end{align}
where the matrix~$A$ is given by,
\begin{eqnarray}
A&=& \left( \begin{array}{cc}
\sqrt{\mu_L}\,s'  & -\mu_L\\
-\lambda_{\mu_L}^2&\sqrt{\mu_L}\,s'  
\end{array} \right)\,.
\label{eqn:A-BC}
\end{eqnarray}
Since~$\mathcal{L}_{\mu_L}$ is a linear operator, it is 
straightforward to show that, after applying this  
operator~$m$ times, we obtain,
\begin{align}
\mathcal{L}_{\mu_L}^m
\left(\begin{array}{c}
\talpha\\D\talpha
\end{array}
\right)=A^m\,\left(\begin{array}{c}
\talpha\\D\talpha
\end{array}
\right)\,,
\end{align}
where the matrix~$A^m$ satisfies
\begin{align}
&A^m=\\
&\frac{1}{2}\,\left(\begin{array}{cc}
a_+^{m}\,+\,a_-^{m}&
-\frac{1}{-\tau'_{\mu_L-}}(a_+^{m}-a_-^{m})\\
-\frac{1}{-\tau'_{\mu_L-}}(a_+^{m}-a_-^{m})&
a_+^{m}\,+\,a_-^{m}
\end{array} \right)\,.\nonumber
\end{align}
Here~$a_{\pm}=\sqrt{\mu_L}\,(s'\mp \sqrt{\mu_L}\,\tau'_{\mu_L-})$ are 
the eigenvalues of the matrix~$A$. Therefore, the BC operator 
in~(\ref{eqn:LFbc_alpha}) can  be brought into the 
form~(\ref{eqn:BC-LF}) with  
\begin{align}
L_\alpha&=\frac{1}{2}\,\left(\begin{array}{cc}
a_+^{L+1}\,+\,a_-^{L+1}\,,&
-\frac{(a_+^{L+1}-a_-^{L+1})}{-\tau'_{\mu_L-}}
\end{array} \right)\,,
\label{eq_algebraicBClapse}
\end{align}
for any integer~$L\geq 1$.

\paragraph*{Longitudinal component of the shift vector:}
After the change of variables~(\ref{eqn:Gtilde_dot}), 
the LF version of the shorthand $B_{\hat{x}}$ is 
$\tilde{B}_{\hat{x}}=2\,\tilde{\Theta}+s'\,\tilde{\Lambda}/2$. The
BC (\ref{eqn:bc_betas}) becomes
\begin{align}
\mathcal{L}_{\mu_{S_L}}^{L-1}
{\tilde{B}}_{\hat{x}}\hateq
s'^{L+1}\,\tilde{g}_{S_L}\,,
\label{eqn:LFbc_betas}
\end{align}
where $\tilde{g}_{S_L}={\mu_{S_L}}^{(L+1)/2}\,\tilde{h}_{S_L}$.  
Notice that since $\tilde{B}_{\hat{x}}$ satisfies the wave equation
we have
\begin{align}
\p^2_x\tilde{B}_{\hat{x}}+{\tau'}_{\mu_{S_L-}}^2\,\tilde{B}_{\hat{x}}=0\,,
\end{align}
then, by using the above procedure,  it is easy to show that the BC
operator~(\ref{eqn:LFbc_betas}) can be written as 
\begin{align}
L_{\beta_{\hat{x}}}=\frac{1}{2}\,\left(\begin{array}{cc}
b_+^{L-1}\,+\,b_-^{L-1}\,,&
-\frac{(b_+^{L-1}-b_-^{L-1})}{-\tau'_{\mu_{S_L-}}}
\end{array} \right)\,,
\label{eq:beta_xSH}
\end{align}
where~$b_{\pm}=\sqrt{\mu_{S_L}}\,(s'\mp \sqrt{\mu_{S_L}}\,\tau'_{\mu_{S_L-}})$ 
for any integer~$L\geq 1$. 


\paragraph*{Transversal components of the shift vector:}
The LF version of the condition~(\ref{eqn:bc_betaA})
is,
\begin{align}
&\mathcal{L}_{\mu_{S}}^{L}
\tilde{B}^A
\hateq s'^{L-1}\,{-\tau'}^2_{\mu_S-}\,\tilde{g}^{A}_S\,,
\label{eqn:LFbc_betaA}
\end{align}
where the shorthand~$\tilde{B}^A$ and the boundary data are
\begin{align}
\tilde{B}^A&=(D\tbeta_{\hat{\omega}}-i\,\omega'\,\tbeta_{\hat{x}})\,
\delta^A_{\hat{\omega}}+
D\tilde{\beta}_{\hat{\nu}}\,\delta^A_{\hat{\nu}}\,,\nonumber\\
\tilde{g}^A_S&=2\,{\mu_{S}^{(L-1)/2}}
\,\big(\tilde{h}^{\hat{\omega}}_S \,\delta^A_{\hat{\omega}}+
\tilde{h}^{{\hat{\nu}}}_S\,\delta^A_{\hat{\nu}}\big)\,.
\end{align}
Once again the combination
$D\tbeta_{\hat{\omega}}-i\,\omega'\,\tbeta_{\hat{x}}$,
which can be written as
\begin{align}
D\tbeta_{\hat{\omega}}-i\,\omega'\,\tbeta_{\hat{x}}&=
\frac{i\,s'\,\omega'\,(\eta_L\,\mu_L-\mu_{S_L})\,
D\talpha}{\lambda_{\mu_S}\,(\mu_L-\mu_{S_L})}
+\frac{s'^2\,D\tbeta_{\hat{\omega}}}{\lambda_{\mu_S}}\nonumber\\&
-\frac{2\,i\,\mu_{S}\,\omega'\,D\tilde{\Theta}}{\lambda_{\mu_S}}
-\frac{i\,s'\,\mu_{S_L}\,\omega'\,D\tilde{\Lambda}}{2\,\lambda_{\mu_S}}\,,
\end{align}
satisfies a wave equation with propagation speed $\mu_S$. Therefore, the
boundary conditions on the transversal components of the shift vector can
be recast in the form
\begin{align}
L^{A}_S&=\frac{1}{2}\,\left(\begin{array}{cc}
c_+^{L}\,+\,c_-^{L}\,,
&-\frac{(c_+^{L}-c_-^{L})}{
-\tau_{\mu_{S-}}'}\,
\end{array} \right)\,,
\label{eq_algebraicBCbetaA}
\end{align}
with $c_{\pm}=\sqrt{{\mu}_{S}}\,(s'\mp \sqrt{{\mu}_{S}}\,\tau'_{{\mu}_{S-}})$ for
any integer~$L\geq 1$. 


\paragraph*{Constraint preserving BCs:} 
The BC on the $\Tilde{\Theta}$ constraint in the LF space is given by,
\begin{eqnarray}
\mathcal{L}^L\tilde{\Theta}\hateq s'^{L+1}\,\tilde{g}_\Theta\,,
\label{eqn:LFbc_betaconstraints}
\end{eqnarray}
with $\tilde{g}_\Theta=\mu_C^{L/2}\,\tilde{h}_\Theta$ and, 
since the constraints satisfy a wave equation the boundary 
operator can be written as
\begin{align}
L_{\textrm{C}}&=\frac{1}{2}\,\left(\begin{array}{cc}
d_+^{L}\,+\,d_-^{L}\,,&-\frac{(d_+^{L}-d_-^{L})}
{-\tau_{\mu_{C-}}'}
\end{array} \right)\,,
\label{eq_algebraicBCcostraints}
\end{align}
with $d_{\pm}=\sqrt{\mu_{C}}\,(s'\mp\ \sqrt{\mu_{C}}\,\tau'_{\mu_C-})$
for~$L\geq 1$. The lowest order BC for $Z_i$ can be specified in the above 
form with $d_{\pm}=s'\mp\tau'_-$. For the remaining conditions
we have
\begin{align}
\mathcal{L}^{L-1}\tilde{X}^i\hateq s'^{L-1}\,\lambda^2\,\tilde{h}^i_Z\,,
\end{align}
where
\begin{align}
\tilde{X}^i&=\big(s'\tilde{Z}_{\hat{x}}-\mu_C\,D\tilde{\Theta}\big)\,
\delta^i_{\hat{x}}\nonumber\\&+
\big(s'\tilde{Z}_{\hat{\omega}}-\mu_C\,i\,\omega\,\tilde{\Theta}\big)\,
\delta^i_{\hat{\omega}}+s'\tilde{Z}_{\hat{\nu}}\,
\delta^i_{\hat{\nu}}\,.
\end{align}
Therefore, the LF version of the boundary operator in the 
condition~(\ref{eqn:CPBCs_Z}) can  be written as
\begin{align}
L_{\textrm{Z}}&=\frac{1}{2}\,\left(\begin{array}{cc}
f_+^{L-1}\,+\,f_-^{L-1}\,,&-\frac{(f_+^{L}-f_-^{L})}
{-\tau_-'}
\end{array} \right)\,,
\label{eq_algebraicBCcostraintZ}
\end{align}
with $f_{\pm}=s'\mp\tau'_-$ for~$L\geq 2$.  

\paragraph*{Radiation controlling BCs:} To obtain the 
LF version of the condition on the incoming gravitational 
radiation~(\ref{eqn:bc_psi0}),  we perform  a LF transformation
in both the orthogonal vectors~$\iota^i$ and~$\upsilon^i$ defined 
in~(\ref{eqn:dyadnull}) and the electric and magnetic parts of 
the Weyl tensor which allows the construction of~$\tilde{\Psi}_0$ 
in the LF space. The resulting basis must be related 
with~$(\hat{\omega},\hat{\nu})$  through an~$SO(2)$-rotation 
of angle~$\theta$, namely,
\begin{align}
\iota=&\,\hat{\nu}\,\cos\theta-\hat{\omega}\,\sin\theta\,,\nonumber\\
\upsilon=&\,\hat{\nu}\,\sin\theta+\hat{\omega}\,\cos\theta\,.
\end{align}
Therefore, the LF version of the radiation controlling condition can
be written  in the form,
\begin{align}
\mathcal{L}^{L-1}&\re(\tilde{\hat{\Psi}}_0)\hateq s'^{L+1}\,
\tilde{g}_{\tiny{\re({\Psi_0})}}\,,\nonumber\\
\mathcal{L}^{L-1}&\im(\tilde{\hat{\Psi}}_0)
\hateq s'^{L+1}\,\tilde{g}_{\tiny{\im({\Psi_0})}}\,,
\label{eq:BCs_radiation}
\end{align}
with
\begin{align}
\re(\tilde{\hat{\Psi}}_0) =\re(\tilde{\Psi}_0)\,, \qquad
\im(\tilde{\hat{\Psi}}_0) =\im(\tilde{\Psi}_0)\,,
\end{align}
for $L=1$. For higher order conditions ($L\geq 2$) we have,
\begin{align}
\re(\tilde{\hat{\Psi}}_0) &=\re(\tilde{\Psi}_0) -
\mu_C\,\omega'^2\,\cos(2\theta)\,\tilde\Theta\,,
\nonumber\\
\im(\tilde{\hat{\Psi}}_0) &=\im(\tilde{\Psi}_0) + \mu_C\,
\omega'^2\,\sin(2\theta)\,\tilde\Theta\,,
\end{align}
where the shorthand $\tilde{\Psi}_0$ is the LF transformation
of $\Psi_0$. The LF transformation of the given  boundary data
is given by
\begin{align}
\tilde{g}_{\tiny{\re({\Psi_0})}}&=\re(\tilde{h}_{\tiny{\Psi_0}})\,,
\qquad
\tilde{g}_{\tiny{\im({\Psi_0})}}=\im(\tilde{h}_{\tiny{\Psi_0}})\,.
\end{align}
As before, for high derivative order, we can rewrite the conditions
(\ref{eq:BCs_radiation}) in algebraic form by using that
$\tilde{\hat{\Psi}}_0$ satisfies a wave equation. One can show that the
boundary operator  can be brought into the form
\begin{align}
L_{\Psi_0}&=\frac{1}{2}\,\left(\begin{array}{cc}
g_+^{L-1}\,+\,g_-^{L-1}\,,&-\frac{(g_+^{L-1}-g_-^{L-1})}
{-\tau_-'}
\end{array} \right)\,,
\label{eq_algebraicPsi0}
\end{align}
with $g_{\pm}=s'\mp\tau'_-$ for~$L\geq 2$.  

\subsection{Well-posedness results using fifth order BCs,
the boundary orthogonality condition and general gauges}

\paragraph*{The solution:} In the following  calculations we employ the final 
shorthand~$\Pi'_\mu=s'-\sqrt{\mu}\,\tau'_{\mu-}$. As before, we write
$\Pi'_1=\Pi'$. In the LF space, the~$L_2$ solution for the gauge variables 
at the boundary with  fifth order BCs ($L=4$) is given by:
\begin{align}
\label{eqn:est_met_first}
\talpha
&\hateq\frac{s'^5}{\Pi_{\mu_L}'^5}\,\tg_\alpha\,,\nonumber\\
\tilde{\beta}_{\hat{x}}&\hateq 
-\frac{\left(\eta_L\,\mu_L-\mu_{S_L}\right)
\,s'^4\,\tau'_{\mu_L-}}{\Pi_{\mu_L}'^5\,\left(\mu_L-\mu_{S_L}\right)}\,\tg_{\alpha}
+\frac{\mu_{S_L}\,s'^3\,\tau_{\mu_{S_L}-}'}{\Pi_{\mu_{S_L}}'^4}\,\tg_{\beta_{\hat{x}}}\nonumber\\
&\quad-\frac{i\,\omega'\,\mu_S^2\,s'\,\tau_{\mu_S-}'^2}{\Pi_{\mu_S}'^4}\,\tg_{\beta{\tomega}}\,,
\nonumber\\
\tbeta_{\hat{\omega}}&\hateq
-\frac{\big(\eta_L\,\mu_L-\mu_{S_L}\big)\,i\,\omega'
\,s'^4}{{\Pi'}_{\mu_L}^5\,\left(\mu_L-\mu_{S_L}\right)}\,\tg_{\alpha}
+\frac{i\,\omega'\,\mu_{S_L}\,s'^3}{\Pi_{\mu_{S_L}}'^4}\,\tg_{\beta_{\hat{x}}}\nonumber\\
&\quad+\frac{i\,\mu_S^2\,s'\,\tau_{\mu_S-}'^3}{\Pi_{\mu_S}'^4}\,\tg_{\beta{\tomega}}\,,\nonumber\\
\tbeta_{\hat{\nu}}&\hateq\frac{\mu_S\,s'^3\,\tau_{\mu_S-}'}
{\Pi_{\mu_L}'^4}\,\tg_{\hat{\nu}}\,.
\end{align}
The expressions for the metric are slightly more complicated. The longitudinal
component of the metric perturbation is
\begin{align}
\tgamma_{\hat{x}\hat{x}}&\hateq
\frac{2\,\left(1-\eta_L\right)\,\mu_L\,s'^3\,
\tau_{\mu_L-}'^2}{\Pi_{\mu_L}'^5\,\left(\mu_L-\mu_{S_L}\right)}\,\tg_{\alpha}
+\frac{2\,s'\,\tau_-'^2}{\Pi'^3}\,\tg_{Z_{\hat{x}}}
\nonumber\\
&\!\!+
\frac{2\,\mu_{S_L}\,s'^2\,\tau_{\mu_{S_L-}}'^2}
{\Pi_{\mu_{S_L}}'^4}\,\tg_{\beta_{\hat{x}}}
+\frac{4\,i\,\omega'\,s'\,\tau_-'^2}{\Pi'^4}\,\tg_{Z_{\tomega}}
\nonumber\\
&\!\!+
\frac{4\,(s'+\tau_-')\,s'^2}{\Pi'^3}\,
\left[\tg_{\tiny{\re(\Psi_0)}}\,\cos(2\,\theta)-
\tg_{\tiny{\im(\Psi_0)}}\,\sin(2\,\theta)\right]
\nonumber\\
&\!\!-
\frac{2\,i\,\omega'\,\mu_S^2\,\tau_{\mu_S-}'^3}
{\Pi_{\mu_S}'^4}\,\tg_{\beta_{\tomega}}
-\frac{4\,\mu_C\,\,s'^2\tau_{\mu_C-}'^2}
{\Pi_{\mu_C}'^4}\,\tg_\Theta\,.
\end{align}
The trace of the metric perturbation at the boundary is
\begin{align}
\tgamma_{qq}&\hateq
\frac{2\,s'^3\,\left(1-\eta_L\right)\,
\left(s'+\sqrt{\mu_L}\,\tau_{\mu_L-}'\right)}{\Pi_{\mu_L}'^4\,\left(\mu_L-\mu_{S_L}\right)}
\,\tg_{\alpha}
-\frac{2\,s'\,\tau_-'^2}{\Pi'^3}\,\tg_{Z_{\hat{x}}}\nonumber\\
&\!\!+
\frac{2\,s'^2\,\left(s'+\sqrt{\mu_{S_L}}\,\tau_{\mu_{S_L-}}'\right)}
{\Pi_{\mu_{S_L}}'^3}\,\tg_{\beta_{\hat{x}}}
-\frac{4\,i\,\omega'\,s'\,\tau_-'^2}{\Pi'^4}\,\tg_{Z_{\tomega}}
\nonumber\\
&\!\!+\frac{4\,(s'+\tau_-')\,s'^2}{\Pi'^3}\,\left[\tg_{\tiny{\im(\Psi_0)}}\,\sin(2\,\theta)-
\tg_{\tiny{\re(\Psi_0)}}\,\cos(2\theta)\right]\nonumber\\
&\!\!+\frac{2\,i\,\omega'\,\mu_S^2\,\tau_{\mu_S-}'^3}{\Pi_{\mu_S}'^4}\,\tg_{\beta_{\tomega}}
+\frac{4\,\mu_C\,s'^2\,\omega'^2}{\Pi_{\mu_C}'^4}\,\tg_{\Theta}\,.
\label{eqn:1/sterm1}
\end{align}
For the mixed longitudinal transverse components of the metric 
perturbation we find
\begin{align}
\tgamma_{\hat{x}\hat{\omega}}&\hateq
-\frac{4\,s'\,\tau_-'^3}{\Pi'^4}\,\tg_{Z_{\tomega}}
+\frac{2\,i\,\omega'\,(1-\eta_L)\,\mu_L\,s'^3
\,\tau_{\mu_L-}'}{\Pi_{\mu_L}'^5\,(\mu_L-\mu_{S_L})}\,\tg_{\alpha}\nonumber\\
&\quad-\frac{\mu_S\,(s'^2-2\,\mu_S\,\tau_{\mu_S-}'^2)\,\tau_{\mu_S-}'^2}
{\Pi_{\mu_S}'^4}\,\tg_{\beta_{\tomega}}
-\frac{2\,i\,\omega'\,
s'\,\tau_-'^2}{{\Pi'}^4}\,\tg_{Z_{\hat{x}}}\nonumber\\
&\quad-
\frac{4\,i\,\omega'\,s'^2\,\tau_-'}{\Pi'^4}
\,[\tg_{\tiny{\re(\Psi_0)}}\,\cos(2\,\theta)-\tg_{\tiny{\im(\Psi_0)}}\,\sin(2\theta)]
\nonumber\\
&\quad+\frac{2\,\mu_{S_L}\,i\,\omega'\,s'^2\,
\tau_{\mu_{S_L-}}'}{\Pi_{\mu_{S_L}}'^4}\,\tg_{\beta_{\hat{x}}}
-\frac{4\,i\,\mu_C\,\omega'\,s'^2\,\tau_{\mu_C-}}
{\Pi_{\mu_C}'^4}\,\tg_{\Theta}\,.
\end{align}
Next,
\begin{align}
\tgamma_{\hat{x}\hat{\nu}}&\hateq
\frac{\mu_S\,s'^2\,\tau_{\mu_S-}'^2}
{\Pi_{\mu_S}'^4}\,\tg_{\beta_{\tnu}}
+\frac{4\,s'^2\,\tau_-'^2}
{\Pi'^4}\,\tg_{Z_{\tnu}}
\nonumber\\
&+
\frac{4\,i\,\omega'\,s'^3}{\Pi'^4}
\,\left[\tg_{\tiny{\im(\Psi_0)}}\,\cos(2\theta)+
\tg_{\tiny{\re(\Psi_0)}}\,\sin(2\theta)\right]\,.
\end{align}
Finally, for the transverse-transverse components of the 
perturbation we have,
\begin{align}
\tgamma_{\hat{\omega}\hat{\nu}}&\hateq
\frac{\mu_S\,i\,\omega'\,s'^2\,\tau_{\mu_S-}'}
{\Pi_{\mu_S}'^4}\,\tg_{\beta_{\tnu}}
-\frac{2\,i\,\omega'\,s'^2\,\tau_-'^2}
{\Pi'^5}\,\tg_{Z_{\tnu}}\nonumber\\&-
\frac{4\,s'^3\,\tau_-'}{\Pi'^4}\,[\tg_{\tiny{\im(\Psi_0)}}\,\cos(2\theta)+
\tg_{\tiny{\re(\Psi_0)}}\,\sin(2\theta)]\,.
\end{align}
and,
\begin{align}
\tgamma_{\hat{\nu}\hat{\nu}}&\hateq
-\frac{2\,s'^2\,\tau_-'^2}{{\Pi'}^5}
\,\left(\Pi'\,\tg_{Z_{\hat{x}}}
+i\,\omega'\,\tg_{Z_{\tomega}}\right)+\\&
\frac{4\,s'^4}{\Pi'^4}\,\left[\tg_{\tiny{\im(\Psi_0)}}\,\sin(2\,\theta)
-\tg_{\tiny{\re(\Psi_0)}}\,\cos(2\,\theta)\right]\,.
\label{eqn:est_met_last}
\end{align}
The reduction~`$d$' reduction variables can be obtained by replacing the
given data  in these expressions according to the rules,
\begin{align}
\tilde{g}_{\alpha}&\to -\tau_{\mu_{L-}}'\tilde{g}_{\alpha}\,,
\quad&\tilde{g}_{\beta_{\hat{x}}}\to-\tau_{\mu_{S_L-}}'
\tilde{g}_{\beta_{\hat{x}}}\,,\nonumber\\
\tilde{g}_{\beta_{\hat{\omega}}}&\to-\tau_{\mu_{S-}}'
\tilde{g}_{\beta_{\hat{\omega}}}\,,\quad&
\tilde{g}_{\beta_{\hat{\nu}}}\to-\tau_{\mu_{S-}}'
\tilde{g}_{\beta_{\hat{\nu}}}\,,\nonumber\\
\tilde{g}_{\tT}&\to-\tau_{\mu_{C-}}'\tilde{g}_{\tT}\,,\quad&
\end{align}
and~$\tilde{g}\to-\tau_-'\,\tilde{g}$ for the remaining given data.
Note that this need not be the case, and is a result of the fact
that our boundary conditions are very carefully chosen so as not
to mix the eigensolutions associated with different speeds. 

\begin{figure}[t!]
\begin{minipage}[t]{0.49\textwidth}
  \includegraphics[clip,trim= 0 0 0 0 0,width=0.92\textwidth,angle=0]{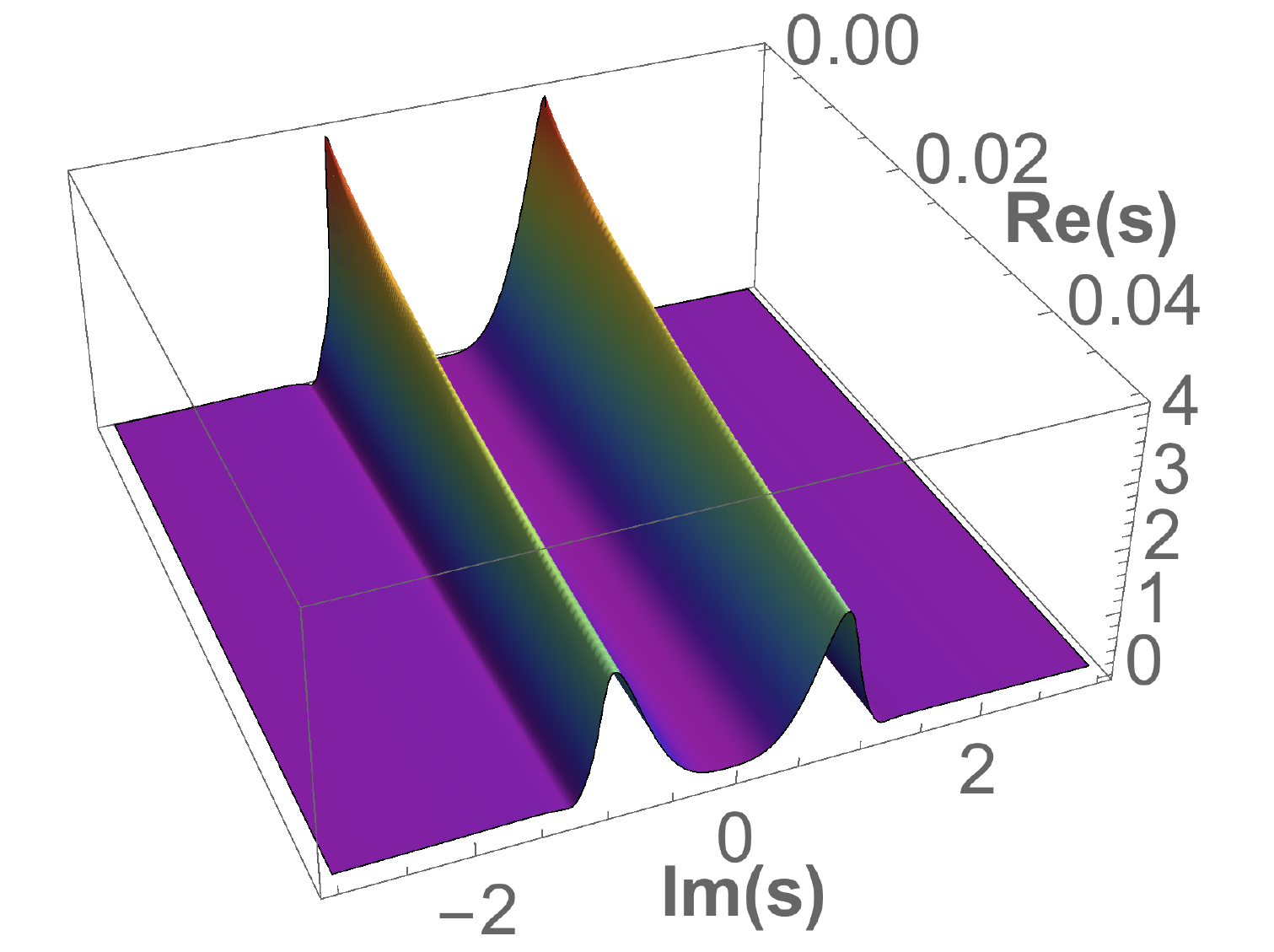}
\medskip
\end{minipage}
\caption{\label{fig:largest_coeff} The magnitude of the coefficient of
  the~$L_2$ solution in Eqs~(\ref{eqn:est_met_first}-\ref{eqn:est_met_last})
  with the largest peak. In this particular plot we
  chose~$\mu_L=2,\mu_{S}=9/4,\mu_{S_L}=3,\eta_L=0$ and
  also~$\omega=1$. We see that the coefficients are bounded at
  this~$\omega$ even as~$\re(s')\rightarrow 0$. The
  estimate~\eqref{eqn:Pi_estimate} shows that this holds at
  every~$\omega$.}
\end{figure}

\paragraph*{Boundary stability:} 
The next step is to show that the above system is boundary  
stable. Examining the right hand sides of the~$L_2$ solution, 
Eqs.~(\ref{eqn:est_met_first}-\ref{eqn:est_met_last}), it is  
clear that we must estimate~$\Pi'_\mu$. But following~\cite{KreWin06,RuiRinSar07}
there is a strictly positive constant~$\delta$ such that,
\begin{align}
\left|\Pi'_\mu\right|&=
\left|s'+\sqrt{s'^2+\mu\,\omega'^2}\right|\geq\,\delta>0\,,
\label{eqn:Pi_estimate}
\end{align}
for all~$\re(s')>0$ and~$\omega'\in \mathbb{R}$ with $|s'|^2+|\omega'|^2=1$. 
Therefore, the solution of the gauge and metric components at the boundaries
are bounded by the given boundary data.  Fig.~\ref{fig:largest_coeff} displays
the largest coefficient of the above~$L_2$ solution for~$\omega'\to 1$ ($|s'|\to 0$).
We note that the solution remains continuously bounded. Thus, there is a
positive constant~$C$ such that,
\begin{align}
|\talpha(s,0,\omega)|\leq\,C\,|{\tg}_\alpha|\,.
\end{align} 
Similar arguments hold for the other components. We conclude 
that the full solution of the system with fifth order BCs 
satisfies the estimate
\begin{equation}
|\tilde{u}(s,x=0,\omega)|\leq\,C'\,|\tilde{g}(s,\omega)|\,,
\label{eqn:bs_fullsystem}
\end{equation}
for all~$\re(s)> 0$ and~$\omega\in\mathbb{R}$ with~$C'>0$ a positive 
constant. We then conclude that the above system is boundary 
stable~\cite{Kre70}.  As we have seen in section~\ref{sec:KA-theory}, 
it implies that there is a symmetrizer~$\hat{R}=\hat{R}(s',\omega')$ such 
that~\cite{RuiRinSar07},
\begin{align}
\p_x\left<\tilde{u},\hat{R}\,\tilde{u}\right>&=
2\,\left<\tilde{u},\hat{R}\,\p_x\tilde{u}\right>\nonumber\\
=&\left<\tilde{u},\left(\hat{R}\,M+M^*\,\hat{R}\right)\,\tilde{u}
\right>\,.
\end{align}
Here we have used the equations of motion. Using the first 
and second properties of~$\hat{R}$, we obtain
\begin{align}
\p_x\left<\tilde{u},\hat{R}\,\tilde{u}\right>\geq C_1\,\eta\,
|\tilde{u}|^2\,,
\end{align}
with~$\eta=\re(s)$. Integrating both sides from~$x=0$ to~$x=\infty$
and using the last property of the symmetrizer~$\hat{R}$, 
it follows that
\begin{align}
\eta\,\int_0^\infty|\tilde{u}|^2\,dx&\leq-\frac{1}{C_1}\,\left.
\left<\tilde{u},\hat{R}\,\tilde{u}\right>\right|_{x=0}\nonumber\\&
\leq\frac{1}{C_1}\,\left(-C_3\,\left.|\tilde{u}|^2\right|_{x=0}+ 
C_2\,|\tilde{g}|^2\right)\,.
\label{eqn:basic_LF_estimate} 
\end{align}
This inequality is the basic estimate in the Laplace-Fourier space, but 
because our boundary conditions contain many derivatives of the primitive 
fields, a little more book-keeping is required to build an estimate that 
can be inverted to give an estimate in an appropriate norm, involving
higher derivatives of the primitive variables in the physical space. Note, 
crucially, that the form of the given data, in particular the choice of 
derivatives, in the boundary conditions is needed to obtain boundary stability. 
This form cancels terms that would otherwise result in singular behavior breaking 
boundary stability. This is what prevents us from choosing for example lower 
order~$L=1$ conditions.

\paragraph*{The final estimate:} Following~\cite{RuiRinSar07}, where the 
estimate~(\ref{eqn:kreiss-estim}) has been generalized in order to estimate 
the $L_2$-norm of the  higher derivatives of the primitive fields in terms 
of the $L_2$-norm of the given boundary data for all $\re(s)> 0$ and all 
smooth solutions~$u$ with the property that its~$L+1$-time derivatives vanish 
identically at~$t=0$, we multiply the inequality~\eqref{eqn:basic_LF_estimate} 
by~$\kappa^{8}$ to obtain an estimate for the tangential derivatives to the 
boundary. For the normal derivatives, namely second or higher derivatives of the  
fields, we use the equations of motion~(\ref{eq:EqoMMatrix}) to obtain,
\begin{align}
&\eta\,\int\limits_0^\infty\,
\sum\limits_{l=0}^{5} 
\Big(\left|\kappa^{5-l}\,\p_x^l\talpha\right|^2+
\sum\limits_{i=1}^{3} \left|\kappa^{5-l}\,\p_x^l\tbeta_{i}\right|^2\nonumber\\
&\qquad\qquad+\left|\kappa^{5-l}\,\p_x^l\tgamma_{ij}\right|^2\Big)
\,\textrm{d}x\nonumber\\
&+\Big(\left|\kappa^{5-l}\,\p_x^l\talpha\right|^2+
\sum\limits_{i=1}^{3} \left|\kappa^{5-l}\,\p_x^l\tbeta_{i}\right|^2\nonumber\\
&\qquad\qquad\left.+\left|\kappa^{5-l}\,\p_x^l\tgamma_{ij}\right|^2\Big)
\right|_{x=0}\nonumber\\
&\leq C\,\left(|\kappa^{5}\,\tilde{h}_\alpha|^2+
\cdots+|\kappa^{5}\,\im(\tilde{h}_{\tiny{\Psi}_0})|^2\right)\,,
\label{eqn:full-estimate}
\end{align}
for some strictly positive constant~${C} > 0$. Finally, integrating
over~$\im(s)$ and over all frequencies~$\omega_A$ and using 
Parseval's relation we obtain~\cite{RuiRinSar07},
\begin{align}
&\eta\,\|\alpha\|^2_{\eta,5,\Omega} + 
\eta\,\sum\limits_{i}\|\beta^i\|^2_{\eta,5,\Omega} + 
\eta\,\sum\limits_{ij}\|\gamma_{ij}\|^2_{\eta,5,\Omega} +\nonumber\\ 
&\eta\,\|\alpha\|^2_{\eta,5,\mathcal{T}} + 
\eta\,\sum\limits_{i}\|\beta^i\|^2_{\eta,5,\mathcal{T}} + 
\eta\,\sum\limits_{ij}\|\gamma_{ij}\|^2_{\eta,5,\mathcal{T}}\nonumber\\ 
&\leq C_{5}\,\left(\|h_\alpha\|^2_{\eta,5,{\mathcal{T}}}+
\cdots+\|h_{\Psi_0}\|^2_{\eta,5,{\mathcal{T}}}
\right)\,,\label{eqn:final_est}
\end{align}
where $C_{5}$ is a positive constant,~$\Omega$ is, as we have mentioned 
before, the domain of integration,~$\mathcal{T}$ is the boundary surface 
and the above $L_2$-norms are defined by,
\begin{align}
&\|u\|^2_{\eta,5,\Omega} =\nonumber\\
&\quad\int_{\Omega} e^{-2\,\eta\,t}
\sum\limits_{|\rho|\leq 5}
| \partial_t^{\rho_t}\,\partial_x^{\rho_x}
\partial_y^{\rho_y}\,
\partial_z^{\rho_z} u(t,x,y,z) |^2\,\textrm{d}\Omega\,,\\
&\|u\|^2_{\eta,5,{\mathcal{T}}}=\nonumber\\ 
&\quad\int_{\mathcal{T}} e^{-2\,\eta\,t}
\sum\limits_{|\rho|\leq 5} |\partial_t^{\rho_t}\,\partial_x^{\rho_x}
\partial_y^{\rho_y}\,\partial_z^{\rho_z}\,u(t,0,y,z)|^2 
\textrm{d}\mathcal{T}\,.\label{eq:final_norm}
\end{align}
Here~$\rho=(\rho_t,\rho_x,\rho_y,\rho_z)$ is a multi-index, and 
we denote~$\textrm{d}\Omega=\,\textrm{d}t\,\textrm{d}x\,
\textrm{d}y\,\textrm{d}z$ and~$\textrm{d}\mathcal{T}=\textrm{d}t\,
\textrm{d}y\,\textrm{d}z$. Adding forcing terms to the equations 
of motion modifies the estimate~\eqref{eqn:final_est} in the 
standard way. Here we have dropped the forcing terms~$F$ from the 
estimates, but these can also be dealt with exactly as 
in~\cite{RuiRinSar07}.

The above result can be easily generalized for high order conditions.
One can show that, once we have a regular $L_2$-solution for given
BCs, i. e. regular coefficients for all $\omega'\in \mathbb{R}$
and $\re(s')>0$, increasing the order the derivatives at the boundary does
not generate singular coefficients. One then can show that the resulting
IBVP is boundary stable and use the above procedure to show that the
problem is well-posed.

\subsection{Why work under the boundary orthogonality condition?
}\label{section:Why_orthogonal?}

We saw in the previous sections that using the boundary orthogonality condition 
results in the simplification that~$\mbeta=0$ in the Laplace-Fourier analysis. Since 
we are able to construct the general~$L_2$ solution even without this restriction 
it is natural to ask why we do so. The reason is that to pick natural 
boundary conditions it is very helpful if the symbol~$M$ has a 
simple eigendecomposition, for then we may look at the left 
eigenvectors of~$M$ contracted with the state vector~$\tilde{u}$
and essentially read off sensible boundary conditions. 
Therefore the deficiency of~$M$ in the special 
case~$s'=-\mbeta\,\omega'$ is a {\it very} serious problem, because 
conditions that work elsewhere in frequency space fail to 
give control at these special points. The special case appears 
because at this particular frequency all of the relevant 
eigenvalues, the different~$\tau_{\mu-}$, many of which are 
generically distinct, clash. When this happens the associated 
eigenspace has to support many more eigenvectors, but can not. Under the boundary 
orthogonality condition with~$\mbeta=0$ 
this breakdown of diagonalizability occurs at~$s'=0$, but is not a problem because 
for boundary stability we are concerned with the solution {\it in the limit}~$s'\to0$.
It may be possible to find boundary conditions that are 
well-behaved also across the bad frequency~$s'=-\mbeta\,\omega'$, 
but doing so will result in several other deficiencies. Such conditions 
will necessarily require more complicated mixing of the eigensolutions
in the analysis. This will result in more complicated absorption
properties and in difficult estimates to perform, for which 
we do not presently have adequate computer algebra tools.

Therefore it is highly desirable to side-step the special case 
completely. Several strategies for this are apparent. The first
of these is to try and choose a formulation for which the special 
case does not appear. Even if we fix the gauge choice, one might 
hope that this is possible by adjusting the constraint subsystem,
and how it is coupled to the gauge variables. We have attempted
this~\cite{Hilwebsite} within the large class of formulations considered 
in~\cite{HilRic13}, but to no avail. The missing eigenvectors 
are associated with the~$\tilde{\Lambda}$ variable but since 
this is not a constraint, such adjustments do not help. Thus the 
next option is to change the gauge conditions, which we do under 
duress, because we would like to show well-posedness for 
arbitrary hyperbolic gauges. The highly restricted 
class containing the harmonic gauge~\eqref{eqn:still_diag} 
suffices. From the PDEs point of view is perhaps not surprising; 
the simple characteristic structure of the restriction 
eradicates nearly all coupling between different metric components,
but this it must do so that hyperbolicity can be achieved with many 
shared speeds. The symbol~$M$ inherits, to a large extent, the same
decoupling. It may be that these gauges then allow for estimates with 
fewer derivatives, and without using the boundary orthogonality condition. 
Certainly this is the case for the harmonic gauge. Throughout we have focused 
on evolved gauge conditions where the time derivative is naturally given in the 
form~$\alpha^{-1}(\p_t-\beta^i\p_i)$. The next option for adjusting 
the gauge, which we have not investigated but which we do think 
may help avoid the special case, is to switch to conditions built
naturally on the time derivative~$\p_t$. In applications one can
easily transition from the first form to the second, and we 
expect that in this way the special case can be cured, at least 
for some gauge conditions. As mentioned earlier on, we also 
expect that once a well-posed IBVP is obtained with a particular 
formulation it will be straightforward to obtain well-posedness 
by employing the dual-foliation formalism~\cite{Hil15}. We 
furthermore expect that in this way one will naturally obtain 
geometric uniqueness. 

The final obvious strategy is to work under the boundary orthogonality condition so 
that the special case simply does not occur. This solution is 
inconvenient the point of view of numerical implementation both 
because of the drifting boundary, and, depending on the gauge choice, 
because of the number of derivatives present in the boundary 
conditions. But this approach is geometrically natural, allowed us 
to demonstrate boundary stability for a wide range of gauge 
conditions {\it and} as shown in section~\ref{section:Conformal_BCs} 
allows reasonable approximations of the desired conditions to be 
implemented straightforwardly.

\section{Conclusion}
\label{section:Conclusion}

To obtain solutions of the Cauchy problem for asymptotically 
flat spacetimes in numerical GR, one option is 
to make the computational domain as large as possible so that 
the boundary remains causally disconnected from the central 
body. Unfortunately the computational cost of this option is 
prohibitive, even if one uses mesh-refinement or compactification
to spatial infinity, because numerical error can travel  
faster than physical effects. A second, much more elegant, 
possibility is to evolve initial data which is hyperboloidal, that 
is, compactified to null 
infinity~\cite{Fra04,CalGunHil05,ZenHus06,Zen08,MonRin08,BucPfeBar09,Rin10,VanHusHil15}, 
or, along similar lines of thought data in which a Cauchy region is 
attached to a null outer zone~\cite{Win98,ReiBisPol09}. Many obstacles 
are still to be overcome before such data can be routinely evolved, 
which means that in the immediate future we are left with one option; 
the specification of {\it improved outer boundary conditions} for 
applications. As greater accuracy is required of numerical data, 
or when the boundary becomes an integral part of the physics of 
the system, as in the case of asymptotically AdS 
spacetimes~\cite{BizRos11,BanPreGub12,CarGuaHer12}, boundary 
conditions must be carefully considered. 

In this paper, we were concerned with boundary conditions appropriate 
for the evolution of asymptotically flat spacetimes with the moving 
puncture method. We considered constraint preserving conditions 
for free-evolution formulations of the Einstein equations coupled to 
a parametrized set of dynamical gauge choices. We derived a new 
class of high order boundary conditions for this family of gauge 
conditions. To reduce the amount of spurious gravitational wave 
reflections, we also employ a high order freezing-$\Psi_0$ 
condition~\cite{BucSar06,BucSar07}. We analyzed well-posedness of 
the resulting initial boundary value problem on a four dimensional 
spacetime with timelike outer boundary by considering high-frequency 
perturbations of a given smooth background solution. Using the 
Laplace-Fourier transform we showed that the resulting IBVP is boundary 
stable. The Kreiss-Agranovich-M\'etivier theory, valid even when the system 
is only strongly hyperbolic of constant multiplicity, guarantees that the 
IBVP is well-posed in the frozen coefficient approximation. By virtue of the 
theory of pseudo-differential operators, the general problem is expected 
to be well-posed too. These results generalize our previous 
study~\cite{RuiHilBer10} in which the constraint absorption properties 
of the CPBCs were considered.

This work could be generalized in a number of ways. Firstly 
one could consider a larger family of dynamical gauge conditions. 
We do not expect such a generalization to be very taxing, 
provided that one is still able to make the necessary 
manipulation of the symbol~$M$ by computer algebra. 
Another possibility is to maintain the same family of gauge 
conditions but to alter the boundary conditions. By construction 
our boundary conditions are those that render the proof of boundary 
stability as close as possible to that of the wave equation. Therefore, 
besides the trivial reflecting case, we expect that other choices 
will rapidly become intractable. One might also consider in what 
approximation, if any, a finite difference approximation to the IBVP 
could be shown to be formally numerically stable. Finally one could 
examine how readily the present calculations could be extended to other 
formulations of GR. 

\acknowledgments

It is a pleasure to thank Bernd Br\"ugmann, Luisa Buchman, Ronny
Richter and especially Olivier Sarbach for helpful discussions 
and comments on the manuscript. This work was supported in part 
by DFG grant SFB/Transregio~7~``Gravitational Wave Astronomy'', 
by Spanish Ministry of Science and Innovation under grants 
CSD2007-00042, CSD2009-00064 and FPA2010-16495, the 
Conselleria d'Economia Hisenda i Innovaci\'o of the
Govern de les Illes Balears and by Colciencias under
program ``Es tiempo de Volver''. We also wish to acknowledge to
the ESI and to the organizers of the ESI  workshop on ``Dynamics 
of General Relativity: Numerical and Analytical Approaches'', 
July - September, 2011, where part of this work was 
developed.

\bibliographystyle{apsrev}           
\bibliography{WP_IBVP.bbl}{}                

\begin{thebibliography}{85}
\expandafter\ifx\csname natexlab\endcsname\relax\def\natexlab#1{#1}\fi
\expandafter\ifx\csname bibnamefont\endcsname\relax
  \def\bibnamefont#1{#1}\fi
\expandafter\ifx\csname bibfnamefont\endcsname\relax
  \def\bibfnamefont#1{#1}\fi
\expandafter\ifx\csname citenamefont\endcsname\relax
  \def\citenamefont#1{#1}\fi
\expandafter\ifx\csname url\endcsname\relax
  \def\url#1{\texttt{#1}}\fi
\expandafter\ifx\csname urlprefix\endcsname\relax\def\urlprefix{URL }\fi
\providecommand{\bibinfo}[2]{#2}
\providecommand{\eprint}[2][]{\url{#2}}

\bibitem[{\citenamefont{Kreiss}(1978)}]{Kre78}
\bibinfo{author}{\bibfnamefont{H.-O.} \bibnamefont{Kreiss}},
  \emph{\bibinfo{title}{Numerical Methods for Solving Time-Dependent Problems
  for Partial Differential Equations}} (\bibinfo{publisher}{Les Presses De
  L'Universit\'{e} de Montr\'{e}al (University of Montreal Press)},
  \bibinfo{address}{Montreal (Canada)}, \bibinfo{year}{1978}), ISBN
  \bibinfo{isbn}{ISBN 0-8405-0430-6}.

\bibitem[{\citenamefont{Kreiss and Lorenz}(1989)}]{KreLor89}
\bibinfo{author}{\bibfnamefont{H.~O.} \bibnamefont{Kreiss}} \bibnamefont{and}
  \bibinfo{author}{\bibfnamefont{J.}~\bibnamefont{Lorenz}},
  \emph{\bibinfo{title}{Initial-boundary value problems and the
  {N}avier-{S}tokes equations}} (\bibinfo{publisher}{Academic Press},
  \bibinfo{address}{New York}, \bibinfo{year}{1989}).

\bibitem[{\citenamefont{Miller et~al.}(2004)\citenamefont{Miller, Gressman, and
  Suen}}]{MilGreSue03a}
\bibinfo{author}{\bibfnamefont{M.}~\bibnamefont{Miller}},
  \bibinfo{author}{\bibfnamefont{P.}~\bibnamefont{Gressman}}, \bibnamefont{and}
  \bibinfo{author}{\bibfnamefont{W.-M.} \bibnamefont{Suen}},
  \bibinfo{journal}{Phys. Rev. D} \textbf{\bibinfo{volume}{69}},
  \bibinfo{pages}{064026} (\bibinfo{year}{2004}),
  \bibinfo{note}{gr-qc/0312030}.

\bibitem[{\citenamefont{Buchman and Sarbach}(2006)}]{BucSar06}
\bibinfo{author}{\bibfnamefont{L.~T.} \bibnamefont{Buchman}} \bibnamefont{and}
  \bibinfo{author}{\bibfnamefont{O.~C.~A.} \bibnamefont{Sarbach}},
  \bibinfo{journal}{Class. Quant. Grav.} \textbf{\bibinfo{volume}{23}},
  \bibinfo{pages}{6709} (\bibinfo{year}{2006}), \eprint{gr-qc/0608051}.

\bibitem[{\citenamefont{Ruiz et~al.}(2011)\citenamefont{Ruiz, Hilditch, and
  Bernuzzi}}]{RuiHilBer10}
\bibinfo{author}{\bibfnamefont{M.}~\bibnamefont{Ruiz}},
  \bibinfo{author}{\bibfnamefont{D.}~\bibnamefont{Hilditch}}, \bibnamefont{and}
  \bibinfo{author}{\bibfnamefont{S.}~\bibnamefont{Bernuzzi}},
  \bibinfo{journal}{Phys. Rev. D} \textbf{\bibinfo{volume}{83}},
  \bibinfo{pages}{024025} (\bibinfo{year}{2011}), \eprint{1010.0523}.

\bibitem[{\citenamefont{Givoli}(1991)}]{Giv91}
\bibinfo{author}{\bibfnamefont{D.}~\bibnamefont{Givoli}},
  \bibinfo{journal}{Journal of Computational Physics}
  \textbf{\bibinfo{volume}{94}}, \bibinfo{pages}{1} (\bibinfo{year}{1991}).

\bibitem[{\citenamefont{Friedrich and Nagy}(1999)}]{FriNag99}
\bibinfo{author}{\bibfnamefont{H.}~\bibnamefont{Friedrich}} \bibnamefont{and}
  \bibinfo{author}{\bibfnamefont{G.}~\bibnamefont{Nagy}},
  \bibinfo{journal}{Commun. Math. Phys.} \textbf{\bibinfo{volume}{201}},
  \bibinfo{pages}{619} (\bibinfo{year}{1999}).

\bibitem[{\citenamefont{Kreiss and Winicour}(2006)}]{KreWin06}
\bibinfo{author}{\bibfnamefont{H.-O.} \bibnamefont{Kreiss}} \bibnamefont{and}
  \bibinfo{author}{\bibfnamefont{J.}~\bibnamefont{Winicour}},
  \bibinfo{journal}{Class. Quantum Grav.} \textbf{\bibinfo{volume}{23}},
  \bibinfo{pages}{S405} (\bibinfo{year}{2006}), \eprint{gr-qc/0602051}.

\bibitem[{\citenamefont{Rinne}(2006)}]{Rin06a}
\bibinfo{author}{\bibfnamefont{O.}~\bibnamefont{Rinne}},
  \bibinfo{journal}{Class. Quant. Grav.} \textbf{\bibinfo{volume}{23}},
  \bibinfo{pages}{6275} (\bibinfo{year}{2006}), \eprint{gr-qc/0606053}.

\bibitem[{\citenamefont{Kreiss et~al.}(2007)\citenamefont{Kreiss, Reula,
  Sarbach, and Winicour}}]{KreReuSar07}
\bibinfo{author}{\bibfnamefont{H.}~\bibnamefont{Kreiss}},
  \bibinfo{author}{\bibfnamefont{O.}~\bibnamefont{Reula}},
  \bibinfo{author}{\bibfnamefont{O.}~\bibnamefont{Sarbach}}, \bibnamefont{and}
  \bibinfo{author}{\bibfnamefont{J.}~\bibnamefont{Winicour}},
  \bibinfo{journal}{Class.Quant.Grav.} \textbf{\bibinfo{volume}{24}},
  \bibinfo{pages}{5973} (\bibinfo{year}{2007}).

\bibitem[{\citenamefont{Ruiz et~al.}(2007)\citenamefont{Ruiz, Rinne, and
  Sarbach}}]{RuiRinSar07}
\bibinfo{author}{\bibfnamefont{M.}~\bibnamefont{Ruiz}},
  \bibinfo{author}{\bibfnamefont{O.}~\bibnamefont{Rinne}}, \bibnamefont{and}
  \bibinfo{author}{\bibfnamefont{O.}~\bibnamefont{Sarbach}},
  \bibinfo{journal}{Class. Quant. Grav.} \textbf{\bibinfo{volume}{24}},
  \bibinfo{pages}{6349} (\bibinfo{year}{2007}), \eprint{0707.2797}.

\bibitem[{\citenamefont{Kreiss et~al.}(2009)\citenamefont{Kreiss, Reula,
  Sarbach, and Winicour}}]{KreReuSar08}
\bibinfo{author}{\bibfnamefont{H.-O.} \bibnamefont{Kreiss}},
  \bibinfo{author}{\bibfnamefont{O.}~\bibnamefont{Reula}},
  \bibinfo{author}{\bibfnamefont{O.}~\bibnamefont{Sarbach}}, \bibnamefont{and}
  \bibinfo{author}{\bibfnamefont{J.}~\bibnamefont{Winicour}},
  \bibinfo{journal}{Commun.Math.Phys.} \textbf{\bibinfo{volume}{289}},
  \bibinfo{pages}{1099} (\bibinfo{year}{2009}), \eprint{0807.3207}.

\bibitem[{\citenamefont{Friedrich}(1985)}]{Fri85}
\bibinfo{author}{\bibfnamefont{H.}~\bibnamefont{Friedrich}},
  \bibinfo{journal}{Comm. Math. Phys.} \textbf{\bibinfo{volume}{100}},
  \bibinfo{pages}{525} (\bibinfo{year}{1985}).

\bibitem[{\citenamefont{Friedrich}(1986)}]{Fri86}
\bibinfo{author}{\bibfnamefont{H.}~\bibnamefont{Friedrich}},
  \bibinfo{journal}{Comm. Math. Phys.} \textbf{\bibinfo{volume}{107}},
  \bibinfo{pages}{587} (\bibinfo{year}{1986}).

\bibitem[{\citenamefont{Garfinkle}(2002)}]{Gar01}
\bibinfo{author}{\bibfnamefont{D.}~\bibnamefont{Garfinkle}},
  \bibinfo{journal}{Phys. Rev. D} \textbf{\bibinfo{volume}{65}},
  \bibinfo{pages}{044029} (\bibinfo{year}{2002}), \eprint{gr-qc/0110013}.

\bibitem[{\citenamefont{Pretorius}(2005{\natexlab{a}})}]{Pre04}
\bibinfo{author}{\bibfnamefont{F.}~\bibnamefont{Pretorius}},
  \bibinfo{journal}{Class. Quant. Grav.} \textbf{\bibinfo{volume}{22}},
  \bibinfo{pages}{425} (\bibinfo{year}{2005}{\natexlab{a}}),
  \eprint{gr-qc/0407110}.

\bibitem[{\citenamefont{Pretorius}(2005{\natexlab{b}})}]{Pre05}
\bibinfo{author}{\bibfnamefont{F.}~\bibnamefont{Pretorius}},
  \bibinfo{journal}{Phys. Rev. Lett.} \textbf{\bibinfo{volume}{95}},
  \bibinfo{pages}{121101} (\bibinfo{year}{2005}{\natexlab{b}}),
  \eprint{gr-qc/0507014}.

\bibitem[{\citenamefont{Lindblom et~al.}(2006)\citenamefont{Lindblom, Scheel,
  Kidder, Owen, and Rinne}}]{LinSchKid05}
\bibinfo{author}{\bibfnamefont{L.}~\bibnamefont{Lindblom}},
  \bibinfo{author}{\bibfnamefont{M.~A.} \bibnamefont{Scheel}},
  \bibinfo{author}{\bibfnamefont{L.~E.} \bibnamefont{Kidder}},
  \bibinfo{author}{\bibfnamefont{R.}~\bibnamefont{Owen}}, \bibnamefont{and}
  \bibinfo{author}{\bibfnamefont{O.}~\bibnamefont{Rinne}},
  \bibinfo{journal}{Class. Quant. Grav.} \textbf{\bibinfo{volume}{23}},
  \bibinfo{pages}{S447} (\bibinfo{year}{2006}), \eprint{gr-qc/0512093}.

\bibitem[{\citenamefont{Boyle et~al.}(2007)\citenamefont{Boyle, Lindblom,
  Pfeiffer, Scheel, and Kidder}}]{BoyLinPfe06}
\bibinfo{author}{\bibfnamefont{M.}~\bibnamefont{Boyle}},
  \bibinfo{author}{\bibfnamefont{L.}~\bibnamefont{Lindblom}},
  \bibinfo{author}{\bibfnamefont{H.}~\bibnamefont{Pfeiffer}},
  \bibinfo{author}{\bibfnamefont{M.}~\bibnamefont{Scheel}}, \bibnamefont{and}
  \bibinfo{author}{\bibfnamefont{L.~E.} \bibnamefont{Kidder}},
  \bibinfo{journal}{Phys. Rev.} \textbf{\bibinfo{volume}{D75}},
  \bibinfo{pages}{024006} (\bibinfo{year}{2007}), \eprint{gr-qc/0609047}.

\bibitem[{\citenamefont{Pfeiffer et~al.}(2007)\citenamefont{Pfeiffer, Brown,
  Kidder, Lindblom, Lovelace, and Scheel}}]{PfeBroKid07}
\bibinfo{author}{\bibfnamefont{H.~P.} \bibnamefont{Pfeiffer}},
  \bibinfo{author}{\bibfnamefont{D.~A.} \bibnamefont{Brown}},
  \bibinfo{author}{\bibfnamefont{L.~E.} \bibnamefont{Kidder}},
  \bibinfo{author}{\bibfnamefont{L.}~\bibnamefont{Lindblom}},
  \bibinfo{author}{\bibfnamefont{G.}~\bibnamefont{Lovelace}}, \bibnamefont{and}
  \bibinfo{author}{\bibfnamefont{M.}~\bibnamefont{Scheel}},
  \bibinfo{journal}{Classical Quantum Gravity} \textbf{\bibinfo{volume}{24}},
  \bibinfo{pages}{S59} (\bibinfo{year}{2007}), \eprint{gr-qc/0702106}.

\bibitem[{\citenamefont{Seiler et~al.}(2008)\citenamefont{Seiler, Szilagyi,
  Pollney, and Rezzolla}}]{SeiSziPol08}
\bibinfo{author}{\bibfnamefont{J.}~\bibnamefont{Seiler}},
  \bibinfo{author}{\bibfnamefont{B.}~\bibnamefont{Szilagyi}},
  \bibinfo{author}{\bibfnamefont{D.}~\bibnamefont{Pollney}}, \bibnamefont{and}
  \bibinfo{author}{\bibfnamefont{L.}~\bibnamefont{Rezzolla}},
  \bibinfo{journal}{Class. Quant. Grav.} \textbf{\bibinfo{volume}{25}},
  \bibinfo{pages}{175020} (\bibinfo{year}{2008}), \eprint{0802.3341}.

\bibitem[{\citenamefont{Hilditch et~al.}(2016)\citenamefont{Hilditch,
  Weyhausen, and Brügmann}}]{HilWeyBru15}
\bibinfo{author}{\bibfnamefont{D.}~\bibnamefont{Hilditch}},
  \bibinfo{author}{\bibfnamefont{A.}~\bibnamefont{Weyhausen}},
  \bibnamefont{and}
  \bibinfo{author}{\bibfnamefont{B.}~\bibnamefont{Brügmann}},
  \bibinfo{journal}{Phys. Rev.} \textbf{\bibinfo{volume}{D93}},
  \bibinfo{pages}{063006} (\bibinfo{year}{2016}), \eprint{1504.04732}.

\bibitem[{\citenamefont{Baumgarte and Shapiro}(1998)}]{BauSha98}
\bibinfo{author}{\bibfnamefont{T.~W.} \bibnamefont{Baumgarte}}
  \bibnamefont{and} \bibinfo{author}{\bibfnamefont{S.~L.}
  \bibnamefont{Shapiro}}, \bibinfo{journal}{Phys. Rev. D}
  \textbf{\bibinfo{volume}{59}}, \bibinfo{pages}{024007}
  (\bibinfo{year}{1998}), \eprint{gr-qc/9810065}.

\bibitem[{\citenamefont{Shibata and Nakamura}(1995)}]{ShiNak95}
\bibinfo{author}{\bibfnamefont{M.}~\bibnamefont{Shibata}} \bibnamefont{and}
  \bibinfo{author}{\bibfnamefont{T.}~\bibnamefont{Nakamura}},
  \bibinfo{journal}{Phys. Rev. D} \textbf{\bibinfo{volume}{52}},
  \bibinfo{pages}{5428} (\bibinfo{year}{1995}).

\bibitem[{\citenamefont{Nakamura et~al.}(1987)\citenamefont{Nakamura, Oohara,
  and Kojima}}]{NakOohKoj87}
\bibinfo{author}{\bibfnamefont{T.}~\bibnamefont{Nakamura}},
  \bibinfo{author}{\bibfnamefont{K.}~\bibnamefont{Oohara}}, \bibnamefont{and}
  \bibinfo{author}{\bibfnamefont{Y.}~\bibnamefont{Kojima}},
  \bibinfo{journal}{Prog. Theor. Phys. Suppl.} \textbf{\bibinfo{volume}{90}},
  \bibinfo{pages}{1} (\bibinfo{year}{1987}).

\bibitem[{\citenamefont{Bona et~al.}(2003{\natexlab{a}})\citenamefont{Bona,
  Ledvinka, Palenzuela, and {\v Z}{\'a}{\v c}ek}}]{BonLedPal03}
\bibinfo{author}{\bibfnamefont{C.}~\bibnamefont{Bona}},
  \bibinfo{author}{\bibfnamefont{T.}~\bibnamefont{Ledvinka}},
  \bibinfo{author}{\bibfnamefont{C.}~\bibnamefont{Palenzuela}},
  \bibnamefont{and} \bibinfo{author}{\bibfnamefont{M.}~\bibnamefont{{\v
  Z}{\'a}{\v c}ek}}, \bibinfo{journal}{Phys. Rev. D}
  \textbf{\bibinfo{volume}{67}}, \bibinfo{pages}{104005}
  (\bibinfo{year}{2003}{\natexlab{a}}), \eprint{gr-qc/0302083}.

\bibitem[{\citenamefont{Bona et~al.}(2003{\natexlab{b}})\citenamefont{Bona,
  Ledvinka, Palenzuela, and {\v Z}{\'a}{\v c}ek}}]{BonLedPal03a}
\bibinfo{author}{\bibfnamefont{C.}~\bibnamefont{Bona}},
  \bibinfo{author}{\bibfnamefont{T.}~\bibnamefont{Ledvinka}},
  \bibinfo{author}{\bibfnamefont{C.}~\bibnamefont{Palenzuela}},
  \bibnamefont{and} \bibinfo{author}{\bibfnamefont{M.}~\bibnamefont{{\v
  Z}{\'a}{\v c}ek}} (\bibinfo{year}{2003}{\natexlab{b}}),
  \bibinfo{note}{gr-qc/0307067}.

\bibitem[{\citenamefont{Bernuzzi and Hilditch}(2010)}]{BerHil09}
\bibinfo{author}{\bibfnamefont{S.}~\bibnamefont{Bernuzzi}} \bibnamefont{and}
  \bibinfo{author}{\bibfnamefont{D.}~\bibnamefont{Hilditch}},
  \bibinfo{journal}{Phys. Rev. D} \textbf{\bibinfo{volume}{81}},
  \bibinfo{pages}{084003} (\bibinfo{year}{2010}), \eprint{0912.2920}.

\bibitem[{\citenamefont{Weyhausen et~al.}(2012)\citenamefont{Weyhausen,
  Bernuzzi, and Hilditch}}]{WeyBerHil11}
\bibinfo{author}{\bibfnamefont{A.}~\bibnamefont{Weyhausen}},
  \bibinfo{author}{\bibfnamefont{S.}~\bibnamefont{Bernuzzi}}, \bibnamefont{and}
  \bibinfo{author}{\bibfnamefont{D.}~\bibnamefont{Hilditch}},
  \bibinfo{journal}{Phys. Rev. D} \textbf{\bibinfo{volume}{85}},
  \bibinfo{pages}{024038} (\bibinfo{year}{2012}), \eprint{1107.5539}.

\bibitem[{\citenamefont{Alic et~al.}(2012)\citenamefont{Alic, Bona-Casas, Bona,
  Rezzolla, and Palenzuela}}]{AliBonBon11}
\bibinfo{author}{\bibfnamefont{D.}~\bibnamefont{Alic}},
  \bibinfo{author}{\bibfnamefont{C.}~\bibnamefont{Bona-Casas}},
  \bibinfo{author}{\bibfnamefont{C.}~\bibnamefont{Bona}},
  \bibinfo{author}{\bibfnamefont{L.}~\bibnamefont{Rezzolla}}, \bibnamefont{and}
  \bibinfo{author}{\bibfnamefont{C.}~\bibnamefont{Palenzuela}},
  \bibinfo{journal}{Phys. Rev. D} \textbf{\bibinfo{volume}{85}},
  \bibinfo{pages}{064040} (\bibinfo{year}{2012}), \eprint{1106.2254}.

\bibitem[{\citenamefont{Cao and Hilditch}(2012)}]{CaoHil11}
\bibinfo{author}{\bibfnamefont{Z.}~\bibnamefont{Cao}} \bibnamefont{and}
  \bibinfo{author}{\bibfnamefont{D.}~\bibnamefont{Hilditch}},
  \bibinfo{journal}{Phys. Rev. D} \textbf{\bibinfo{volume}{85}},
  \bibinfo{pages}{124032} (\bibinfo{year}{2012}), \eprint{1111.2177}.

\bibitem[{\citenamefont{Alic et~al.}(2013)\citenamefont{Alic, Kastaun, and
  Rezzolla}}]{AliKasRez13}
\bibinfo{author}{\bibfnamefont{D.}~\bibnamefont{Alic}},
  \bibinfo{author}{\bibfnamefont{W.}~\bibnamefont{Kastaun}}, \bibnamefont{and}
  \bibinfo{author}{\bibfnamefont{L.}~\bibnamefont{Rezzolla}},
  \bibinfo{journal}{Phys. Rev.} \textbf{\bibinfo{volume}{D88}},
  \bibinfo{pages}{064049} (\bibinfo{year}{2013}), \eprint{1307.7391}.

\bibitem[{\citenamefont{Bona et~al.}(1995{\natexlab{a}})\citenamefont{Bona,
  Mass{\'o}, Seidel, and Stela}}]{BonMasSei94a}
\bibinfo{author}{\bibfnamefont{C.}~\bibnamefont{Bona}},
  \bibinfo{author}{\bibfnamefont{J.}~\bibnamefont{Mass{\'o}}},
  \bibinfo{author}{\bibfnamefont{E.}~\bibnamefont{Seidel}}, \bibnamefont{and}
  \bibinfo{author}{\bibfnamefont{J.}~\bibnamefont{Stela}},
  \bibinfo{journal}{Phys. Rev. Lett.} \textbf{\bibinfo{volume}{75}},
  \bibinfo{pages}{600} (\bibinfo{year}{1995}{\natexlab{a}}),
  \eprint{gr-qc/9412071}.

\bibitem[{\citenamefont{Alcubierre}(2003)}]{Alc02}
\bibinfo{author}{\bibfnamefont{M.}~\bibnamefont{Alcubierre}},
  \bibinfo{journal}{Class. Quantum Grav.} \textbf{\bibinfo{volume}{20}},
  \bibinfo{pages}{607} (\bibinfo{year}{2003}), \eprint{gr-qc/0210050}.

\bibitem[{\citenamefont{Baker et~al.}(2006)\citenamefont{Baker, Centrella,
  Choi, Koppitz, and van Meter}}]{BakCenCho05}
\bibinfo{author}{\bibfnamefont{J.~G.} \bibnamefont{Baker}},
  \bibinfo{author}{\bibfnamefont{J.}~\bibnamefont{Centrella}},
  \bibinfo{author}{\bibfnamefont{D.-I.} \bibnamefont{Choi}},
  \bibinfo{author}{\bibfnamefont{M.}~\bibnamefont{Koppitz}}, \bibnamefont{and}
  \bibinfo{author}{\bibfnamefont{J.}~\bibnamefont{van Meter}},
  \bibinfo{journal}{Phys. Rev. Lett.} \textbf{\bibinfo{volume}{96}},
  \bibinfo{pages}{111102} (\bibinfo{year}{2006}), \eprint{gr-qc/0511103}.

\bibitem[{\citenamefont{Campanelli et~al.}(2006)\citenamefont{Campanelli,
  Lousto, Marronetti, and Zlochower}}]{CamLouMar05}
\bibinfo{author}{\bibfnamefont{M.}~\bibnamefont{Campanelli}},
  \bibinfo{author}{\bibfnamefont{C.~O.} \bibnamefont{Lousto}},
  \bibinfo{author}{\bibfnamefont{P.}~\bibnamefont{Marronetti}},
  \bibnamefont{and}
  \bibinfo{author}{\bibfnamefont{Y.}~\bibnamefont{Zlochower}},
  \bibinfo{journal}{Phys. Rev. Lett.} \textbf{\bibinfo{volume}{96}},
  \bibinfo{pages}{111101} (\bibinfo{year}{2006}), \eprint{gr-qc/0511048}.

\bibitem[{\citenamefont{van Meter et~al.}(2006)\citenamefont{van Meter, Baker,
  Koppitz, and Choi}}]{MetBakKop06}
\bibinfo{author}{\bibfnamefont{J.~R.} \bibnamefont{van Meter}},
  \bibinfo{author}{\bibfnamefont{J.~G.} \bibnamefont{Baker}},
  \bibinfo{author}{\bibfnamefont{M.}~\bibnamefont{Koppitz}}, \bibnamefont{and}
  \bibinfo{author}{\bibfnamefont{D.-I.} \bibnamefont{Choi}},
  \bibinfo{journal}{Phys. Rev. D} \textbf{\bibinfo{volume}{73}},
  \bibinfo{pages}{124011} (\bibinfo{year}{2006}), \eprint{gr-qc/0605030}.

\bibitem[{\citenamefont{Gundlach and Martin-Garcia}(2006)}]{GunGar06}
\bibinfo{author}{\bibfnamefont{C.}~\bibnamefont{Gundlach}} \bibnamefont{and}
  \bibinfo{author}{\bibfnamefont{J.~M.} \bibnamefont{Martin-Garcia}},
  \bibinfo{journal}{Phys. Rev. D} \textbf{\bibinfo{volume}{74}},
  \bibinfo{pages}{024016} (\bibinfo{year}{2006}), \eprint{gr-qc/0604035}.

\bibitem[{\citenamefont{Alcubierre}(2008)}]{Alc08}
\bibinfo{author}{\bibfnamefont{M.}~\bibnamefont{Alcubierre}},
  \emph{\bibinfo{title}{Introduction to 3+1 Numerical Relativity}}
  (\bibinfo{publisher}{Oxford University Press}, \bibinfo{address}{Oxford},
  \bibinfo{year}{2008}).

\bibitem[{\citenamefont{Beyer and Sarbach}(2004)}]{BeySar04}
\bibinfo{author}{\bibfnamefont{H.}~\bibnamefont{Beyer}} \bibnamefont{and}
  \bibinfo{author}{\bibfnamefont{O.}~\bibnamefont{Sarbach}},
  \bibinfo{journal}{Phys. Rev. D} \textbf{\bibinfo{volume}{70}},
  \bibinfo{pages}{104004} (\bibinfo{year}{2004}), \eprint{gr-qc/0406003}.

\bibitem[{\citenamefont{Nunez and Sarbach}(2010)}]{NunSar09}
\bibinfo{author}{\bibfnamefont{D.}~\bibnamefont{Nunez}} \bibnamefont{and}
  \bibinfo{author}{\bibfnamefont{O.}~\bibnamefont{Sarbach}},
  \bibinfo{journal}{Phys. Rev.} \textbf{\bibinfo{volume}{D81}},
  \bibinfo{pages}{044011} (\bibinfo{year}{2010}), \eprint{0910.5763}.

\bibitem[{\citenamefont{Ruiz et~al.}(2012)\citenamefont{Ruiz, Degollado,
  Alcubierre, Nunez, and Salgado}}]{RuDeAlNuSa12}
\bibinfo{author}{\bibfnamefont{M.}~\bibnamefont{Ruiz}},
  \bibinfo{author}{\bibfnamefont{J.~C.} \bibnamefont{Degollado}},
  \bibinfo{author}{\bibfnamefont{M.}~\bibnamefont{Alcubierre}},
  \bibinfo{author}{\bibfnamefont{D.}~\bibnamefont{Nunez}}, \bibnamefont{and}
  \bibinfo{author}{\bibfnamefont{M.}~\bibnamefont{Salgado}},
  \bibinfo{journal}{Phys. Rev.} \textbf{\bibinfo{volume}{D86}},
  \bibinfo{pages}{104044} (\bibinfo{year}{2012}), \eprint{1207.6142}.

\bibitem[{\citenamefont{Alcubierre and Torres}(2015)}]{AlToe14}
\bibinfo{author}{\bibfnamefont{M.}~\bibnamefont{Alcubierre}} \bibnamefont{and}
  \bibinfo{author}{\bibfnamefont{J.~M.} \bibnamefont{Torres}},
  \bibinfo{journal}{Class. Quant. Grav.} \textbf{\bibinfo{volume}{32}},
  \bibinfo{pages}{035006} (\bibinfo{year}{2015}), \eprint{1407.8529}.

\bibitem[{\citenamefont{Sarbach and Tiglio}(2012)}]{SarTig12}
\bibinfo{author}{\bibfnamefont{O.}~\bibnamefont{Sarbach}} \bibnamefont{and}
  \bibinfo{author}{\bibfnamefont{M.}~\bibnamefont{Tiglio}},
  \bibinfo{journal}{Living Reviews in Relativity} \textbf{\bibinfo{volume}{15}}
  (\bibinfo{year}{2012}), \eprint{1203.6443},
  \urlprefix\url{http://www.livingreviews.org/lrr-2012-9}.

\bibitem[{\citenamefont{Bona et~al.}(2005)\citenamefont{Bona, Ledvinka,
  Palenzuela-Luque, and Zacek}}]{BonLedLuq04a}
\bibinfo{author}{\bibfnamefont{C.}~\bibnamefont{Bona}},
  \bibinfo{author}{\bibfnamefont{T.}~\bibnamefont{Ledvinka}},
  \bibinfo{author}{\bibfnamefont{C.}~\bibnamefont{Palenzuela-Luque}},
  \bibnamefont{and} \bibinfo{author}{\bibfnamefont{M.}~\bibnamefont{Zacek}},
  \bibinfo{journal}{Class. Quantum Grav.} \textbf{\bibinfo{volume}{22}},
  \bibinfo{pages}{2615} (\bibinfo{year}{2005}), \eprint{gr-qc/0411110}.

\bibitem[{\citenamefont{Bona and Bona-Casas}(2010)}]{BonBon10}
\bibinfo{author}{\bibfnamefont{C.}~\bibnamefont{Bona}} \bibnamefont{and}
  \bibinfo{author}{\bibfnamefont{C.}~\bibnamefont{Bona-Casas}},
  \bibinfo{journal}{Phys. Rev.} \textbf{\bibinfo{volume}{D82}},
  \bibinfo{pages}{064008} (\bibinfo{year}{2010}), \eprint{1003.3328}.

\bibitem[{\citenamefont{Hilditch et~al.}(2013)\citenamefont{Hilditch, Bernuzzi,
  Thierfelder, Cao, Tichy, and Br{\"u}gmann}}]{HilBerThi12}
\bibinfo{author}{\bibfnamefont{D.}~\bibnamefont{Hilditch}},
  \bibinfo{author}{\bibfnamefont{S.}~\bibnamefont{Bernuzzi}},
  \bibinfo{author}{\bibfnamefont{M.}~\bibnamefont{Thierfelder}},
  \bibinfo{author}{\bibfnamefont{Z.}~\bibnamefont{Cao}},
  \bibinfo{author}{\bibfnamefont{W.}~\bibnamefont{Tichy}}, \bibnamefont{and}
  \bibinfo{author}{\bibfnamefont{B.}~\bibnamefont{Br{\"u}gmann}},
  \bibinfo{journal}{Phys. Rev. D} \textbf{\bibinfo{volume}{88}},
  \bibinfo{pages}{084057} (\bibinfo{year}{2013}), \eprint{1212.2901}.

\bibitem[{\citenamefont{Hilditch and Richter}(2016)}]{HilRic13}
\bibinfo{author}{\bibfnamefont{D.}~\bibnamefont{Hilditch}} \bibnamefont{and}
  \bibinfo{author}{\bibfnamefont{R.}~\bibnamefont{Richter}},
  \bibinfo{journal}{Phys. Rev.} \textbf{\bibinfo{volume}{D94}},
  \bibinfo{pages}{044028} (\bibinfo{year}{2016}), \eprint{1303.4783}.

\bibitem[{\citenamefont{Eskin}(1981)}]{Es73}
\bibinfo{author}{\bibfnamefont{G.~I.} \bibnamefont{Eskin}},
  \emph{\bibinfo{title}{Boundary value problems for elliptic pseudodifferential
  equations; Translations of mathematical monographs, V. 52}}
  (\bibinfo{publisher}{American Mathematical Society},
  \bibinfo{address}{Providence, R.I.}, \bibinfo{year}{1981}).

\bibitem[{\citenamefont{Hilditch}(2015)}]{Hil15}
\bibinfo{author}{\bibfnamefont{D.}~\bibnamefont{Hilditch}}
  (\bibinfo{year}{2015}), \eprint{1509.02071}.

\bibitem[{\citenamefont{Gundlach et~al.}(2005)\citenamefont{Gundlach,
  Martin-Garcia, Calabrese, and Hinder}}]{GunGarCal05}
\bibinfo{author}{\bibfnamefont{C.}~\bibnamefont{Gundlach}},
  \bibinfo{author}{\bibfnamefont{J.~M.} \bibnamefont{Martin-Garcia}},
  \bibinfo{author}{\bibfnamefont{G.}~\bibnamefont{Calabrese}},
  \bibnamefont{and} \bibinfo{author}{\bibfnamefont{I.}~\bibnamefont{Hinder}},
  \bibinfo{journal}{Class. Quantum Grav.} \textbf{\bibinfo{volume}{22}},
  \bibinfo{pages}{3767} (\bibinfo{year}{2005}), \eprint{gr-qc/0504114}.

\bibitem[{\citenamefont{Bona et~al.}(1995{\natexlab{b}})\citenamefont{Bona,
  Mass{\'o}, Seidel, and Stela}}]{BonMas94}
\bibinfo{author}{\bibfnamefont{C.}~\bibnamefont{Bona}},
  \bibinfo{author}{\bibfnamefont{J.}~\bibnamefont{Mass{\'o}}},
  \bibinfo{author}{\bibfnamefont{E.}~\bibnamefont{Seidel}}, \bibnamefont{and}
  \bibinfo{author}{\bibfnamefont{J.}~\bibnamefont{Stela}},
  \bibinfo{journal}{Phys. Rev. Lett.} \textbf{\bibinfo{volume}{75}},
  \bibinfo{pages}{600} (\bibinfo{year}{1995}{\natexlab{b}}),
  \eprint{gr-qc/9412071}.

\bibitem[{\citenamefont{Alcubierre et~al.}(2003)\citenamefont{Alcubierre,
  Br{\"u}gmann, Diener, Koppitz, Pollney, Seidel, and Takahashi}}]{AlcBruDie02}
\bibinfo{author}{\bibfnamefont{M.}~\bibnamefont{Alcubierre}},
  \bibinfo{author}{\bibfnamefont{B.}~\bibnamefont{Br{\"u}gmann}},
  \bibinfo{author}{\bibfnamefont{P.}~\bibnamefont{Diener}},
  \bibinfo{author}{\bibfnamefont{M.}~\bibnamefont{Koppitz}},
  \bibinfo{author}{\bibfnamefont{D.}~\bibnamefont{Pollney}},
  \bibinfo{author}{\bibfnamefont{E.}~\bibnamefont{Seidel}}, \bibnamefont{and}
  \bibinfo{author}{\bibfnamefont{R.}~\bibnamefont{Takahashi}},
  \bibinfo{journal}{Phys. Rev. D} \textbf{\bibinfo{volume}{67}},
  \bibinfo{pages}{084023} (\bibinfo{year}{2003}), \eprint{gr-qc/0206072}.

\bibitem[{\citenamefont{Schnetter}(2010)}]{Sch10}
\bibinfo{author}{\bibfnamefont{E.}~\bibnamefont{Schnetter}},
  \bibinfo{journal}{Class. Quant. Grav.} \textbf{\bibinfo{volume}{27}},
  \bibinfo{pages}{167001} (\bibinfo{year}{2010}), \eprint{1003.0859}.

\bibitem[{\citenamefont{M{\"u}ller and Br{\"u}gmann}(2010)}]{MueBru09}
\bibinfo{author}{\bibfnamefont{D.}~\bibnamefont{M{\"u}ller}} \bibnamefont{and}
  \bibinfo{author}{\bibfnamefont{B.}~\bibnamefont{Br{\"u}gmann}},
  \bibinfo{journal}{Class. Quant. Grav.} \textbf{\bibinfo{volume}{27}},
  \bibinfo{pages}{114008} (\bibinfo{year}{2010}), \eprint{0912.3125}.

\bibitem[{\citenamefont{Alic et~al.}(2010)\citenamefont{Alic, Rezzolla, Hinder,
  and Mosta}}]{AliRezHin10}
\bibinfo{author}{\bibfnamefont{D.}~\bibnamefont{Alic}},
  \bibinfo{author}{\bibfnamefont{L.}~\bibnamefont{Rezzolla}},
  \bibinfo{author}{\bibfnamefont{I.}~\bibnamefont{Hinder}}, \bibnamefont{and}
  \bibinfo{author}{\bibfnamefont{P.}~\bibnamefont{Mosta}},
  \bibinfo{journal}{Class. Quant. Grav.} \textbf{\bibinfo{volume}{27}},
  \bibinfo{pages}{245023} (\bibinfo{year}{2010}), \eprint{1008.2212}.

\bibitem[{\citenamefont{Rinne}(2005)}]{Rin06}
\bibinfo{author}{\bibfnamefont{O.}~\bibnamefont{Rinne}}, Ph.D. thesis,
  \bibinfo{school}{University of {C}ambridge}, \bibinfo{address}{Cambridge,
  England} (\bibinfo{year}{2005}), \bibinfo{note}{gr-qc/0601064}.

\bibitem[{\citenamefont{Hilditch}(2017)}]{Hil17}
\bibinfo{author}{\bibfnamefont{D.}~\bibnamefont{Hilditch}}
  (\bibinfo{year}{2017}).

\bibitem[{\citenamefont{Sarbach and Tiglio}(2005)}]{SarTig04}
\bibinfo{author}{\bibfnamefont{O.}~\bibnamefont{Sarbach}} \bibnamefont{and}
  \bibinfo{author}{\bibfnamefont{M.}~\bibnamefont{Tiglio}},
  \bibinfo{journal}{Journal of Hyperbolic Differential Equations}
  \textbf{\bibinfo{volume}{2}}, \bibinfo{pages}{839} (\bibinfo{year}{2005}),
  \eprint{gr-qc/0412115}.

\bibitem[{\citenamefont{Rinne et~al.}(2007)\citenamefont{Rinne, Lindblom, and
  Scheel}}]{RinLinSch07}
\bibinfo{author}{\bibfnamefont{O.}~\bibnamefont{Rinne}},
  \bibinfo{author}{\bibfnamefont{L.}~\bibnamefont{Lindblom}}, \bibnamefont{and}
  \bibinfo{author}{\bibfnamefont{M.~A.} \bibnamefont{Scheel}},
  \bibinfo{journal}{Class. Quant. Grav.} \textbf{\bibinfo{volume}{24}},
  \bibinfo{pages}{4053} (\bibinfo{year}{2007}), \eprint{0704.0782}.

\bibitem[{\citenamefont{Buchman and Sarbach}(2007)}]{BucSar07}
\bibinfo{author}{\bibfnamefont{L.~T.} \bibnamefont{Buchman}} \bibnamefont{and}
  \bibinfo{author}{\bibfnamefont{O.~C.} \bibnamefont{Sarbach}},
  \bibinfo{journal}{Class.Quant.Grav.} \textbf{\bibinfo{volume}{24}},
  \bibinfo{pages}{S307} (\bibinfo{year}{2007}), \eprint{gr-qc/0703129}.

\bibitem[{\citenamefont{Witek et~al.}(2011)\citenamefont{Witek, Hilditch, and
  Sperhake}}]{WitHilSpe10}
\bibinfo{author}{\bibfnamefont{H.}~\bibnamefont{Witek}},
  \bibinfo{author}{\bibfnamefont{D.}~\bibnamefont{Hilditch}}, \bibnamefont{and}
  \bibinfo{author}{\bibfnamefont{U.}~\bibnamefont{Sperhake}},
  \bibinfo{journal}{Phys. Rev.} \textbf{\bibinfo{volume}{D83}},
  \bibinfo{pages}{104041} (\bibinfo{year}{2011}), \eprint{1011.4407}.

\bibitem[{\citenamefont{Pollney et~al.}(2011)\citenamefont{Pollney, Reisswig,
  Schnetter, Dorband, and Diener}}]{PolReiSch11}
\bibinfo{author}{\bibfnamefont{D.}~\bibnamefont{Pollney}},
  \bibinfo{author}{\bibfnamefont{C.}~\bibnamefont{Reisswig}},
  \bibinfo{author}{\bibfnamefont{E.}~\bibnamefont{Schnetter}},
  \bibinfo{author}{\bibfnamefont{N.}~\bibnamefont{Dorband}}, \bibnamefont{and}
  \bibinfo{author}{\bibfnamefont{P.}~\bibnamefont{Diener}},
  \bibinfo{journal}{Phys. Rev. D} \textbf{\bibinfo{volume}{83}},
  \bibinfo{pages}{044045} (\bibinfo{year}{2011}), \eprint{0910.3803}.

\bibitem[{\citenamefont{{Bayliss} and {Turkel}}(1980)}]{BayTur80}
\bibinfo{author}{\bibfnamefont{A.}~\bibnamefont{{Bayliss}}} \bibnamefont{and}
  \bibinfo{author}{\bibfnamefont{E.}~\bibnamefont{{Turkel}}},
  \bibinfo{journal}{Communications in Pure Applied Mathematics}
  \textbf{\bibinfo{volume}{33}}, \bibinfo{pages}{707} (\bibinfo{year}{1980}).

\bibitem[{\citenamefont{Rinne et~al.}(2009)\citenamefont{Rinne, Buchman,
  Scheel, and Pfeiffer}}]{RinBucSch08}
\bibinfo{author}{\bibfnamefont{O.}~\bibnamefont{Rinne}},
  \bibinfo{author}{\bibfnamefont{L.~T.} \bibnamefont{Buchman}},
  \bibinfo{author}{\bibfnamefont{M.~A.} \bibnamefont{Scheel}},
  \bibnamefont{and} \bibinfo{author}{\bibfnamefont{H.~P.}
  \bibnamefont{Pfeiffer}}, \bibinfo{journal}{Class. Quant. Grav.}
  \textbf{\bibinfo{volume}{26}}, \bibinfo{pages}{075009}
  (\bibinfo{year}{2009}), \eprint{0811.3593}.

\bibitem[{\citenamefont{Gustafsson et~al.}(1995)\citenamefont{Gustafsson,
  Kreiss, and Oliger}}]{GusKreOli95}
\bibinfo{author}{\bibfnamefont{B.}~\bibnamefont{Gustafsson}},
  \bibinfo{author}{\bibfnamefont{H.-O.} \bibnamefont{Kreiss}},
  \bibnamefont{and} \bibinfo{author}{\bibfnamefont{J.}~\bibnamefont{Oliger}},
  \emph{\bibinfo{title}{Time dependent problems and difference methods}}
  (\bibinfo{publisher}{Wiley}, \bibinfo{address}{New York},
  \bibinfo{year}{1995}).

\bibitem[{Hil()}]{Hilwebsite}
\bibinfo{note}{{\url{https://www.tpi.uni-jena.de/tiki-view_tracker_item.php?itemId=254}}}.

\bibitem[{\citenamefont{Nagy et~al.}(2004)\citenamefont{Nagy, Ortiz, and
  Reula}}]{NagOrtReu04}
\bibinfo{author}{\bibfnamefont{G.}~\bibnamefont{Nagy}},
  \bibinfo{author}{\bibfnamefont{O.~E.} \bibnamefont{Ortiz}}, \bibnamefont{and}
  \bibinfo{author}{\bibfnamefont{O.~A.} \bibnamefont{Reula}},
  \bibinfo{journal}{Phys. Rev. D} \textbf{\bibinfo{volume}{70}},
  \bibinfo{pages}{044012} (\bibinfo{year}{2004}).

\bibitem[{\citenamefont{Gundlach and Mart{\'\i}n-Garc{\'\i}a}(2006)}]{GunGar05}
\bibinfo{author}{\bibfnamefont{C.}~\bibnamefont{Gundlach}} \bibnamefont{and}
  \bibinfo{author}{\bibfnamefont{J.~M.} \bibnamefont{Mart{\'\i}n-Garc{\'\i}a}},
  \bibinfo{journal}{Class. Quantum Grav.} \textbf{\bibinfo{volume}{23}},
  \bibinfo{pages}{S387} (\bibinfo{year}{2006}), \eprint{gr-qc/0506037}.

\bibitem[{\citenamefont{Kreiss}(1970)}]{Kre70}
\bibinfo{author}{\bibfnamefont{H.-O.} \bibnamefont{Kreiss}},
  \bibinfo{journal}{Comm. Pure Appl. Math.} \textbf{\bibinfo{volume}{23}},
  \bibinfo{pages}{277} (\bibinfo{year}{1970}).

\bibitem[{\citenamefont{Agranovich}(1972)}]{Agr72}
\bibinfo{author}{\bibfnamefont{M.~S.} \bibnamefont{Agranovich}},
  \bibinfo{journal}{Functional Analysis and Its Applications}
  \textbf{\bibinfo{volume}{6}}, \bibinfo{pages}{85} (\bibinfo{year}{1972}),
  ISSN \bibinfo{issn}{0016-2663}.

\bibitem[{\citenamefont{M\'etivier}(2000)}]{Met00}
\bibinfo{author}{\bibfnamefont{G.}~\bibnamefont{M\'etivier}},
  \bibinfo{journal}{Bulletin of the London Mathematical Society}
  \textbf{\bibinfo{volume}{32}}, \bibinfo{pages}{689} (\bibinfo{year}{2000}).

\bibitem[{\citenamefont{Frauendiener}(2004)}]{Fra04}
\bibinfo{author}{\bibfnamefont{J.}~\bibnamefont{Frauendiener}},
  \bibinfo{journal}{Living Rev. Relativity} \textbf{\bibinfo{volume}{7}}
  (\bibinfo{year}{2004}),
  \bibinfo{note}{http://www.livingreviews.org/lrr-2004-1}.

\bibitem[{\citenamefont{Calabrese et~al.}(2006)\citenamefont{Calabrese,
  Gundlach, and Hilditch}}]{CalGunHil05}
\bibinfo{author}{\bibfnamefont{G.}~\bibnamefont{Calabrese}},
  \bibinfo{author}{\bibfnamefont{C.}~\bibnamefont{Gundlach}}, \bibnamefont{and}
  \bibinfo{author}{\bibfnamefont{D.}~\bibnamefont{Hilditch}},
  \bibinfo{journal}{Class.Quant.Grav.} \textbf{\bibinfo{volume}{23}},
  \bibinfo{pages}{4829} (\bibinfo{year}{2006}), \eprint{gr-qc/0512149}.

\bibitem[{\citenamefont{Zengino{\u g}lu and Husa}(2008)}]{ZenHus06}
\bibinfo{author}{\bibfnamefont{A.}~\bibnamefont{Zengino{\u g}lu}}
  \bibnamefont{and} \bibinfo{author}{\bibfnamefont{S.}~\bibnamefont{Husa}},
  \bibinfo{journal}{Class. Quantum Grav.} \textbf{\bibinfo{volume}{25}},
  \bibinfo{pages}{19} (\bibinfo{year}{2008}), \eprint{gr-qc/0612161}.

\bibitem[{\citenamefont{Zenginoglu}(2008)}]{Zen08}
\bibinfo{author}{\bibfnamefont{A.}~\bibnamefont{Zenginoglu}},
  \bibinfo{journal}{Class. Quant. Grav.} \textbf{\bibinfo{volume}{25}},
  \bibinfo{pages}{195025} (\bibinfo{year}{2008}), \eprint{0808.0810}.

\bibitem[{\citenamefont{Moncrief and Rinne}(2009)}]{MonRin08}
\bibinfo{author}{\bibfnamefont{V.}~\bibnamefont{Moncrief}} \bibnamefont{and}
  \bibinfo{author}{\bibfnamefont{O.}~\bibnamefont{Rinne}},
  \bibinfo{journal}{Class.Quant.Grav.} \textbf{\bibinfo{volume}{26}},
  \bibinfo{pages}{125010} (\bibinfo{year}{2009}), \eprint{0811.4109}.

\bibitem[{\citenamefont{Buchman et~al.}(2009)\citenamefont{Buchman, Pfeiffer,
  and Bardeen}}]{BucPfeBar09}
\bibinfo{author}{\bibfnamefont{L.~T.} \bibnamefont{Buchman}},
  \bibinfo{author}{\bibfnamefont{H.~P.} \bibnamefont{Pfeiffer}},
  \bibnamefont{and} \bibinfo{author}{\bibfnamefont{J.~M.}
  \bibnamefont{Bardeen}}, \bibinfo{journal}{Phys.Rev.}
  \textbf{\bibinfo{volume}{D80}}, \bibinfo{pages}{084024}
  (\bibinfo{year}{2009}), \eprint{0907.3163}.

\bibitem[{\citenamefont{Rinne}(2010)}]{Rin10}
\bibinfo{author}{\bibfnamefont{O.}~\bibnamefont{Rinne}},
  \bibinfo{journal}{Class.Quant.Grav.} \textbf{\bibinfo{volume}{27}},
  \bibinfo{pages}{035014} (\bibinfo{year}{2010}), \eprint{0910.0139}.

\bibitem[{\citenamefont{Va\~n\'o Vi\~nuales et~al.}(2015)\citenamefont{Va\~n\'o
  Vi\~nuales, Husa, and Hilditch}}]{VanHusHil15}
\bibinfo{author}{\bibfnamefont{A.}~\bibnamefont{Va\~n\'o Vi\~nuales}},
  \bibinfo{author}{\bibfnamefont{S.}~\bibnamefont{Husa}}, \bibnamefont{and}
  \bibinfo{author}{\bibfnamefont{D.}~\bibnamefont{Hilditch}},
  \bibinfo{journal}{Class. Quant. Grav.} \textbf{\bibinfo{volume}{32}},
  \bibinfo{pages}{175010} (\bibinfo{year}{2015}), \eprint{1412.3827}.

\bibitem[{\citenamefont{Winicour}(1998)}]{Win98}
\bibinfo{author}{\bibfnamefont{J.}~\bibnamefont{Winicour}},
  \bibinfo{journal}{Living Rev. Relativity} \textbf{\bibinfo{volume}{1}},
  \bibinfo{pages}{5} (\bibinfo{year}{1998}), \bibinfo{note}{[Online article]},
  \urlprefix\url{http://www.livingreviews.org/lrr-1998-5}.

\bibitem[{\citenamefont{Reisswig et~al.}(2009)\citenamefont{Reisswig, Bishop,
  Pollney, and Szilagyi}}]{ReiBisPol09}
\bibinfo{author}{\bibfnamefont{C.}~\bibnamefont{Reisswig}},
  \bibinfo{author}{\bibfnamefont{N.~T.} \bibnamefont{Bishop}},
  \bibinfo{author}{\bibfnamefont{D.}~\bibnamefont{Pollney}}, \bibnamefont{and}
  \bibinfo{author}{\bibfnamefont{B.}~\bibnamefont{Szilagyi}},
  \bibinfo{journal}{Phys. Rev. Lett.} \textbf{\bibinfo{volume}{103}},
  \bibinfo{pages}{221101} (\bibinfo{year}{2009}), \eprint{0907.2637}.

\bibitem[{\citenamefont{Bizon and Rostworowski}(2011)}]{BizRos11}
\bibinfo{author}{\bibfnamefont{P.}~\bibnamefont{Bizon}} \bibnamefont{and}
  \bibinfo{author}{\bibfnamefont{A.}~\bibnamefont{Rostworowski}},
  \bibinfo{journal}{Phys. Rev. Lett.} \textbf{\bibinfo{volume}{107}},
  \bibinfo{pages}{031102} (\bibinfo{year}{2011}), \eprint{1104.3702}.

\bibitem[{\citenamefont{Bantilan et~al.}(2012)\citenamefont{Bantilan,
  Pretorius, and Gubser}}]{BanPreGub12}
\bibinfo{author}{\bibfnamefont{H.}~\bibnamefont{Bantilan}},
  \bibinfo{author}{\bibfnamefont{F.}~\bibnamefont{Pretorius}},
  \bibnamefont{and} \bibinfo{author}{\bibfnamefont{S.~S.}
  \bibnamefont{Gubser}}, \bibinfo{journal}{Phys. Rev.}
  \textbf{\bibinfo{volume}{D85}}, \bibinfo{pages}{084038}
  (\bibinfo{year}{2012}), \eprint{1201.2132}.

\bibitem[{\citenamefont{Cardoso et~al.}(2012)\citenamefont{Cardoso, Gualtieri,
  Herdeiro, Sperhake, Chesler et~al.}}]{CarGuaHer12}
\bibinfo{author}{\bibfnamefont{V.}~\bibnamefont{Cardoso}},
  \bibinfo{author}{\bibfnamefont{L.}~\bibnamefont{Gualtieri}},
  \bibinfo{author}{\bibfnamefont{C.}~\bibnamefont{Herdeiro}},
  \bibinfo{author}{\bibfnamefont{U.}~\bibnamefont{Sperhake}},
  \bibinfo{author}{\bibfnamefont{P.~M.} \bibnamefont{Chesler}},
  \bibnamefont{et~al.} (\bibinfo{year}{2012}), \eprint{1201.5118}.

\end{thebibliography}

\end{document}